\documentclass[aps,twocolumn,showpacs,preprintnumbers,nofootinbib,float,superscriptaddress]{revtex4-1}

\usepackage{graphicx}
\usepackage[figuresright]{rotating}
\usepackage{makecell}
\usepackage{bm,amsmath,amssymb}
\usepackage[mathscr]{eucal}
\usepackage{multirow}
\usepackage{xcolor}
\usepackage{mathrsfs}
\usepackage{mathtools}
\usepackage{slashed}
\usepackage{graphics}
\usepackage{graphicx}
\usepackage{subfigure}
\usepackage{dsfont}
\usepackage{longtable}
\usepackage{bbm} 
\usepackage{float}
\usepackage{tcolorbox}
\usepackage{lipsum}
\usepackage{cancel}
\usepackage{enumerate}
\usepackage{diagbox}
\usepackage{ulem}
\usepackage[breaklinks,colorlinks,citecolor=citcolor,urlcolor=blue,linkcolor=lcolor]{hyperref}
\definecolor{lcolor}{rgb}{0.5,0,0}
\definecolor{citcolor}{rgb}{0,0.3,0.0}

\newcommand{\hats}{\hat{s}}
\newcommand{\hatt}{\hat{t}}
\newcommand{\hatu}{\hat{u}}
\newcommand{\Hcal}{\mathcal{H}}
\newcommand{\Ocal}{\mathcal{O}}
\newcommand{\Mcal}{\mathcal{M}}

\newcommand{\Acal}{\mathcal{A}}
\newcommand{\Ncal}{\mathcal{N}}

\newcommand{\Scal}{\mathcal{S}}

\newcommand{\Pcal}{\mathcal{P}}

\newcommand{\etU}[1]{\boldsymbol{\epsilon}_{\perp}^{#1}}

\newcommand{\ltU}[1]{\boldsymbol{l}_{\perp}^{#1}}

\newcommand{\LtL}[1]{\boldsymbol{L}_{\perp #1}}
\newcommand{\LtU}[1]{\boldsymbol{L}_{\perp}^{#1}}

\newcommand{\ltCL}[1]{\boldsymbol{l}'_{\perp #1}}
\newcommand{\ltCU}[1]{\boldsymbol{l}_{\perp}^{'#1}}

\newcommand{\ytCL}[1]{\boldsymbol{y}'_{\perp #1}}

\newcommand{\ytU}[1]{\boldsymbol{y}_{\perp}^{#1}}
\newcommand{\ytL}[1]{\boldsymbol{y}_{\perp #1}}

\newcommand{\ytone}{\boldsymbol{y}_{1\perp}}
\newcommand{\yttwo}{\boldsymbol{y}_{2\perp}}

\newcommand{\ptU}[1]{\boldsymbol{p}_{\perp}^{#1}}

\newcommand{\gammatL}[1]{\boldsymbol{\gamma}_{\perp #1}}
\newcommand{\gammatU}[1]{\boldsymbol{\gamma}_{\perp}^{#1}}

\newcommand{\vect}[1]{\boldsymbol{#1}_{\perp}}

\newcommand{\pt}{\vect{p}}

\newcommand{\lt}{\vect{l}}
\newcommand{\ellt}{\vect{\ell}}
\newcommand{\ltC}{\vect{l}'}

\newcommand{\xt}{\vect{x}}
\newcommand{\yt}{\vect{y}}
\newcommand{\ytC}{\vect{y'}}

\newcommand{\Lt}{\vect{L}}

\newcommand{\pgammat}{\boldsymbol{p_{\gamma\perp}}}

\newcommand{\pgammatL}[1]{\boldsymbol{p}_{\gamma\perp #1}}
\newcommand{\pgammatU}[1]{\boldsymbol{p}_{\gamma\perp}^{#1}}

\newcommand{\GammatLU}[2]{\boldsymbol{\Gamma}_{\perp #1}^{#2}}

\newcommand{\GammatCLU}[2]{\boldsymbol{\overline{\Gamma}}_{\perp #1}^{#2}}

\newcommand{\Nc}{N_c}
\newcommand{\der}{\mathrm{d}}

\newcommand{\Tr}{\mathrm{Tr}}

\renewcommand{\arraystretch}{1.2}

\begin{document}

\title{Color Glass Condensate meets High Twist Expansion}

\author{Yu Fu}
\affiliation{Department of Physics, Duke University,
Durham, NC 27708, USA}
\affiliation{Key Laboratory of Quark and Lepton Physics (MOE) \& Institute of Particle Physics, Central China Normal University, Wuhan 430079, China}

\author{Zhong-Bo Kang}
\affiliation{Department of Physics and Astronomy, University of California, Los Angeles, CA 90095, USA}
\affiliation{Mani L. Bhaumik Institute for Theoretical Physics, University of California, Los Angeles, CA 90095, USA}
\affiliation{Center for Frontiers in Nuclear Science, Stony Brook University, Stony Brook, NY 11794, USA}

\author{Farid Salazar}
\affiliation{Institute for Nuclear Theory, University of Washington, Seattle, WA 98195, USA}
\affiliation{Nuclear Science Division, Lawrence Berkeley National Laboratory, Berkeley, CA 94720, USA}
\affiliation{Physics Department, University of California, Berkeley, CA 94720, USA}
\affiliation{Department of Physics and Astronomy, University of California, Los Angeles, CA 90095, USA}
\affiliation{Mani L. Bhaumik Institute for Theoretical Physics, University of California, Los Angeles, CA 90095, USA}

\author{Xin-Nian Wang}
\affiliation{Key Laboratory of Quark and Lepton Physics (MOE) \& Institute of Particle Physics, Central China Normal University, Wuhan 430079, China}
\affiliation{Nuclear Science Division, Lawrence Berkeley National Laboratory, Berkeley, CA 94720, USA}

\author{Hongxi Xing}
\affiliation{State Key Laboratory of Nuclear Physics and Technology, Institute of Quantum Matter, South China Normal University, Guangzhou 510006, China}
\affiliation{Guangdong Basic Research Center of Excellence for Structure and Fundamental Interactions of Matter, Guangdong Provincial Key Laboratory of Nuclear Science, Guangzhou 510006, China}
\affiliation{Southern Center for Nuclear-Science Theory (SCNT), Institute of Modern Physics, Chinese Academy of Sciences, Huizhou 516000, China}

\begin{abstract}

We establish the correspondence between two well-known frameworks for QCD multiple scattering in nuclear media: the Color Glass Condensate (CGC) and the High-Twist (HT) expansion formalism. We argue that a consistent matching between both frameworks, in their common domain of validity, is achieved by incorporating the sub-eikonal longitudinal momentum phase in the CGC formalism, which mediates the transition between coherent and incoherent scattering. We perform a detailed calculation and analysis of direct photon production in proton-nucleus scattering as a concrete example to establish the matching between HT and CGC up to twist-4, including initial- and final-state interactions, as well as their interferences. The techniques developed in this work can be adapted to other processes in electron-nucleus and proton-nucleus collisions, and they provide a potential avenue for a unified picture of dilute-dense dynamics in nuclear media.

\end{abstract}
\maketitle


\section{Introduction}

In recent years, significant experimental progress has been made in studying particle production within the nuclear media, revealing many intriguing nuclear-dependent effects \cite{PHENIX:2004nzn,PHENIX:2019gix,ALICE:2021est,PHOBOS:2004fsu,BRAHMS:2004xry,ATLAS:2016xpn,PHENIX:2017caf,LHCb:2021vww,LHCb:2022rlh,ALICE:2018vuu,STAR:2021fgw,CMS:2023snh,Braidot:2010zh,STAR:2006dgg,PHENIX:2011puq}. Multiple parton scattering has been crucial in understanding these novel effects. Therefore, to better extract the properties of QCD matter from experimental measurements, it is essential to clarify the differences and connections among various QCD theoretical frameworks used to study multiple parton scattering in nuclear media. Two widely used theoretical frameworks are the Color Glass Condensate (CGC) effective theory and the collinear factorization at high-twist (HT) or high-twist expansion formalism.  This paper aims to elucidate their differences and establish a connection between these two formalisms.

The CGC effective field theory and the high-twist expansion formalism are applicable depending on the kinematics of the scattering process. Two variables classify the kinematic regions in scattering processes: the momentum fraction $x$ carried by the parton with respect to the nucleon, and the hard scale $Q^2$ of the partonic scattering process. These variables determine the probed longitudinal momentum and the resolution scale. Accordingly, the ``phase diagram'' of parton density can be expressed as a function of  $x$ and $Q^2$. Depending on whether the nuclear medium is dilute or dense, these two aforementioned theoretical frameworks have been extensively used to describe the dynamics in different regions of the phase diagram of parton density.

In the dilute region where $x\sim \mathcal{O}(1)$ or the intermediate region where $x \lesssim \mathcal{O}(1)$, the energetic parton interacts with the medium incoherently. For a given hard scale $Q^2$, the high-twist expansion is applicable, and the hard process can be factorized according to the collinear factorization formalism at different levels of twist. In particular, in the dilute region where $x\sim \mathcal{O}(1)$, the leading twist collinear factorization~\cite{Collins:1989gx} has been very successful and set as a benchmark theory for high-energy physics. In the relatively dense region where $x \lesssim \mathcal{O}(1)$, the high-twist (HT) expansion approach based on the QCD collinear factorization theorem~\cite{Qiu:1990xxa,Qiu:1990xy} provides a robust framework to describe multiple scatterings in nuclear medium order by order in the number of scatterings, which appear as power corrections $1/Q^n$ to the leading twist cross-section. The multiple scattering processes generally involve high-twist multi-parton correlations in analogy to the leading twist parton distribution operators. Although they are suppressed by powers of $1/Q^2$, higher twist corrections are enhanced by a factor of the large nuclear radius $\sim A^{1/3}$. The collinear factorization-based high twist approach has been successfully applied to calculate the incoherent multiple scattering at the next-to-leading power~\cite{Kang:2013ufa,Kang:2014hha}, and to the study of jet quenching in cold nuclei~\cite{Guo:2000nz,Wang:2001ifa}.

In the dense region, $x \sim Q^2/s \to 0$, the gluon density proliferates resulting in a high gluon occupation number. The rapid growth due to parton splitting carrying a small momentum fraction $x$ is eventually tamed by recombination effects, resulting in the saturation of partons \cite{Gribov:1984tu,Mueller:1985wy}. An effective theory for this saturated regime of nuclear matter is the CGC \cite{McLerran:1993ni,McLerran:1993ka,McLerran:1994vd,Ayala:1995kg,Ayala:1995hx,Gelis:2010nm,Kovchegov:2012mbw}. In this formalism, the degrees of freedom are separated according to the momentum fraction $x$ they carry. Large-$x$ partons are treated as stochastic, static, and localized color sources, which generate a color current. On the other hand, small-$x$ partons are treated as dynamical gluon fields obtained as solutions to the Yang-Mills equations in the presence of the current generated by large-$x$ partons. The interaction of the energetic parton with the background gluon field of the nucleus is encoded in a light-like Wilson line which resums multiple eikonal scatterings to all orders. The eikonal approximation implies that the interaction of the probe with the nucleus is coherent. The saturated parton density provides a new dimensionful energy-dependent and nuclear size-dependent transverse momentum scale $Q_s$. Thanks to this emergent scale, the CGC naturally follows a transverse momentum-dependent factorization. The CGC has been successfully applied to calculate various observables across different collider experiments \cite{Kharzeev:2004yx,Marquet:2007vb,Lappi:2012nh,Albacete:2018ruq,Zheng:2014vka,Lappi:2013zma,JalilianMarian:2012bd,Ducloue:2017kkq,Ducloue:2015gfa,Shi:2021hwx,Tong:2022zwp,Benic:2022ixp,Al-Mashad:2022zbq,Liu:2022ijp,Liu:2023aqb,Caucal:2023fsf,Morreale:2021pnn}.

One of the main differences between the HT expansion and the CGC formalism is their respective QCD evolution equations. In the dilute limit, the HT expansion coincides with conventional collinear factorization, thus resuming large logs of $Q^2$ in parton distribution functions via Dokshitzer-Gribov-Lipatov-Altarelli-Parisi (DGLAP) evolution equation\cite{Altarelli:1977zs,Gribov:1972ri,Gribov:1972rt,Dokshitzer:1977sg}. Similarly, at higher twists, the HT formalism encodes multiple scattering of the medium in the multi-parton quantum correlation functions satisfying the DGLAP-type evolution \cite{Kang:2013raa,Kang:2014ela,Kang:2016ron}. In contrast, in the CGC formalism, multiple scattering with the nuclear medium is encoded in correlators of light-like Wilson lines, which satisfy the Jalilian-Marian-Iancu-McLerran-Weigert-Leonidov-Kovner/Balitsky-Kovchegov nonlinear evolution \cite{Balitsky:1995ub,Kovchegov:1999ua,Jalilian-Marian:1997qno,Jalilian-Marian:1997jhx,Jalilian-Marian:1997ubg,Kovner:2000pt,Iancu:2000hn,Iancu:2001ad, Ferreiro:2001qy} resumming all the leading logs in $1/x$. These two approaches also treat the multiple scattering differently: additional soft rescatterings are considered order by order in a power series and the hard scattering in the twist expansion approach. In the CGC, which relies on the eikonal approximation, all scatterings are resummed and exponentiated into the light-like Wilson line. 

Across the last two decades, several suggestive signatures of gluon saturation have been observed in collider experiments (for a recent review see \cite{Morreale:2021pnn}). Most recently, measurements of dihadron correlations by STAR \cite{STAR:2021fgw} and $J/\psi$ photoproduction by CMS \cite{CMS:2023snh} suggest possible signatures of nonlinear gluon dynamics arising at high parton densities. Nevertheless, a systematic framework that allows for the transition between dilute and dense regimes of QCD in nuclear media is missing. Establishing the correspondence between different underlying theoretical frameworks for multiple scattering is key to interpreting the experimental data properly. To this end there have been various efforts to extend the applicability of CGC from small-$x$ (dense) to large-$x$ (dilute) region which include: the sub-eikonal corrections to the parton propagators \cite{Altinoluk:2014oxa,Altinoluk:2015gia,Altinoluk:2015xuy,Agostini:2019avp,Agostini:2019hkj,Agostini:2022ctk,Altinoluk:2020oyd,Altinoluk:2021lvu, Chirilli:2018kkw,Chirilli:2021lif,Altinoluk:2022jkk}, the rapidity evolution of unintegrated gluon distributions and their interplay with collinear QCD evolution \cite{Balitsky:2015qba,Balitsky:2016dgz,Balitsky:2017flc,Balitsky:2017gis,Balitsky:2019ayf,Mukherjee:2023snp}, as well as novel semi-classical approaches \cite{Boussarie:2020fpb,Boussarie:2021wkn,Boussarie:2023xun,Jalilian-Marian:2017ttv,Jalilian-Marian:2018iui,Jalilian-Marian:2019kaf,Kovner:2023vsy}. It is also worth noting that sub-eikonal corrections are necessary to describe the physics of spin at small-$x$ \cite{Kovchegov:2015pbl,Kovchegov:2016weo,Kovchegov:2017lsr,Kovchegov:2018znm,Kovchegov:2020hgb,Cougoulic:2022gbk,Borden:2023ugd,Adamiak:2023okq,Li:2023tlw}, as well as the propagation of jets and medium-induced emissions in QCD media \cite{Zakharov:2004vm,Caron-Huot:2010qjx,Feal:2018sml,Ke:2018jem,Andres:2020vxs,Schlichting:2020lef,Mehtar-Tani:2019tvy,Mehtar-Tani:2019ygg,Barata:2020sav,Barata:2020rdn,Barata:2021wuf,Isaksen:2022pkj}.

The goal of this manuscript is to establish the correspondence between high-twist expansion formalism and CGC formalisms in their common domain of validity; namely, the low-$x$ regime with intermediate values of the hard scale $Q^2 \gtrsim Q_s^2$. To this end, we use direct photon production in $pA$ collisions as an example.  This process is the simplest process for our purpose because the final state photons are colorless and thus do not interact with the QCD medium, which tremendously reduced the complexity of the calculation. We present a complete analysis of the high-twist expansion for this process, including initial, final, and interference contributions. Interestingly, our analysis suggests that the small-$x$ limit of the result obtained in the high-twist expansion does not coincide with the twist-4 expansion of the result obtained in the CGC, demonstrating that the collinear and eikonal limits do not commute with each other. We show that demanding the consistency between the CGC formalism and the HT expansion in their common domain of validity demands the inclusion of longitudinal sub-eikonal phases in the CGC, which mediate the transition between coherent and incoherent scattering, the so-called Pomeranchuk-Migdal (LPM) interference effect \cite{Landau:1953um,Migdal:1956tc}. The main results of the present study were already summarized as a short letter in Ref.~\cite{Fu:2023jqv}. We provide a detailed derivation and discussion in this manuscript. In Sec.~\ref{sec:high-twist formalism}, we provide the general framework of collinear factorization, taking direct photon production in $pA$ collisions as an example. We introduce the kinematic variables, and briefly review the single scattering or leading twist contribution to the direct photon production in $pA$ collisions. We then provide a comprehensive description of direct photon production at the twist-4 level, providing a detailed derivation for the cross-section of the initial state scattering with the central cut. In Sec.~\ref{CGC-formalism}, we review the calculation of direct photon production in $pA$ in the CGC formalism and perform the leading twist and next-to-leading twist (twist-4) expansion. We show that the results of this naive expansion are inconsistent with the high-twist expansion unless some stringent constraints are satisfied by the twist-4 distributions. In Sec.~\ref{sec:CGC-sub-eik} we modify the CGC calculation by bringing back the sub-eikonal phases in the effective interaction with the background field and compute the contributions from single scattering and double scattering to the cross-section for direct-photon production. By taking the twist expansion of these results we prove the consistency between the high-twist formalism and the CGC framework with sub-eikonal phases. In Sec.~\ref{sec:conclusion-perspectives} we summarize our findings and discuss future work.  Our manuscript is supplemented by various appendices. Useful identities to evaluate the cross-section and to perform the twist expansion are shown in appendix \ref{app:useful-identities}. In appendix \ref{app:CGC-collinear-dist-correspondance} we show the correspondence between moments of the dipole correlator in momentum space and the collinear gluon distributions (twist-2 and twist-4). Detailed calculations of the triple-single scattering contribution are provided in \ref{app:triple-single-interference}. Lastly, in appendix \ref{app:Hard_factors} we show the complete expressions for the perturbative factors for double and single-triple contributions, and in appendix \ref{app:Hard_factors_collinear} we show their collinear expansion.


\begin{widetext}

    \begin{figure}[H]
    \centering
     \includegraphics[width=0.95\textwidth]{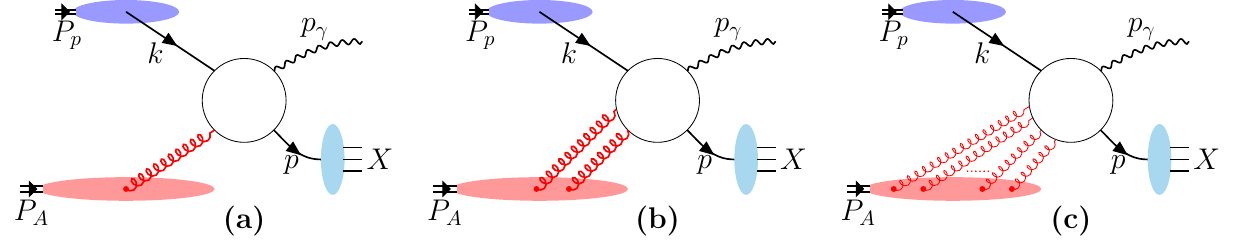}
    \caption{Schematic diagrams for single (a), double (b), and multiple (c) scatterings for direct photon production in pA collisions, the circles represent quark-gluon(s) partonic hard interaction.}
     \label{fig:pA}
\end{figure}
\end{widetext}

\section{The twist expansion formalism \label{sec:high-twist formalism}}
The main purpose of this paper is to study the multiple scattering effect in the presence of large nuclei. In particular, we consider a real scattering process of direct photon production in proton-nucleus ($pA$) collision, 
\begin{align}
    p(P_p)+A(P_A)\to\gamma (p_{\gamma}) +X,
\end{align}
where $P_p$ and $P_A$ are the momentum for the incoming proton and the average momentum per nucleon inside the nucleus respectively, and $p_{\gamma}$ is the momentum of the observed photon. In this process, the incident parton from the proton may encounter multiple scatterings with the partons from the nucleus before eventually producing the direct photon. Such multiple scatterings can be expressed as a sum of contributions from single, double, and higher multiple scatterings \cite{Qiu:2001hj},
 \begin{align}
\der\sigma_{pA\to\gamma X}=  \der\sigma_{pA\to\gamma X}^{\rm LT}+ \der\sigma_{pA\to\gamma X}^{\rm T4}+\cdots,
\end{align}
where the superscript ${\rm LT}$ indicates the leading twist (i.e. twist-2) contribution from single scattering, while ${\rm T4}$ stands for twist-4 contribution from double scattering and the single-triple interference\footnote{ The interference diagram between one gluon and two gluon exchange amplitudes (twist-3) contributes in the polarized case. We neglect this contribution as our focus in the unpolarized differential cross-section.}. The ellipsis represents the contributions beyond the twist-4 level, which involve higher multiple scattering processes. 

To show explicitly the comparison between two commonly used frameworks: the HT expansion and the CGC, we focus on photon production in proton going direction, i.e. forward region, in $pA$ collisions. In this region, the dominant contribution comes from the interaction between a quark from the proton side and gluons from the nucleus side. In the case of a single scattering as illustrated in Fig.\ref{fig:pA}(a), the quark from the proton interacts with a single gluon from the nucleus to produce the final observed direct photon. In this case, the nuclear effect is encoded into the $A$ dependence of non-perturbative nuclear parton distribution functions. On the other hand, as shown in Fig.\ref{fig:pA}(b) for double scattering, the incoming or outgoing quark may experience one additional scattering with the gluon from the nucleus. As compared to the single scattering, the hard coefficient function for the double scattering is suppressed by an additional power $1/\pgammatU{2}$, with $\pgammat$ the transverse momentum of the produced photon. However, the corresponding matrix element is enhanced by a factor of $A^{1/3}$ when the two gluons come from different nucleons inside the nucleus. Therefore, the double scattering contribution can be significant for large nuclei. 

Likewise, there could also be higher multiple scatterings between the incoming or outgoing quark with the gluons from the nucleus. Higher multiple scattering processes are illustrated in Fig.\ref{fig:pA}(c). These higher multiple scatterings are further suppressed by the hard scale $p_{\gamma\perp}^2$ compared to the leading nuclear effect from double scattering. Therefore, we will neglect these suppressed higher multiple scatterings and will focus on the double scattering in this paper.

\subsection{Single scattering contribution: leading twist}
We start this section by specifying the reference frame for the direct photon production in $pA$ collision  
\begin{align}
    P_p^\mu = \left(0, P_p^-, 0 \right)\,, \\ 
    P_A^\mu = \left(P_A^+, 0, 0 \right)\,,
\end{align}
where we neglected the mass of the nucleon, and
\begin{align}
    k^{\mu}&=(0,k^-,\vect{0})=(0,x_q P_p^-,\vect{0})\,, \nonumber \\
    p^\mu &= \left( \frac{\pt^2 }{2p^-}, p^- , \boldsymbol{p_{\perp}} \right) = \left( \frac{\pt^2 }{2 (1-\xi) k^-}, (1-\xi) k^- , \boldsymbol{p_{\perp}} \right) \,, \nonumber \\
    p_\gamma^\mu &= \left( \frac{\pgammatU{2}}{2p_{\gamma}^-}, p_{\gamma}^- , \boldsymbol{p_{\gamma\perp}} \right) = \left( \frac{\pgammatU{2}}{2 \xi k^-}, \xi k^- , \boldsymbol{p_{\gamma\perp}} \right) \,,
    \label{eq-notation}
\end{align}
where $k$ and $p$ denote the momenta for the incoming and outgoing massless quark, respectively. We introduce $\xi$, the longitudinal momentum fraction of the incoming quark carried by the final observed photon.

\begin{figure}[H]
    \centering
    \includegraphics[width=0.45\textwidth]{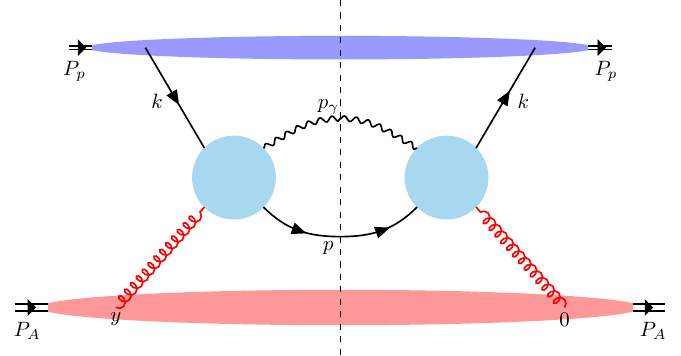}
    \caption{Cutting diagram for single scattering in direct photon production in $pA$ collisions, the blob represents the leading order partonic interaction of $q+g\to\gamma+q$ shown in Fig.\ref{fig:blob}.}
    \label{fig:T2}
\end{figure}

\begin{figure}[H]
    \centering
     \includegraphics[width=0.45\textwidth]{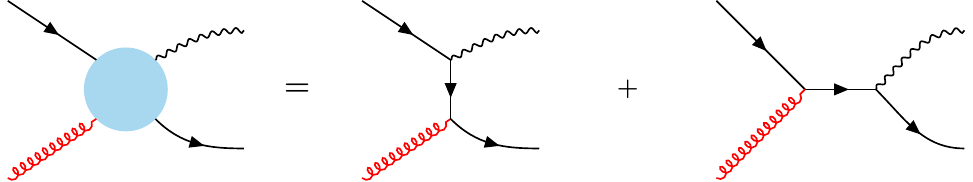}
    \caption{The representation of hard $2\to 2$ processes for quark-gluon interaction, including $``t$-$channel"$ and $``s$-$channel"$.}
     \label{fig:blob}
\end{figure}

The result for single scattering at leading order (LO), shown in Fig. \ref{fig:T2}, can be derived using the standard leading twist QCD collinear factorization \cite{Collins:1989gx}
\begin{align}
       E_{\gamma} \frac{\der\sigma^{\rm LT}_{pA\to\gamma X}}{\der ^3 \boldsymbol{p_{\gamma}}}
        =&\frac{\alpha_{em}\alpha_{s}}{s} \sum_q e_q^2\int \frac{\der x_q}{x_q}f_{q/p}(x_q)\int \frac{\der x_g}{x_g}  \nonumber\\
        \times& f_{g/A}(x_g) H_{qg\to q\gamma}(\hats,\hatt,
        \hatu)\> \delta(\hats+\hatt+\hatu),
        \label{eq:twist-2-coll-fact-stu}
\end{align}
where $\alpha_{em}$ and $\alpha_s$ stand for the fine structure constant and strong coupling constant, respectively, and $e_q$ refers to the electric charge fraction of the quark. The partonic Mandelstam variables are defined as 
\begin{align}
\hat{s}=&(x_qP_p+x_gP_A)^2,\nonumber \\
\hat{t}=&(x_qP_p- p_{\gamma})^2,\nonumber \\
\hat{u}=&(x_gP_A- p_{\gamma})^2, 
\end{align}
and $s=(P_p+P_A)^2$. The hard part at LO is given by \cite{Owens:1986mp}
\begin{align}
     H_{qg\to q\gamma}(\hats,\hatt,
        \hatu)= \frac{1}{\Nc}\left(-\frac{\hatt}{\hats}-\frac{\hats}{\hatt}\right),
    \label{eq::H_qg}
\end{align}
where $N_c$ is the number of colors, $f_{q/p}(x_q)$ is the standard leading twist quark distribution function inside the proton, and $f_{g/A}(x_g)$ is the leading twist nuclear gluon distribution function defined as
\begin{align}
     f_{g/A}(x)=\frac{1}{x P_A^+ } \int\frac{\der  y^-}{2\pi}e^{-ix P^+_A y^-}\langle P_A|F^{+\alpha}(0)F^+_{\alpha}(y^-)|P_A\rangle.
\end{align}
Integrating over $x_g$ in Eq.~(\ref{eq:twist-2-coll-fact-stu}) with the help of $\delta(\hats+\hatt+\hatu)$, one can rewrite Eq.~(\ref{eq:twist-2-coll-fact-stu}) as 
\begin{align}
       E_{\gamma} \frac{\der \sigma^{\rm LT}_{pA\to\gamma X}}{\der ^3 \boldsymbol{p_{\gamma}}}
        =&\alpha_{em}\alpha_{s}\frac{1}{N_c}\sum_qe_q^2\int \der  x_q f_{q/p}(x_q) x f_{g/A}(x) \nonumber \\
        &\times\frac{\xi^2\left[1+(1-\xi)^2\right]}{\pgammatU{4}},
        \label{eq:twist-2-coll-fact-xi}
\end{align}
where 
\begin{align}
 x=\frac{\pgammatU{2}}{\xi(1-\xi)x_q s} \,.
 \label{eq::x-definition}
\end{align}

\subsection{Double scattering contribution: twist-4}
\begin{figure}[]
    \centering  \includegraphics[width=0.45\textwidth]{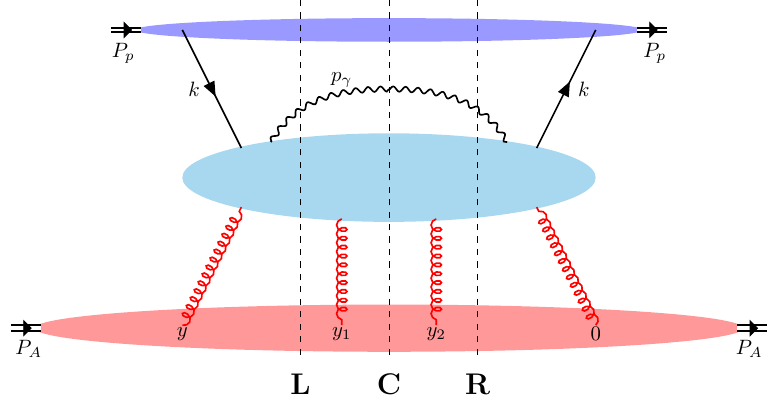}
    \caption{Schematic diagram representing the cross-section of direct photon production in $pA$ collisions at twist-4. The central dashed line ``C" represents the contribution from pure double scattering, while dashed lines ``L" and ``R" represent the interference between single and triple scatterings.}
    \label{fig:T4}
\end{figure}

In the presence of a large nucleus, multiple scatterings become important, where the leading contribution comes from twist-4. As clarified before, our main focus is on the forward photon production in $pA$ collisions, which is dominated by the interactions between a quark from the proton and gluons from the nucleus. Shown in Fig.~\ref{fig:T4} is a schematic diagram representing such a particular set of twist-4 contributions involving four-gluon correlation in the nucleus. 
Fig.\ref{fig:T4} contains both symmetric and asymmetric cuts. The symmetric (central) cut labeled with ``C" corresponds to double scattering, while the left (right) cut labeled with ``L" (``R") represents the interference between single and triple scatterings. On the other hand, for processes with twist-4 contribution, the quark from the proton undergoes the soft scattering before and/or after the hard scattering with a gluon from the nuclear medium, which allows us to categorize the processes with twist-4 contribution into initial state scattering, final state scattering, and initial-final (or finial-initial) interference. Based on the classification method introduced above, we present several exemplary diagrams: shown in Fig.\ref{fig:T4-inital} are the left, central, and right cuts for initial state scattering, and the Fig.\ref{fig:T4-central}, together with Fig.\ref{fig:T4-inital}(b) are the four possible diagrams for tcentral cut. Again, the $``$blob" in these diagrams represents $``q+g\to q+\gamma"$ hard processes that contain $``t$-$channel"$ and $``s$-$channel"$ as illustrated in Fig.~\ref{fig:blob}. 

\begin{widetext}

\begin{figure}[H]
    \centering
     \includegraphics[width=0.95\textwidth]{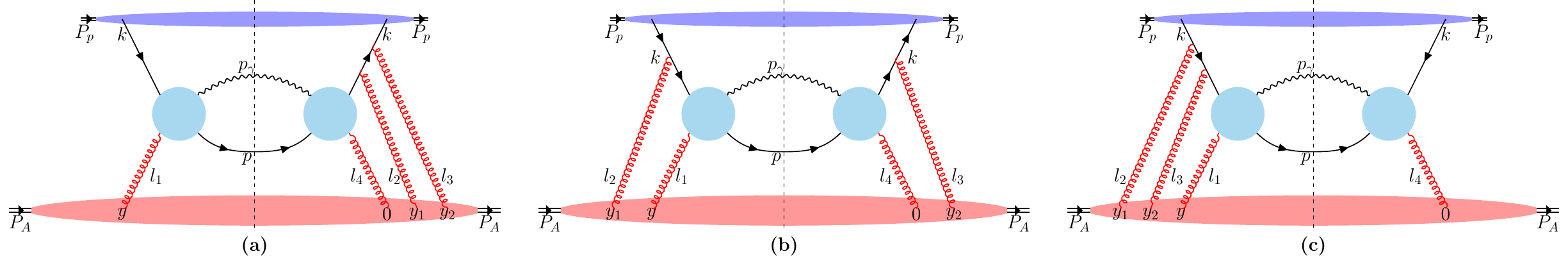}
    \caption{The diagrams with (a)left, (b)central, and (c)right cut for initial state scattering diagrams in direct photon production at twist-4 level in $pA$ collision.}
    \label{fig:T4-inital}
\end{figure}
\begin{figure}[H]
    \centering     \includegraphics[width=0.95\textwidth]{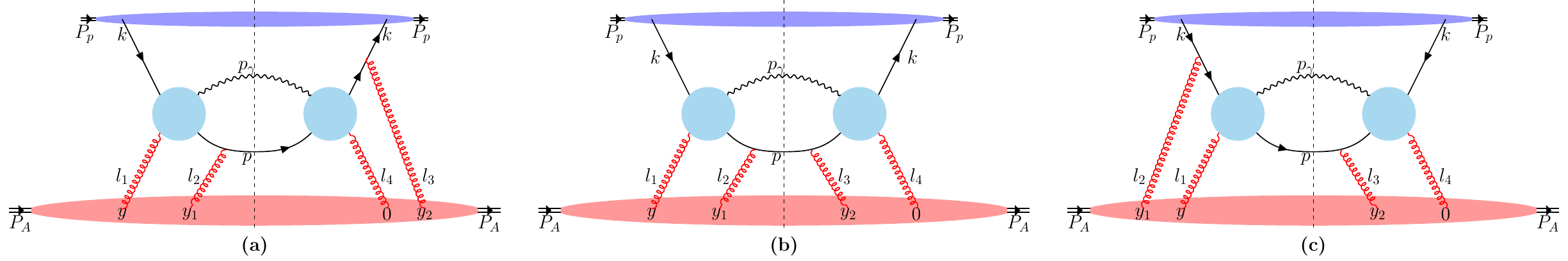}
    \caption{ The central cut diagrams for (a)final-initial interference, (b)final state scattering, and (c)initial-final interference with central cut in direct photon production at twist-4 level in $pA$ collision.}
    \label{fig:T4-central}
\end{figure}
\end{widetext}
To simplify the calculation, we follow the methodology in Ref. \cite{Luo:1994np} to use light-cone gauge $A^+=0$ for the hard gluons, and implement covariant gauge for the soft gluons, which can avoid the subtle interplay of soft-poles and zeros encountered in using light cone gauge for soft gluons \cite{Fries:2002mu}
\footnote{ It is worth noting that although all the medium fields in the case under consideration are gluon degrees of freedom, we chose different gauges for them as these gluons have different kinematical properties, either hard or soft. Our choice does not violate Ward's identity and can be verified by comparing the result with those obtained from choosing the same gauge for gluons.}. We denote the momenta of the hard gluons from the nucleus as $l_1$ and $l_4$, which are localized at $y$ and $0$, respectively. The soft gluons localized at $y_1$ and $y_2$ have momenta $l_2$ and $l_3$, respectively. The distinction between hard and soft refers to the longitudinal momentum. Taking into account the momentum conservation, these momenta can be defined as 
\begin{align}
    l_1&=x_1P_A,\\ 
    l_2&=x_2P_A+l_{T}
\end{align}
\begin{align}
    l_3&=(x_2-x_3)P_A+l_{T},\\
    l_4&=(x_1+x_3)P_A,
\end{align}
where $x_1,x_2,x_3$ are independent collinear momentum fractions and $l_T=(0,0,\vect{l})$ is the transverse momentum of soft gluons \footnote{We neglected the transverse momenta of the hard gluons since these will not contribute in the collinear limit due to our choice of light-cone gauge $A^+=0$. On the other hand, for the soft gluons in covariant gauge, we must keep track of the transverse momentum when performing the collinear expansion (cf. Eq.\,\eqref{eq:collinear-expansion-soft}).}. For later convenience, we define the following dimensionless variables
\begin{align}
    x_{c}=\frac{x}{-\hat{t}}(2p_{\gamma}\cdot l_{T}-l_{T}^2), \ \ \ \
     x_{d}=\frac{x}{\hat{s}} l_{T}^2,
\end{align}
where $x$ is defined in Eq. (\ref{eq::x-definition}). With this setup, one can write down the general expression for the cross-section of direct photon production
\begin{widetext}
\begin{equation}
\begin{split}
        \sigma^{\rm T4}_{pA\to\gamma q} =& \sum_{q} e_q^2
        \int \der x_q f_{q/p}(x_q) \frac{1}{2x_qs}\int\frac{\der ^4 p_{\gamma}}{(2\pi)^4}(2\pi)\delta(p_{\gamma}^2) \int\frac{\der ^4 p}{(2\pi)^4}(2\pi)\delta(p^2)\  (2\pi)^4\delta^4(P_{in}-p_{\gamma}-p)
        \int\frac{\der ^4 l_1}{(2\pi)^4}\frac{\der ^4 l_2}{(2\pi)^4}\frac{\der ^4 l_3}{(2\pi)^4}
        \\
        &\times\int \der ^4 y \der ^4 y_1 \der ^4 y_2 \ e^{il_1\cdot y} e^{il_2\cdot y_1} e^{il_3\cdot y_2} \langle P_A|A^{\alpha}(0) A^{\rho}(y_2) A^{\sigma}(y_1) A^{\beta}(y)|P_A\rangle \  \hat{H}_{\alpha\rho\sigma\beta}(\{x_i\},l_{T}),
\end{split}
\end{equation}
\end{widetext}
where $\frac{1}{2x_q s}=\frac{1}{2(k+P_A)^2}$ is the flux factor between the incoming quark and nucleus, and the $\hat{H}_{\alpha\rho\sigma\beta}(\{x_i\},l_T)$ represents the partonic hard part of the subprocess $q+gg\to \gamma+q$, and $P_{in}=k+l_1+l_2, k+l_1, k+l_4$ for central, left, right cut, respectively. We also denote $\{x_i\} = x_1, x_2, x_3$, and the subsequent text follows the same convention.

\begin{figure}[H]
    \centering
     \includegraphics[width=0.45\textwidth]{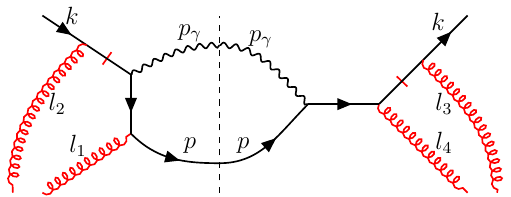}
    \caption{The partonic diagram with the interference between $s-$ and $t- channel$ in the initial state scattering processes. Short bars indicate the so-called ``pole” propagators.}
    \label{fig:st}
\end{figure}

In the following, we take initial state scattering with the central cut, which is illustrated in Fig.\ref{fig:T4-inital}(b), as an example to provide a detailed derivation for the cross-section. Notice that the hard scattering contains the amplitudes $\mathcal{M}^t$ and $\mathcal{M}^s$, due to the $``t$-$channel"$ and $``s$-$channel"$, respectively. Therefore, the partonic scattering contribution to the cross-section is given by $(\mathcal{M}^s+\mathcal{M}^t)(\mathcal{M}^s+\mathcal{M}^t)^{\dagger}$. We illustrate one of such a partonic scattering contribution, the interference between $``t$-$channel"$ and $``s$-$channel"$ in Fig.\ref{fig:st}, 
For the initial state scattering with central cut, the on-shell condition for the unobserved quark is\\
\begin{align}
    \delta(p^2)&=\delta([k+l_1+l_2-p_{\gamma}]^2)\\ \nonumber
    &=-\frac{x}{\hat{t}}\delta(x_1+x_2-x-x_c),
\label{eq:onshell-central}
\end{align}
which can be used to integrate over one of the longitudinal momentum fractions and leads to the relation between the remaining momentum fractions, i.e., $x_1+x_2=x+x_c$. Meanwhile, among the quark propagators in the partonic scattering processes, there are ``pole" propagators (for example, the propagators marked by a short bar in Fig. \ref{fig:st}.), which will be used to perform contour integrals to fix the remaining two momentum fractions. They are given by the following expressions
\begin{align}
    \frac{1}{(k+l_2)^2+i\epsilon}&=\frac{x}{\hats}\frac{1}{x_2+x_{d}+i\epsilon},\\
     \frac{1}{(k+l_3)^2-i\epsilon}&=\frac{x}{\hats}\frac{1}{x_2-x_3+x_{d}-i\epsilon}.
     \label{eq:pole-initial-central}
\end{align}
Then the differential cross-section can be expressed as follows
\begin{widetext}
\begin{equation}
\begin{split}
       &E_{\gamma} \frac{\der \sigma^{\rm T4}_{qA\to\gamma X}}{\der ^3 \boldsymbol{p_{\gamma}}}= \sum_{q} e_q^2 \int \der x_q f_{q/p}(x_q) \frac{1}{2x_q s} \frac{1}{(2\pi)^2
       } \int \der x_1 \der x_2 \der x_3 \ e^{ix_1 P_A^+ y^-}  e^{ix_2 P_A^+ (y_1^- - y_2^-)}  e^{ix_3 P_A^+ y_2^-} (P_A^+)^3 \int \frac{ \der  y^-}{2\pi} \frac{ \der  y_1^-}{2\pi} \frac{ \der  y_2^-}{2\pi} \\
        &\times \int \der ^2 \vect{l}  \int \frac{\der ^2 \vect{y}}{(2\pi)^2}   e^{-i \vect{l}\cdot \vect{y}} 
        \langle P_A|A^{\rho}(y_2^-,\vect{0}) A^{\alpha}(0^-) A^{\beta}(y^-) A^{\sigma}(y_1^-,\vect{y}) |P_A\rangle \  \hat{H}_{\rho\alpha\beta\sigma}(\{x_i\},l_T) \ \frac{x}{-\hatt}\delta(x_1+x_2-x-x_{c}).
        \label{eq-t42}
\end{split}
\end{equation}
Here, the partonic hard part of the subprocess is denoted as 
\begin{align}
    \hat{H}_{\rho\alpha\beta\sigma}
    &=g^4C \frac{x}{\hats}\frac{1}{x_2+x_{d}+i\epsilon}  \frac{x}{\hats}\frac{1}{x_2-x_3+x_{d}-i\epsilon} (\hat{H}_{\rho\alpha\beta\sigma}^{ss}+\hat{H}_{\rho\alpha\beta\sigma}^{tt}+\hat{H}_{\rho\alpha\beta\sigma}^{st}+\hat{H}_{\rho\alpha\beta\sigma}^{ts}),
\end{align}
 where $\hat{H}_{\rho\alpha\beta\sigma}^{ss}$, $\hat{H}_{\rho\alpha\beta\sigma}^{tt}$, $\hat{H}_{\rho\alpha\beta\sigma}^{st}$, $\hat{H}_{\rho\alpha\beta\sigma}^{ts}$  are hard scattering contribution due to $``s/t$-$channel"$ and their interference. $C$ is the color factor given by $C=\frac{1}{N_c(N_c-1)^2}\Tr[t^a t^b t^b t^a]=\frac{1}{4N_c^2}$, with $t^{a}$ and $t^{b}$ the generators of SU(3) in the fundamental representation. For example, the interference shown in Fig.\ref{fig:st} reads
\begin{align}
    \hat{H}_{\rho\alpha\beta\sigma}^{ts}&= \frac{1}{4}\Tr[\slashed{k}\gamma_{\rho}(\slashed{k}+\slashed{l}_3)\gamma_{\alpha} (\slashed{k}+\slashed{l}_3+\slashed{l}_4) \gamma_{\mu} \slashed{P}_A \gamma_{\beta} (\slashed{k}+\slashed{l}_2-\slashed{p}_{\gamma}) \gamma_{\nu}  (\slashed{k}+\slashed{l}_2)  \gamma_{\sigma}]\frac{1}{(k+l_2-p_{\gamma})^2} \frac{1}{(k+l_3+l_4)^2}(-g^{\mu\nu}) \,.
\end{align}
\end{widetext}
Notice that Eq. (\ref{eq-t42}) is not well defined, and one needs to convert the gluon field $A$ to field strength $F$. Regarding the hard gluons that locate at $y$ and $0$, we perform integration by part to rewrite the gluon fields $A^{\alpha}(0^-)$ and $A^{\beta}(y^-)$ as $\frac{\partial^+A^{\alpha}(0^-) }{-i(x_1+x_3)P_A^+} $ and $\frac{ \partial^+ A^{\beta}(y^-) }{ix_1P_A^+} $. Under light-cone gauge for hard gluons $A^+(0^-)=A^+(y^-)=0$, the dominant components of the gauge fields are their transverse components. For soft gluons that locate at $y_1$ and $y_2$, we use covariant gauge and the leading contribution from the gluon field is $A^{\rho}A^{\sigma}\approx (\frac{A^+}{P_A^+})^2P_A^{\rho}P_A^{\sigma}$ \cite{Luo:1993ui}. Therefore, we have approximately
\begin{align}
   &\langle P_A|F^{+\omega}(0^-) A^{\rho}(y_2^-,\vect{0}) A^{\sigma}(y_1^-,\vect{y}) F^+_{\ \ \omega}(y^-)  |P_A\rangle \approx\\ \nonumber
   & \frac{P^{\rho}P^{\sigma}}{(P_A^+)^2} \langle P_A|F^{+\omega}(0^-) A^{+}(y_2^-,\vect{0}) A^{+}(y_1^-,\vect{y}) F^+_{\ \ \omega}(y^-)  |P_A\rangle,
\end{align}
\begin{widetext}
which yields
\begin{equation}
\begin{split}       
       E_{\gamma} \frac{\der \sigma^{\rm T4}_{qA\to\gamma X}}{\der ^3 \boldsymbol{p_{\gamma}}}
         =&\sum_{q} e_q^2  \int \der x_q f_{q/p}(x_q) \frac{1}{2x_q s} \frac{1}{-\hat{t}}\frac{1}{(2\pi)^2
       } \int \der x_1 \der x_2 \der x_3 \ e^{ix_1 P_A^+ y^-}  e^{ix_2 P_A^+ (y_1^- - y_2^-)}  e^{ix_3 P_A^+ y_2^-} \frac{1}{x P_A^+} \int \frac{ \der  y^-}{2\pi} \frac{ \der  y_1^-}{2\pi} \frac{ \der  y_2^-}{2\pi}\\
        & \times \int \der ^2 \vect{l}  \int \frac{\der ^2 \vect{y}}{(2\pi)^2}   e^{-i \vect{l}\cdot \vect{y}} 
         \langle P_A|F^{+\omega}(0^-) A^{+}(y_2^-,\vect{0}) A^{+}(y_1^-,\vect{y}) F^+_{\ \ \omega}(y^-)  |P_A\rangle    \ \overline{H}(\{x_i\},l_{T}),
         \label{eq:twist-4-before-coll-expansion}
\end{split}
\end{equation}
\end{widetext}
where
\begin{align}
&\overline{H}(\{x_i\},l_{T})\equiv  \frac{x^2}{x_1(x_1+x_3)} \frac{-g^{\alpha\beta}}{2} P_A^{\rho}P_A^{\sigma} \nonumber \\
& \times \hat{H}_{\rho\alpha\beta\sigma}(\{x_i\},l_{T})\  \delta(x_1+x_2-x-x_{c}).
\end{align}
Expanding the partonic hard part $\overline{H}(\{x_i\},l_{T})$ at $l_T=0$, we obtain
\begin{align}
    &\overline{H}(\{x_i\},l_{T}) \nonumber \\ 
    =&\overline{H}(l_{T}=0)+\frac{\partial \overline{H}}{\partial l_{T}^{\lambda}}\Bigg{|}_{l_T=0}l_{T}^{\lambda}+\frac{1}{2}\frac{\partial^2 \overline{H}}{\partial l_{T}^{\lambda}\partial l_{T}^{\tau}}\Bigg{|}_{l_T=0}l_{T}^{\lambda}l_{T}^{\tau}+\cdots.
    \label{eq:collinear-expansion-soft}
\end{align}
The term $\overline{H}(l_T=0)$ corresponds to the hard matrix element $H_{q+g \to q +\gamma}$ in Eq.\,\eqref{eq::H_qg}. After inserting the collinear expansion Eq.\,\eqref{eq:collinear-expansion-soft} into Eq.\,\eqref{eq:twist-4-before-coll-expansion}, the first term, proportional to $\overline{H}(l_T=0)$, will account for the gauge link of the twist-2 gluon distribution as has been proved in Ref. [115]. The second term vanishes after integrating over $l_T$. The third term will give a finite contribution to the multiple scattering process. Taking advantage of this collinear expansion and integrating over $l_T$, we obtain
\begin{align}      
       &E_{\gamma} \frac{\der \sigma^{\rm T4}_{qA\to\gamma X}}{\der ^3 \boldsymbol{p_{\gamma}}}
         = \int \der x_q f_{q/p}(x_q) \frac{1}{2x_q s} \frac{1}{-\hat{t}}\frac{1}{(2\pi)^2
       } \nonumber \\
       &\times \int \der x_1 \der x_2 \der x_3  T(\{x_i\}) \frac{-g^{\lambda\tau}}{2}    \ \frac{1}{2}\frac{\partial^2 \overline{H}}{\partial l_{T}^{\lambda}\partial l_{T}^{\tau}}\Bigg{|}_{l_T=0}
       \label{eq:expansion_form},
\end{align}
where $T(\{x_i\})$ is a twist-4 four-gluon correlation function defined as
\begin{align}
   & T(\{x_i\})=\int \frac{d y^-}{2\pi} \frac{d y_1^-}{2\pi} \frac{d y_2^-}{2\pi}  e^{ix_1P^+_A y^-}  e^{ix_2P^+_A (y_1^- - y_2^-)}  e^{ix_3 P^+_A y_2^-}
      \nonumber  \\
        &\times \frac{1}{x P_A^+ }  \Big\langle P_A|F^{+\alpha}(0^-) F^{+\beta}(y_2^-) F^{+}_{\ \ \beta}(y_1^-) F^+_{\ \  \alpha}(y^-) |P_A\Big\rangle.
\end{align}
Before doing the collinear expansion in Eq. (\ref{eq:expansion_form}), we first integrate over the gluon momentum fractions $x_1, x_2,x_3$ with the help of on-shell condition for unobserved quark Eq. (\ref{eq:onshell-central}) and ``pole" propagators Eq. (\ref{eq:pole-initial-central}),
which leads to the time order functions, i.e. $ \theta(y^--y_1^-)\theta(-y^-_2)$, and fixes the independent momentum fractions as follows,
\begin{align}
    x_1=x+x_c+x_d,\,\,\,\,\,\,\,\,x_2=-x_d,\,\,\,\,\,\,\,\,x_3=0.
\end{align}
\begin{widetext}
Then we can define the relevant twist-4 correlation function for initial state scattering with the central cut as
\begin{align}
    T_{C,I}(x_1,x_2,x_3) 
    =& \int \frac{ \der  y^-}{2\pi} \frac{ \der  y_1^- \der  y_2^-}{2\pi}  e^{ix_1 P^+_A y^-}  e^{ix_2 P^+_A (y_1^- - y_2^-)} e^{ix_3 P^+_A y_2^-}  \nonumber \\
     &\times\frac{1}{P_A^+}\langle P_A|F^{+\omega}(0^-) F^{+\kappa}(y_2^-) F^{+}_{\ \ \kappa}(y_1^-) F^+_{\ \ \omega}(y^-)  |P_A\rangle  \theta(y^--y_1^-)\theta(-y^-_2).
     \label{eq:T4-initial-state-distr}
\end{align}
Now we perform the collinear expansion and arrive at
\begin{equation}
\begin{split}       
       E_{\gamma} \frac{\der \sigma^{\rm T4}_{qA\to\gamma X}}{\der ^3 \boldsymbol{p_{\gamma}}}
         =&\sum_{q} e_q^2  \int \der x_q f_{q/p}(x_q)\frac{4\pi^2\alpha_{s}^2\alpha_{em}}{\Nc^2} \frac{\left[1+(1-\xi)^2\right]}{\pgammatU{6}} \Big[\xi^4x^2  \frac{\partial^2  T_{C,I}(x_1,x_2,x_3)}{\partial x_1^2 }\\
    &-3\xi^4x\frac{\partial  
       T_{C,I}(x_1,x_2,x_3)}{\partial x_1}+(1-\xi)\xi^3 x \frac{\partial 
       T_{C,I}(x_1,x_2,x_3)}{\partial x_2} +4\xi^4T_{C,I}(x_1,x_2,x_3) \Big]_{x_1=x,x_2=x_3=0}.
\end{split}
\label{eq:initial-central}
\end{equation}
Following the same procedure, we can calculate the differential cross-section for other types of scattering. The complete final result at twist-4 can be summarized as the following compact form
\begin{align}
E_{\gamma}\frac{\der^3\sigma^{\rm T4}_{qA\to\gamma X}}{\der^3 \boldsymbol{p_{\gamma}}} 
=&\sum_{q} e_q^2  \int \der x_q f_{q/p}(x_q)  \frac{4\pi^2e_q^2\alpha_{em}\alpha_{s}^2}{N_c^2} \frac{\left[1 + (1-\xi)^2 \right]}{\pgammatU{6}}  \Big[ \mathcal{D}_{X} T_{X}(x_1,x_2,x_3) \Big]_{x_1=x,x_2=x_3=0},
\label{eq::full-T4}
\end{align}
with
\begin{equation}
    T_{X}(x_1,x_2,x_3)= \int\frac{ \der  y^- \der  y_1^-}{2\pi} \frac{ \der  y_2^-}{2\pi} e^{ix_1P_A^+ y^-}  e^{ix_2P_A^+ (y_1^- - y_2^-)}  e^{ix_3 P_A^+ y_2^-}
         \frac{1}{ P_A^+ }  F^4_X \Theta_X .
     \label{eq::TX}
\end{equation}
\end{widetext}
Here, we use $T_{X=A,B}$ with $A$ and $B$ stand for different cut and initial/final state multiple scattering, respectively. In particular, $A=C,L,R$ stands for central-cut (double scattering), left-cut (single-triple interference), and right-cut (triple-single interference), respectively. And $B=I,F,IF,FI$ represent multiple scattering for initial state, final state, initial-final interference, and final-initial interference, respectively. The detailed expressions for  $\mathcal{D}_{X}$, $F^4_{X}$ and $\Theta_X$ are presented in Table.\,\ref{tb::xsection-HT}.
We emphasize that Eq. (\ref{eq::full-T4}) is the first complete result of direct photon production in $pA$ collision at the twist-4 level. In previous literature \cite{Guo:1995zk,Kang:2013ufa}, only the initial state scattering was considered.  We also note that our results recover the previous result for direct photon production using high twist expansion formalism in a kinematic region where the gluon momentum fraction from the nucleus $x\sim \mathcal{O}(1)$. To demonstrate this, let us consider the final state scattering in this kinetic region. We notice that the $\mathcal{D}_{C,F}$, $\mathcal{D}_{L,F}$ and $\mathcal{D}_{R,F}$ are the same up to a minus sign. Considering the combination of three different cuts in final state scattering, the contribution to cross-section from the first-derivative term  is proportional to 
\begin{widetext}
\begin{equation}
\begin{split}
\int\frac{ \der  y^- \der  y_1^-}{2\pi} \frac{ \der  y_2^-}{2\pi} e^{ixP_A^+ y^-}  & (y_1^- - y_2^-) \langle P_A| F^{+\beta}(0^-) F^{+\alpha}(y_2^-) F^+_{\ \  \alpha}(y_1^-) F^{+}_{\ \ \beta}(y^-) |P_A\rangle  \\
&\times [\theta(y_1^- - y_2^-)\theta(y_2^- ) + \theta(y_2^- - y_1^-)\theta(y_1^- - y^-) - \theta(y_1^-  - y^-)\theta(y_2^- )].
\end{split}
\end{equation}
\begin{figure}[H]
    \centering
     \includegraphics[width=0.95\textwidth]{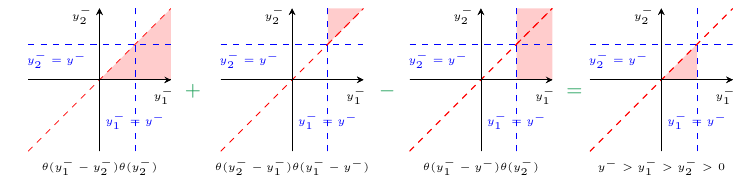}
    \caption{Diagrammatic representation for the combination of theta functions in the final state scattering.}
    \label{fig:timeorder}
\end{figure}

\begin{table*}[htbp]
\renewcommand{\arraystretch}{1.32}
\centering
\caption{$\mathcal{D}_X$ and $T_X$ for different types of scattering with different cut}
\resizebox{\textwidth}{!}{
\begin{tabular}{|c|c|l|}
\hline
\multirow{9}{*}{\makecell{Initial\\ state\\ scattering}} 
  & Central Cut 
    & $\mathcal{D}_{C,I} = \xi^4 x^2 \frac{\partial^2}{\partial x_1^2} - 3 \xi^4 x \frac{\partial}{\partial x_1} + (1-\xi)\xi^3 x \frac{\partial}{\partial x_2} + 4 \xi^4$ \\
  & 
    & $F_{C,I}^4 = \langle P_A| F^{+\alpha}(y_2^-) F^{+\beta}(0^-) F^+_{\ \beta}(y^-) F^+_{\ \alpha}(y_1^-) |P_A \rangle$ \\
  & 
    & $\Theta_{C,I} = \theta(y^- - y_1^-) \theta(-y_2^-)$ \\
\cline{2-3}
  & Left Cut 
    & $\mathcal{D}_{L,I} = -(1-\xi)\xi^3 x \frac{\partial}{\partial x_2}$ \\
  & 
    & $F_{L,I}^4 = \langle P_A| F^{+\alpha}(y_2^-) F^{+\alpha}(y_1^-) F^{+\beta}(0^-) F^+_{\ \beta}(y^-) |P_A \rangle$ \\
  & 
    & $\Theta_{L,I} = \theta(y_1^- - y_2^-) \theta(-y_1^-)$ \\
\cline{2-3}
  & Right Cut 
    & $\mathcal{D}_{R,I} = -(1-\xi)\xi^3 x \frac{\partial}{\partial x_2}$ \\
  & 
    & $F_{R,I}^4 = \langle P_A| F^{+\beta}(0^-) F^{+\beta}(y^-) F^{+\alpha}(y_2^-) F^+_{\ \alpha}(y_1^-) |P_A \rangle$ \\
  & 
    & $\Theta_{R,I} = \theta(y^- - y_2^-) \theta(y_2^- - y_1^-)$ \\
\hline

\multirow{9}{*}{\makecell{Final\\ state\\ scattering}} 
  & Central Cut 
    & $\mathcal{D}_{C,F} = \xi^4 x^2 \frac{\partial^2}{\partial x_2^2} + \xi^3 x \frac{\partial}{\partial x_2}$ \\
  & 
    & $F_{C,F}^4 = \langle P_A| F^{+\beta}(0^-) F^{+\alpha}(y_2^-) F^{+\alpha}(y_1^-) F^+_{\ \beta}(y^-) |P_A \rangle$ \\
  & 
    & $\Theta_{C,F} = \theta(y^- - y_1^-) \theta(y_2^-)$ \\
\cline{2-3}
  & Left Cut 
    & $\mathcal{D}_{L,F} = -\xi^4 x^2 \frac{\partial^2}{\partial x_2^2} - \xi^3 x \frac{\partial}{\partial x_2}$ \\
  & 
    & $F_{L,F}^4 = \langle P_A| F^{+\beta}(0^-) F^{+\alpha}(y_2^-) F^{+\alpha}(y_1^-) F^+_{\ \beta}(y^-) |P_A \rangle$ \\
  & 
    & $\Theta_{L,F} = \theta(y_1^- - y_2^-) \theta(y_2^-)$ \\
\cline{2-3}
  & Right Cut 
    & $\mathcal{D}_{R,F} = -\xi^4 x^2 \frac{\partial^2}{\partial x_2^2} - \xi^3 x \frac{\partial}{\partial x_2}$ \\
  & 
    & $F_{R,F}^4 = \langle P_A| F^{+\beta}(0^-) F^{+\alpha}(y_2^-) F^{+\alpha}(y_1^-) F^+_{\ \beta}(y^-) |P_A \rangle$ \\
  & 
    & $\Theta_{R,F} = \theta(y_2^- - y_1^-) \theta(y_1^- - y^-)$ \\
\hline

\multirow{6}{*}{\makecell{Initial-\\Final\\interference}} 
  & Central Cut 
    & $\mathcal{D}_{C,IF} = \xi^4 x^2 \left(\frac{\partial}{\partial x_1} - \frac{\partial}{\partial x_3} \right)^2 - \xi^4 x \left(\frac{\partial}{\partial x_1} - \frac{\partial}{\partial x_3} \right) + (1 - \xi)\xi^3 x \frac{\partial}{\partial x_2}$ \\
  & 
    & $F_{C,IF}^4 = \langle P_A| F^{+\beta}(0^-) F^{+\alpha}(y_2^-) F^{+\alpha}(y_1^-) F^+_{\ \beta}(y^-) |P_A \rangle$ \\
  & 
    & $\Theta_{C,IF} = \theta(y^- - y_1^-) \theta(y_2^-)$ \\
\cline{2-3}
  & Right Cut 
    & $\mathcal{D}_{R,IF} = -\xi^4 x^2 \left(\frac{\partial}{\partial x_1} - \frac{\partial}{\partial x_3} \right)^2 + \xi^4 x \left(\frac{\partial}{\partial x_1} - \frac{\partial}{\partial x_3} \right) - (1 - \xi)\xi^3 x \frac{\partial}{\partial x_2}$ \\
  & 
    & $T_{R,IF} = \langle P_A| F^{+\beta}(0^-) F^{+\alpha}(y_2^-) F^+_{\ \beta}(y^-) F^{+\alpha}(y_1^-) |P_A \rangle$ \\
  & 
    & $\Theta_{R,IF} = \theta(y_2^- - y^-) \theta(y^- - y_1^-)$ \\
\hline

\multirow{6}{*}{\makecell{Final-\\Initial\\interference}} 
  & Central Cut 
    & $\mathcal{D}_{C,FI} = \xi^4 x^2 \left( \frac{\partial}{\partial x_2} - \frac{\partial}{\partial x_3} \right)^2 + (1 - 2\xi)\xi^3 x \frac{\partial}{\partial x_2} - \xi^4 x \frac{\partial}{\partial x_3}$ \\
  & 
    & $F_{C,FI}^4 = \langle P_A| F^{+\alpha}(y_2^-) F^{+\beta}(0^-) F^{+\alpha}(y_1^-) F^+_{\ \beta}(y^-) |P_A \rangle$ \\
  & 
    & $\Theta_{C,FI} = \theta(y_1^- - y^-) \theta(-y_2^-)$ \\
\cline{2-3}
  & Left Cut 
    & $\mathcal{D}_{L,FI} = -\xi^4 x^2 \left( \frac{\partial}{\partial x_2} - \frac{\partial}{\partial x_3} \right)^2 - (1 - 2\xi)\xi^3 x \frac{\partial}{\partial x_2} + \xi^4 x \frac{\partial}{\partial x_3}$ \\
  & 
    & $F_{L,FI}^4 = \langle P_A| F^{+\alpha}(y_2^-) F^{+\beta}(0^-) F^{+\alpha}(y_1^-) F^+_{\ \beta}(y^-) |P_A \rangle$ \\
  & 
    & $\Theta_{L,FI} = \theta(y_1^-) \theta(-y_2^-)$ \\
\hline
\end{tabular}
}
\label{tb::xsection-HT}
\end{table*}

\end{widetext}
The combination of theta functions can be expressed diagrammatically in Fig. \ref{fig:timeorder}. We notice that it is equivalent to the constrain
\begin{equation}
    y^- > y_2^- > y_1^- >0.
\end{equation}
Therefore the integration $\int  \der  y_1^- \der  y_2^- $ is an ordered integral limited by the value of $y^-$. In the region where the gluon momentum fraction in the nucleus $x\sim \mathcal{O}(1)$, the rapidly oscillating exponential phase $e^{ixP_A^+ y^-}$ restricts $y^- \sim \frac{1}{xP_A^+}$ to $0$, and thereby $y_1^-$ and $y_2^-$ to $0$. The physical meaning is that the scattering points are localized, and there is no nuclear size enhancement for double scattering contribution. Such a contribution is usually referred to as a ``contact" term. Similarly, one can prove that the contributions from the final/initial interference are contact terms, which can be all neglected when compared to other nuclear-enhanced contributions in a large nucleus. In the end, the differential cross-section is dominated by the initial state scattering and can be reduced to the following compact form
\begin{widetext}
\begin{align}
E_{\gamma} \frac{\der \sigma^D_{pA\to\gamma X}}{\der ^3 \boldsymbol{p_{\gamma}}}
  = \sum_{q} e_q^2  \int \der x_q f_{q/p}(x_q)\frac{4\pi^2\alpha_{s}^2\alpha_{em}}{\Nc^2} \frac{\left[1+(1-\xi)^2\right]}{\pgammatU{6}}
\xi^4 x \Big[x^2  \frac{\partial^2 T^{I}(x)}{\partial x^2 }-x\frac{\partial  
       T^{I}(x)}{\partial x} + T^{I}(x)\Big] \,,
\end{align}
consistent with those in Refs. \cite{Guo:1995zk,Kang:2013ufa}, where the reduced initial state scattering four-gluon correlation is given by
\begin{align}
     T^{I}(x)
        =\int\frac{ \der  y^- \der  y_1^-}{2\pi} \frac{ \der  y_2^-}{2\pi} e^{ix P^+_A y^-}  
         \frac{1}{xP_A^+ }  \Big\langle P_A|F^{+\alpha}(0^-) F^{+\beta}(y_2^-) F^{+}_{\ \ \beta}(y_1^-) F^+_{\ \  \alpha}(y^-) |P_A\Big\rangle\ {\theta}(y^- - y_1^-)\theta(-y_2^-) \,.
\end{align} 
\end{widetext}


\section{The Color Glass Condensate formalism}\label{CGC-formalism}
In this section, we review the computation for the differential cross-section of direct photon production in proton-nucleus collision in the Color Glass Condensate effective field theory and within the hybrid factorization formalism \cite{Gelis:2002ki,Gelis:2002fw,Dumitru:2005gt,Ducloue:2017kkq}. We first compute the semi-inclusive production of a quark-photon pair and obtain the direct photon contribution by integrating over the phase-space of the quark and considering the limit in which the photon transverse momentum $p_{\gamma\perp}$ is much larger than the typical momentum transfer from the nucleus characterized by the saturation scale $Q_s$.

\begin{figure}[H]
    \centering
    \includegraphics[width=0.2\textwidth]{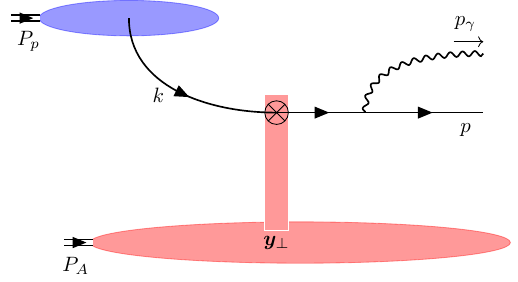}
    \includegraphics[width=0.2\textwidth]{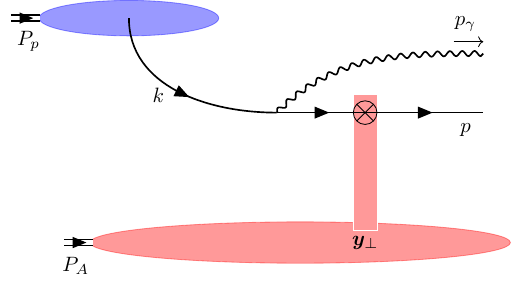}
    \caption{Leading order diagrams for quark + photon production in proton-nucleus collisions in the CGC EFT within the hybrid factorization. The incoming collinear quark from the proton undergoes multiple eikonal scattering with the strong color field of the nucleus. The red rectangle represents the multiple scattering interaction between the quark and the nucleus.}
    \label{eq:Fig-LO-diagram-CGC}
\end{figure}

\subsection{Direct photon production in proton-nucleus collision in the CGC}

At leading order in the CGC, there are two diagrams (at amplitude level) that contribute to the forward production of a photon+quark pair as shown in Fig.\,\ref{eq:Fig-LO-diagram-CGC}. The collinear quark to the proton multiply scatters off of the background field representing the small-$x$ gluon content of the nucleus. The photon can be emitted either before or after quark interaction with the nucleus. The effective vertex for the interactions of a quark with the small-$x$ background field is given by \cite{McLerran:1998nk}
\begin{align}
    \Gamma_q(l) &=  (2\pi) \delta(l^-) \gamma^-  \int \der^2 \yt e^{-i \lt \cdot \vect{y}} V(\vect{y})\,,
    \label{eq:CGC_effective_vertex}
\end{align}
where $l$ is the momentum transfer from the background field to the quark. The light-like Wilson line in the fundamental representation appearing in the effective CGC vertex is given by
\begin{align}
    V(\vect{y}) &= \Pcal \exp{ \left( ig \int_{-\infty}^\infty \der y^- A^{+}  (y^-,\vect{y})  \right)}\,,
\end{align}
where $A^{+}  (y^-,\vect{y}) = A^{+,c}  (y^-,\vect{y}) t_c$, and $t_c$ are the generators of SU(3) in the fundamental representation. $A^+$ is the back-ground gauge field of the classical small-$x$ gluons in Lorenz gauge $\partial_\mu A^\mu = 0$ generated by the fast moving nucleus. Here $\Pcal$ stands for path ordering such that the operator at $y^-=-\infty$ is in the rightmost position, while that at $y^-=+\infty$ is in the leftmost position.

The amplitude $\Mcal$ for quark+photon production in quark-nucleus scattering can be expressed as the product of the Fourier transform of light-like Wilson line $V(\yt)$ in the fundamental representation, encoding the multiple eikonal scattering of the quark with the nucleus, and a perturbative factor $\Ncal$ for the emission of the photon:
\begin{align}
\Mcal^{\lambda \sigma \sigma'}_{ij}& = e e_q \Ncal^{\lambda \sigma \sigma'}(p,p_\gamma) \nonumber \\
    & \times \int \der^2 \yt e^{-i (\pt + \pgammat) \cdot \yt} \left[ V(\yt) -\mathbbm{1} \right]_{ij} \,,
    \label{eq:amplitude_CGC}
\end{align}
where $p$ and $p_\gamma$ are the momenta of produced quark and photon, respectively. The colors of the incoming and outgoing quarks are denoted by $i$, $j$, and their helicities by $\sigma$,$\sigma'$, respectively. The polarization of the produced photon is denoted as $\lambda$. It is convenient to work in the light-cone gauge $A^-=0$ for the photon, where the perturbative factor reads:
\begin{align}
    \Ncal^{\lambda \sigma \sigma'}
    & = \left\{\frac{\left[ \xi  (\ptU{\alpha} + \pgammatU{\alpha})-\pgammatU{\alpha} \right] }{\left[ \xi  (\pt + \pgammat)- \pgammat \right]^2} + \frac{ \pgammatU{\alpha} }{\pgammatU{2}} \right\} \GammatLU{\alpha}{\lambda\sigma\sigma'}  \,. \label{eq:perturbative_qactor_CGC}
\end{align}
Here we introduced the longitudinal momentum fraction of the photon relative to the incoming quark $\xi = p_\gamma^-/k^-$ defined in Eq. (\ref{eq-notation}), and we defined the spinor structure:
\begin{align}
    \GammatLU{\alpha}{\lambda\sigma\sigma'}  &=  \bar{u}(p,\sigma)\left[  \gammatL{\alpha} \gammatL{\beta} + (1-\xi) \gammatL{\beta} \gammatL{\alpha} \right] \nonumber \\
    & \times \frac{\gamma^-}{k^-} u(k,\sigma') \etU{\lambda*,\beta}  \,.
    \label{eq:GammaStructure}
\end{align}
The differential cross-section for semi-inclusive quark-photon production is then obtained from Eq.\,\eqref{eq:qgamma-diff-Xsec-CGC} and convoluting with the parton distribution function $f_{q/p}(x_q)$ and summing over light-quarks:
\begin{align}
     E_q E_{\gamma} \frac{\der \sigma^{\rm CGC}_{pA \to q\gamma}}{\der^3 \boldsymbol{p_{\gamma}} \der^3 \boldsymbol{p} } &=\sum_{q} \frac{x_q f_{q/p}(x_q) }{8 (2\pi)^5} \nonumber \\
    & \times \frac{1}{2 N_c}\sum_{\lambda \sigma \sigma' ij} \left\langle   \overline{\Mcal}^{\lambda\sigma\sigma'}_{ij} \Mcal^{\lambda\sigma\sigma'}_{ij}   \right \rangle_x \,,
    \label{eq:A2_to_XSec_photonquark}
\end{align}
where $x_q = q^-/ P_p^-$ is the momentum fraction of the incoming quark relative to the proton which at this order is fixed by kinematics $q^- = p^- + p_\gamma^-$. The expectation value $\langle ... \rangle_x$ in Eq.\,\eqref{eq:A2_to_XSec_photonquark} represents the CGC average over different color source configurations of the background field, which represents the large-$x$ partons that have been integrated out in the CGC effective theory.

The sum of helicities and polarization can be carried out using Eq.\,\eqref{eq:identity1}, we then find
\begin{align}
     E_q E_{\gamma} \frac{\der \sigma^{\rm CGC}_{pA \to q\gamma}}{\der^3 \boldsymbol{p_{\gamma}} \der^3 \boldsymbol{p} } &= \frac{\alpha_{em}}{2\pi^2}  \sum_q e_q^2 x_q f_{q/p}(x_q) \frac{F(x,\lt)}{{(2\pi)^2}} \nonumber \\
    &   \times \frac{(1-\xi) \xi^2 \left[  1+ (1-\xi)^2 \right] \lt^2}{\left( \xi  \lt - \pgammat \right)^2 \pgammatU{2}} 
    \label{eq:quark+photon_prod} \,,
\end{align}
where $\lt=\pt + \pgammat$, and we introduced the Fourier transform of the dipole correlator (two-point function of light-like Wilson lines)
\begin{align}
    F(x,\lt)&=\int \der^2 \yt \der^2 \yt'  e^{-i \lt \cdot (\yt -\yt')} \nonumber \\
    & \times \frac{1}{N_c} \left \langle \Tr\left[ V^\dagger(\yt') V(\yt)\right] \right \rangle_x \,.
\end{align}
The saturation scale $Q_s$ is implicit in the dipole correlator and corresponds to the $|\lt|$ value at which the distribution $\lt^2F(x,\lt)$ peaks. Physically, it can be interpreted as the typical momentum transfer imparted from the nucleus to the quark-photon pair. The differential cross-section for inclusive photon production is obtained simply by integrating over the quark-phase space:
\begin{align}
    & E_{\gamma} \frac{\der^3\sigma_{pA \to \gamma X}^{\mathrm{CGC}}}{\der^3 \boldsymbol{p_{\gamma}}} 
    =\frac{\alpha_{em} }{2\pi^2} \sum_{q} e_q^2 \int \der x_q f_{q/p}(x_q)  \nonumber \\
    & \times \xi^2 \left[  1+ (1-\xi)^2 \right] \int \frac{\der^2 \lt}{(2\pi)^2}  \frac{ \lt^2 F(x,\lt)}{\left( \xi  \lt - \pgammat \right)^2 \pgammatU{2}} \,,
    \label{eq:CGC_photon_prod}
\end{align}
where we used the change of variables \footnote{When performing this change of variables one should note $p_\gamma^-$ remains fixed, so that $\der p^-/p^- = \der x_q / ((1-\xi) x_q) $. }
\begin{align}
    \frac{\der^3 \boldsymbol{p}}{E_q} = \frac{\der x_q \der^2 \pt}{(1-\xi) x_q} = \frac{\der x_q \der^2 \lt}{(1-\xi) x_q} \,. 
\end{align}
The value of $x$ is taken as in Eq.\,\eqref{eq::x-definition} \cite{Jalilian-Marian:2005tod}:
\begin{align}
    x = \frac{\pgammatU{2}}{\xi(1-\xi) x_q s} \,,
\end{align}
and the value of $x_q$ is bounded below by the relation:
\begin{align}
    x_q \xi = p_\gamma^-/P_p^- = \frac{p_{\gamma\perp} e^{\eta_{\gamma}}}{2 E_p} \,,
\end{align}
where $E_p$ is the energy of the incoming proton, $\eta_{\gamma}$ is the pseudo-rapidity of the photon, and the constrain $\xi \leq 1$.

As it stands, Eq.\,\eqref{eq:CGC_photon_prod} is not well-defined as it contains divergences when the quark and photon are collinear:  $\frac{\pt}{1-\xi} = \frac{\pgammat}{\xi}$ (photon collinear to outgoing quark), and $\pgammat=0$ (photon collinear to incoming quark). Since our focus is to study direct photon production, this divergence can be systematically isolated by subtracting the fragmentation contribution by imposing an isolation cone around the photon (see e.g. \cite{Ducloue:2017kkq}). Alternatively, we note that direct photon production is dominant in the limit in which the photon transverse momentum is hard; thus we can perform a Taylor expansion in inverse powers of $p_{\gamma\perp}$. We will follow this latter approach.

\subsection{Leading twist expansion in the eikonal limit}
In the high $p_{\gamma\perp}$ limit, one can expand the denominator $1/\left( \xi  \lt - \pgammat \right)^2 \approx 1/ \pgammatU{2}$ in Eq.\,\eqref{eq:CGC_photon_prod}. This expansion is justified when the transverse momentum of the photon is much larger than the saturation scale, and then the differential cross-section for direct photon production in $pA$ collisions reads
\begin{align}
    &E_{\gamma} \frac{\der^3\sigma_{pA \to \gamma X}^{\mathrm{CGC}}}{\der^3 \boldsymbol{p_{\gamma}}} \Bigg |_{\mathrm{LT}}
    =\frac{\alpha_{em}}{2\pi^2}  \sum_q e_q^2 \int \der x_q f_{q/p}(x_q)   \nonumber \\
    & \times \frac{\xi^2 \left[  1+ (1-\xi)^2 \right]}{\pgammatU{4}} \int \frac{\der^2 \lt}{(2\pi)^2}  \lt^2 F(x,\lt)  \,.
    \label{eq:CGC_photon_prod_LT}
\end{align}
The leading contribution in Eq.\,\eqref{eq:CGC_photon_prod_LT} has the characteristic collinear perturbative behavior $1/\pgammatU{4}$. In this limit, the photon acquires its transverse momentum by recoiling off the quark which has been integrated out. One can establish the correspondence between the leading twist expansion shown in Eq.\,\eqref{eq:CGC_photon_prod_LT} and the collinear result in Eq.\,\eqref{eq:twist-2-coll-fact-xi} by noting the relation between the collinear gluon distribution in the nucleus $xf_{g/A}(x)$ and the second momentum of the dipole distribution $F(x,\lt)$~\cite{Baier:2004tj}:
\begin{align}
    \lim_{x\to0} xf_{g/A}(x) \simeq \frac{N_c}{2\pi^2 \alpha_{s}} \int \frac{\der^2 \lt}{(2\pi)^2} \lt^2 F(x,\lt) \,.
    \label{eq:gpdf-dipole}
\end{align}
This relation is well-known and its proof is reviewed in Appendix \ref{app:CGC-collinear-dist-correspondance}. Beyond the leading order, one-loop corrections introduce renormalization scale dependence to the relation in Eq.\,\eqref{eq:gpdf-dipole} due to their respective evolution equations. The interplay between different QCD evolution equations is beyond the scope of this manuscript (for some recent attempts, see \cite{Boussarie:2021wkn,Mukherjee:2023snp}).

Thus one can cast the result in Eq.\,\eqref{eq:CGC_photon_prod_LT} as
\begin{align}
    & E_{\gamma} \frac{\der^3\sigma_{pA \to \gamma X}^{\mathrm{CGC}}}{\der^3 \boldsymbol{p_{\gamma}}} \Bigg |_{\mathrm{LT}}
    =  \frac{\alpha_{em} \alpha_{s}}{ N_c} \sum_q e_q^2 \int \der x_q f_{q/p}(x_q)   \nonumber \\
    &  \times \frac{\xi^2 \left[ 1 + (1-\xi)^2 \right]}{\pgammatU{4} } x f_{g/A}(x)  \Big |_{x \to 0}  \,.
    \label{eq:CGC_photon_prod_LT_2}
\end{align}
This result is in agreement with the leading twist result in Eq.\,\eqref{eq:twist-2-coll-fact-xi} in the strict $x \to 0$ limit, or more precisely:
\begin{align}
    e^{ix P_A^+ \Delta y} \sim 1 \to x A^{1/3} \ll 1.
\end{align}
In the next section, we shall show the necessary conditions to establish the matching at the next-to-leading twist (twist-4) will require a more stringent constraint than neglecting the phase $e^{ix P_A^+ \Delta y}$ in the corresponding twist-4 distribution.

\subsection{Next-to-leading twist expansion in the eikonal limit}
To obtain the twist-4 contribution, we consider the next term in the expansion of Eq.\,\eqref{eq:CGC_photon_prod}:
\begin{align}
    \frac{1}{\left( \xi  \lt - \pgammat \right)^2} = \frac{1}{ \pgammatU{2}} + \frac{\xi^2\lt^2}{\pgammatU{4}} + \dots \,,
\end{align}
thus the twist-4 contribution reads \cite{Fu:2023jqv}
\begin{align}
    & E_{\gamma} \frac{\der^3\sigma_{pA \to \gamma X}^{\mathrm{CGC}}}{\der^3 \boldsymbol{p_{\gamma}}} \Bigg |_{\mathrm{T4}}
     = \frac{\alpha_{em} }{2\pi^2}  \sum_q e_q^2 \int \der x_q f_{q/p}(x_q) \nonumber \\
    & \times \frac{ \xi^4 \left[  1+ (1-\xi)^2 \right]}{\pgammatU{6}} \int \frac{\der^2 \lt}{(2\pi)^2} \lt^4 F(x,\lt) \Bigg{|}_{\mathrm{T4}} \,.
    \label{eq:CGC_photon_prod_NLT}
\end{align}
In analogy to the leading twist case, we identify the fourth moment of the dipole correlator to the small-$x$ twist-4 gluon distribution:
\begin{align}
    \lim_{x \to 0} T_{gg}(x,0,0) \simeq \frac{N_c^2}{2(2\pi)^4 \alpha_{s}^2 }  \int \frac{\der^2 \lt}{(2\pi)^2} \lt^4  F(x,\lt) \Bigg{|}_{\mathrm{T4}} \,,
\end{align}
where $T_{gg} = \frac{1}{4} (T_{C,I}+T_{C,IF}+T_{C,FI}+T_{C,F})$. This identification should be understood as originating from the twist-4 contribution. In Appendix \ref{app:CGC-collinear-dist-correspondance} we show that this relation is consistent with the operator definition of the twist-4 gluon distributions in the limit $x \to 0$. Hence, we find that the twist-4 contribution to direct photon production reads
\begin{align}
   & E_{\gamma} \frac{\der^3\sigma_{pA \to \gamma X}^{\mathrm{CGC}}}{\der^3 \boldsymbol{p_{\gamma}}}  \Bigg |_{\mathrm{T4}}
=  \frac{(2\pi)^2 \alpha_{em} \alpha_{s}^2}{N_c^2} 
 \sum_q e_q^2 \int \der x_q f_{q/p}(x_q) \nonumber \\
 & \times \frac{ 4 \xi^4 [1+ (1-\xi)^2]}{\pgammatU{6}}  T_{gg}(x,0,0) \Big |_{x \to 0}  \,.
\end{align}
Provided all twist-4 distributions have the same behavior at small-$x$ ($ \lim_{x \to 0} T_{gg}(x,0,0) = T_{C,I}(x,0,0) = T_{C,IF}(x,0,0) = T_{C,FI}(x,0,0) = T_{C,F}(x,0,0)$), this result only matches the term proportional to $T_{C,I}(x,0,0)$ in Eq.\,\eqref{eq::full-T4} in the high-twist expansion, and misses the terms that contain derivatives of the twist-4 distributions. This is not very surprising as these terms arise from taking derivatives with respect to the longitudinal phases, which are neglected in the light-like Wilson lines. One could argue that powers of $x$ accompany these derivative terms and thus should be neglected in the small-$x$ limit. However, this assumes that 
\begin{align}
    \lim_{x \to 0} x \frac{\partial T_X(x_1,x_2,x_3)}{\partial x_i} \Big |_{\substack{x_1=x \\ x_2=0,x_3=0}} \ll \lim_{x \to 0} T_X (x,0,0) \,,
    \label{eq:derivative-twist-4-condition}
\end{align}
for $x_i = x_1, x_2, x_3$, which can not be validated without sufficient experimental measurements.


\section{The Color Glass Condensate beyond the eikonal approximation}

\label{sec:CGC-sub-eik}

The starting point of our 
CGC computation in the previous section was the effective vertex in Eq.\,\eqref{eq:CGC_effective_vertex} which resums multiple eikonal scattering. Let us examine this expression by expanding the light-like Wilson line in powers of the gauge field:
\begin{align}
    \Gamma_q(l) & \approx (2\pi) \delta(l^-) \gamma^- \!\! \int \der^2 \yt e^{-i \lt \cdot \yt} \nonumber \\
    & \times \int \der y^- ig A^{+}(y^-,\yt) \,,
\label{eq:effective_vertex_expansion}
\end{align}
where we have subtracted the non-scattering contribution which corresponds to setting the light-like Wilson line to unity. 

We recognize the gauge field in momentum space:
\begin{align}
    & \tilde{A}^{+}(l^+ = 0, l^-, \lt ) \nonumber \\
    & = (2\pi) \delta(l^-) \int \der^2 \yt e^{-i \lt \cdot \yt} \int \der y^- ig A^{+}(y^-,\yt) \,.
\end{align}
Thus Eq.\,\eqref{eq:effective_vertex_expansion} corresponds to scattering off the background field with momentum transfer with transverse component $\lt$. To restore the non-zero longitudinal momentum transfer $l^+$, we must keep track of the phase $e^{i l^+ y^-}$. As we will see keeping this phase will be sufficient to reproduce the results from the high-twist expansion formalism. However, we no longer can exponentiate the multiple scattering into the usual light-like Wilson lines. Thus we no longer employ the CGC effective vertices, and instead, we use the full QCD vertices:
\begin{align}
    &(2\pi) \delta(l^-) \gamma^- \int_{y}  e^{-i \lt \cdot \yt} e^{i l^+ y^-} ig A^{+}(y^-,\yt)\,,
    \label{eq:fullQCD_vertex}
\end{align}
for the coupling of the quark with the background field
$A^{\mu}(y) = \delta^{\mu+} A^{+}(y^-,\yt)$. Here we defined the shorthand $\int_y \equiv \int \der^2 \yt \int \der y^-$. 

The main goal of this section is to compute the direct photon production with single and double contributions using the vertex in Eq.\,\eqref{eq:fullQCD_vertex}. Our results will take the following form:
\begin{align}
    & E_{\gamma} \frac{\der^3\sigma_{S}^{\mathrm{CGC_{sub}}}}{\der^3 \boldsymbol{p_{\gamma}}} = \frac{\alpha_{em} \alpha_{s}}{N_c}\sum_q e_q^2 \int \der x_q f_{q/p}(x_q)   \nonumber \\
    & \times \int_{y,y'} \!\!\!\! \Hcal_{S} \  \langle \Tr[A^+(y) A^+(y')] \rangle_x \,,
\end{align}
for the single scattering contribution, and 
\begin{align}
    &E_{\gamma} \frac{\der^3\sigma_{D}^{\mathrm{CGC_{sub}}}}{\der^3 \boldsymbol{p_{\gamma}}} = \frac{\alpha_{em} \alpha_{s}^2}{N_c}\sum_q e_q^2 \int \der x_q f_{q/p}(x_q) \nonumber \\
    & \times \int_{y,y',y_1,y_2} \!\!\!\!\!\!\!\!\!\!\!\! \Theta_{D} \Hcal_{D} \  \langle \Tr[A^+(y_2) A^+(y') A^+(y) A^+(y_1)] \rangle_x \,.
\end{align}
for the double scattering contribution. Here $\Hcal_{S/D}$ are perturbative factors, which will be computed in \ref{sec:single_scattering_phase} and \ref{sec:double_scattering_central_phase} respectively, and $\Theta_{D}$ is a product of step functions which accounts the ordering of the scatterings. In Sec.\,\ref{sec:CGC-HT-correspondence} we will demonstrate the consistency between the CGC with sub-eikonal phases and the high-twist formalism by expanding the hard factor $\Hcal$ in inverse powers of $\pgammatU{2}$. The results will have the form:
\begin{align}
    &\mathcal{H}_{S} = \frac{2 }{\pi} \frac{\xi^2 \left[1 + (1-\xi)^2 \right]}{\pgammatU{4}} e^{i x P_A^+ (y^- - y'^-)}    \nonumber \\
    &\times \delta^{(2)}(\yt-\vect{y'}) (\partial_{\yt} \cdot \partial_{\ytC}) + \mathcal{O}(1/\pgammatU{6})    \,,
    \label{eq:hard_collinear_single_scattering}
\end{align}
for the single scattering contribution, and
\begin{align}
    &\mathcal{H}_{D}=   \frac{8\left[1 + (1-\xi)^2 \right]}{\pgammatU{4}} e^{i x P_A^+ (y^- - y'^-)}  \nonumber \\
    & \times  \delta^{(2)}(\yt-\ytone ) \delta^{(2)}(\ytC-\yttwo ) \delta^{(2)}(\ytone-\yttwo )  \nonumber \\
    & \times  \left[1 + \frac{\mathcal{D}_{X}}{\pgammatU{2}} \ (\partial_{ \ytone} \cdot \partial_{\yttwo}) \right] (\partial_{\yt} \cdot \partial_{\ytC}) + \mathcal{O}(1/\pgammatU{8})
    \label{eq:hard_collinear_double_scattering}
    \,,
\end{align}
for the double scattering contribution. The derivation of Eqs.\,\eqref{eq:hard_collinear_single_scattering} and \eqref{eq:hard_collinear_double_scattering} is one of the key results of this manuscript, as it will allow us to readily reconcile the CGC beyond the eikonal approximation and the high-twist formalism in their common domain of validity. 

\subsection{Single scattering contribution}
\label{sec:single_scattering_phase}

\begin{figure}[H]
    \centering
    \includegraphics[width=0.2\textwidth]{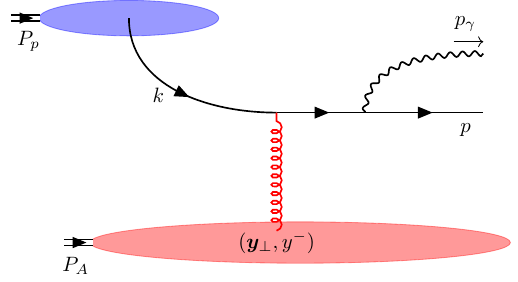}
    \includegraphics[width=0.2\textwidth]{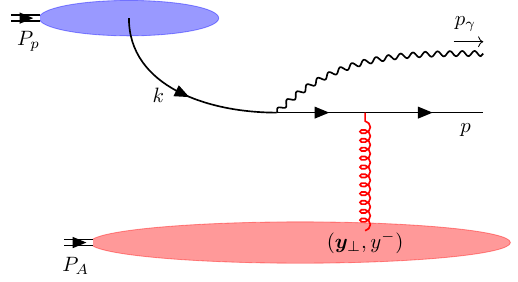}
    \caption{Single scattering diagrams to quark + photon production in proton-nucleus collisions.}
    \label{fig:single-scattering}
\end{figure}
There are two diagrams for the single scattering contribution as shown in Fig.\,\ref{fig:single-scattering}. The calculation is identical to that in Eq.\,\eqref{eq:amplitude_CGC}, and it amounts to the replacement:
\begin{align}
    V(\yt) - \mathbbm{1} \to ig \int \der y^- e^{i x P_A^+ y^-} A^+(y) \,,
\end{align}
which results in
\begin{align}
    \Mcal_{S}^{\lambda \sigma \sigma'} = i e e_q g  \int_{y}  A^+(y) \Acal_{S}^{\lambda \sigma \sigma'}  \,,
\end{align}
where the perturbative factor (in light-cone gauge $A^-=0$ for the photon) reads
\begin{align}
    &\Acal_{S}^{\lambda \sigma \sigma'}
    = e^{-i (\pt + \pgammat) \cdot \yt} e^{i x P_A^+ y^-} \nonumber \\
    & \times \left\{\frac{\left[ \xi (\ptU{\alpha} + \pgammatU{\alpha})-\pgammatU{\alpha} \right] }{\left[ \xi  (\pt + \pgammat)- \pgammat \right]^2} + \frac{ \pgammatU{\alpha} }{\pgammatU{2}} \right\} \GammatLU{\alpha}{\lambda\sigma\sigma'}  \,. \label{eq:perturbative_qactor_amplitude_singlescattering}
\end{align}
where $\GammatLU{\alpha}{\lambda\sigma\sigma'}$ was defined in Eq.\,\eqref{eq:GammaStructure}. Employing Eq.\,\eqref{eq:A2_to_XSec}, the single scattering contribution to the differential cross-section for direct photon production in proton-nucleus collisions reads:
\begin{align}
    & E_{\gamma} \frac{\der^3\sigma^{\mathrm{CGC_{sub}}}_{S}}{\der^3 \boldsymbol{p_{\gamma}}} 
    = \frac{\alpha_{em} \alpha_{s}}{N_c}\sum_q e_q^2 \int \der x_q f_{q/p}(x_q)   \nonumber \\ 
    & \times \int_{y,y'} \Hcal_{S} \left \langle  \Tr\left[ A^{+}(y')  A^{+}(y) \right] \right \rangle_x \,,
    \label{eq:Xsec_single_scattering}
\end{align}
with the perturbative factor:
\begin{align}
    & \Hcal_{S} = \frac{1}{4\pi}  \int \frac{\der^2 \pt}{(2\pi)^2} \frac{1}{1-\xi} \frac{1}{2} \sum_{\lambda \sigma \sigma'} \overline{\Acal}_{S}^{\lambda\sigma\sigma'}\Acal_{S}^{\lambda\sigma\sigma'}  \,. \label{eq:hard_factor_singlescattering}
\end{align}
The explicit expression for Eq.\,\eqref{eq:hard_factor_singlescattering} can be easily obtained from Eq.\,\eqref{eq:perturbative_qactor_amplitude_singlescattering} and Eq.\,\eqref{eq:identity1}, we find
\begin{align}
    & \Hcal_{S}  = \frac{2}{\pi}   \xi^2 \left[  1+ (1-\xi)^2 \right]  e^{i x P_A^+ (y^- - y'^-)} \nonumber \\
    & \times \int \frac{\der^2 \lt}{(2\pi)^2}  \frac{e^{-i \lt \cdot (\yt-\yt')} \lt^2}{(\xi \lt -\pgammat)^2 \pgammatU{2}} \,,
    \label{eq:hard_factor_singlescattering_2}
\end{align}
where we performed the change of variables $\pt \to \lt -\pgammat$ in the integration.

\subsection{Double scattering contribution}
\label{sec:double_scattering_central_phase}

We now proceed to evaluate the double scattering contribution. We begin by computing the amplitudes in Sec.\,\ref{subsec:double-amplitudes} and then construct the differential cross-section in Sec.\,\ref{subsec:double-cross-section}. As with the high-twist formalism, we will classify our results into four contributions (initial, final, initial-final interference, and final-initial interference scattering). We explicitly evaluate the contribution corresponding to initial state scattering (see Eqs.\,\eqref{eq:double_scattering_XSec_Initial} and \eqref{eq:hard_factor_CGC_sub_CI}). We present the results for the other three contributions in the appendix. 

\begin{widetext}

   \begin{figure}[H]
    \centering
    \includegraphics[width=0.3\textwidth]{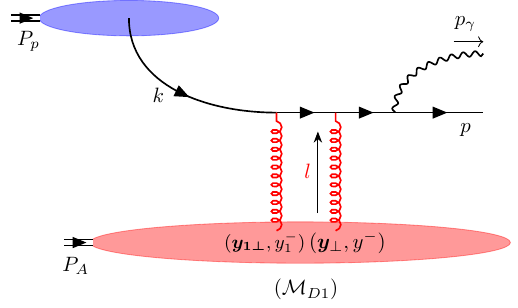}
    \includegraphics[width=0.3\textwidth]{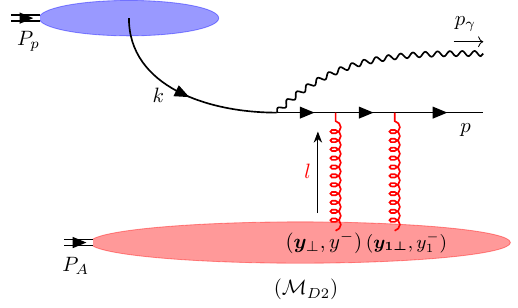}
    \includegraphics[width=0.3\textwidth]
    {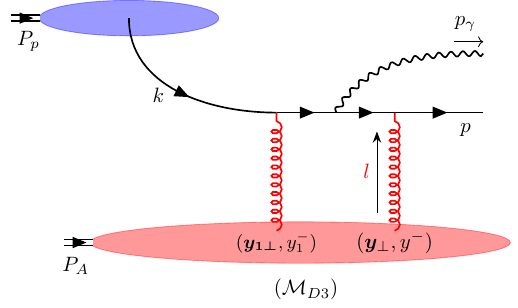}
    \caption{Double scattering diagrams to quark + photon production in proton-nucleus collisions. The momentum $l$ of one of the gluons is unconstrained and must be integrated over. By momentum conservation, the other gluon carries momentum $p+p_\gamma -k -l$.}
    \label{fig:central_cut_diagrams}
\end{figure} 
\end{widetext}
\subsubsection{Amplitudes for double scattering}
\label{subsec:double-amplitudes}
At the level of the amplitude, we must evaluate three diagrams which correspond to photon emission after the scatterings $\Mcal_{D1}$, photon emission before scatterings $\Mcal_{D2}$, and photon emission between the two scatterings $\Mcal_{D3}$ as shown in Fig.\,\ref{fig:central_cut_diagrams}. Unlike the usual CGC calculation, this third contribution is non-vanishing since we keep track of the sub-eikonal phase. The calculations of the amplitudes involve an integral over internal loop momenta $l$ (see diagrams in Fig.\,\ref{fig:central_cut_diagrams}). The integration over $l^-$ has been carried out easily due to the delta function $\delta(l^-)$ in the vertex in Eq.\eqref{eq:fullQCD_vertex}, and the $l^+$ integral is performed via contour integration using Cauchy's residue theorem. The $l^+$ integration sets the quark propagator between scatterings on-shell in the amplitudes corresponding to Eqs.\,\eqref{eq:M1_CentralCut} and \eqref{eq:M2_CentralCut}. On the other hand, for the amplitude in Eq.\,\eqref{eq:M3_CentralCut}, there are two poles in $l^+$ corresponding to setting quark propagator before (or after) the photon emission on-shell. The results for these amplitudes are:
\begin{align}
    & \Mcal_{D1}^{\lambda \sigma \sigma'} 
    = - e e_q g^2 \int_{y,y_1} A^{+}(y)  A^{+}(y_1) \theta(y^- - y_1^-) \nonumber \\
    & \times  \int \frac{\der^2 \lt}{(2\pi)^2} e^{-i\lt \cdot \yt} e^{-i (\pt + \pgammat -\lt) \cdot \ytone} \Ncal^{\lambda\sigma\sigma'}_{D1} \nonumber \\
    & \times  e^{i\left[ \frac{ \xi \pt^2 + (1-\xi)\pgammatU{2} - \xi (1-\xi)(\pt + \pgammat -\lt)^2}{\pgammatU{2}}\right] y^- x P_A^+}  \nonumber \\
    & \times e^{i \left[\frac{\xi (1-\xi) (\pt + \pgammat -\lt)^2}{\pgammatU{2}}\right] y_1^- x P_A^+}   \,,
    \label{eq:M1_CentralCut}
\end{align}
\begin{align}
    & \Mcal_{D2}^{\lambda \sigma \sigma'} = - e e_q g^2 \int_{y_1,y} A^{+}(y_1)  A^{+}(y) \theta(y_1^- - y^-) \nonumber \\
    & \times  \int \frac{\der^2 \lt}{(2\pi)^2} e^{-i (\pt + \pgammat -\lt) \cdot \ytone}  e^{-i \lt  \cdot \yt} \Ncal^{\lambda\sigma\sigma'}_{D2} \nonumber \\
    & \times   e^{i \left[\frac{\xi \pt^2 - \xi( \pgammat-\lt)^2}{\pgammatU{2}}\right] y_1^- x P_A^+  } \nonumber \\
    & \times e^{i\left[ \frac{(1-\xi)\pgammatU{2} + \xi (\pgammat-\lt)^2}{\pgammatU{2}} \right] y^- x P_A^+  }  \,,
    \label{eq:M2_CentralCut}
\end{align}
\begin{align}
    & \Mcal_{D3}^{\lambda \sigma \sigma'} = - e e_q g^2 \int_{y,y_1} A^{+}(y)  A^{+}(y_1) \theta(y^- - y_1^-) \label{eq:M3_CentralCut} \\
    & \times \int \frac{\der^2 \lt}{(2\pi)^2}  e^{-i \lt \cdot \yt} e^{-i (\pt + \pgammat -\lt)  \cdot \ytone} \Ncal^{\lambda\sigma\sigma'}_{D3} \nonumber \\
    & \times  \Bigg\{ e^{i \left[ \frac{\xi \pt^2 + (1-\xi) \pgammatU{2} - \xi (1-\xi) (\pt+\pgammat-\lt)^2}{\pgammatU{2}} \right] y^- x P_A^+ } \nonumber \\
    & \times e^{i \left[ \frac{\xi (1-\xi) (\pt+\pgammat-\lt)^2}{\pgammatU{2}} \right] y_1^- x P_A^+ }  \nonumber \\
    & -  e^{i \left\{ \left[\frac{\xi \pt^2 - \xi (\pt-\lt)^2}{\pgammatU{2}}\right] y^- + \left[\frac{(1-\xi)\pgammatU{2} + \xi (\pt-\lt)^2}{\pgammatU{2}}\right] y_1^- \right\} x P_A^+ }  \Bigg \}  \,. \nonumber 
\end{align}
The corresponding perturbative factors (in light-cone gauge $A^-=0$ for the photon) are:
\begin{align}
    \Ncal^{\lambda\sigma\sigma'}_{D1} &= \frac{\left[ \xi \ptU{\alpha} -(1-\xi)\pgammatU{\alpha}\right]}{\left[ \xi \pt -(1-\xi)\pgammat\right]^2} \GammatLU{\alpha}{\lambda\sigma\sigma'} \nonumber \,, \\
    \Ncal^{\lambda\sigma\sigma'}_{D2} &=\frac{ \pgammatU{\alpha}}{ \pgammatU{2}} \GammatLU{\alpha}{\lambda\sigma\sigma'} \nonumber \,, \\
    \Ncal^{\lambda\sigma\sigma'}_{D3} &= \frac{\left[ \xi \lt^{\alpha}- (\xi \ptU{\alpha} -(1-\xi)\pgammatU{\alpha} )\right] }{  \left[ \xi\lt - (\xi\pt - (1-\xi)\pgammat )\right]^2} \GammatLU{\alpha}{\lambda\sigma\sigma'} \,.
    \label{eq:double-pert-factor-N}
\end{align}
Eq.\,\eqref{eq:M3_CentralCut} displays two different phases, each of which is associated with one of the two poles. In the strict high-energy limit $s \to \infty $ ($x \to 0$), there is perfect destructive interference and this amplitude vanishes. This cancellation is expected in the eikonal approximation employed in shock-wave formalism as emissions between the scattering with the nuclei are kinematically forbidden. It is enlightening to express the phases as:
\begin{align}
    & \Mcal_{D3}^{\lambda \sigma \sigma'} \propto e^{i \left\{ \left[\frac{\xi \pt^2 + (1-\xi) \pgammatU{2} - \xi (1-\xi) \ellt^2}{\pgammatU{2}} \right] y^- + \frac{\xi(1-\xi) \ellt^2}{\pgammatU{2}} y_1^-\right\} x P_A^+}  \nonumber \\
    & \times \Big[ 1 -  e^{-i\frac{(y^- - y_1^-)}{\tau_{\gamma,\mathrm{form}}}}  \Big] \,,
    \label{eq:LPM-double-scattering}
\end{align}
where $\ellt = \pt+\pgammat-\lt$ is the transverse momentum carried by the first scattering. We identified the inverse of the formation time \cite{Zhang:2021tcc} for the photon production:
\begin{align}
    \tau_{\gamma,\mathrm{form}}^{-1} =  \frac{\left[ \pgammat - \xi \ellt \right]^2}{\pgammatU{2}} x P_A^+ =  \frac{\left[ \pgammat - \xi \ellt \right]^2}{2 k^- \xi(1-\xi)} \,.
\end{align}
Eq.\,\eqref{eq:LPM-double-scattering} displays the characteristic Landau-Pomeranchuk-Migdal (LPM) effect \cite{Landau:1953um,Migdal:1956tc}. In the limit $ \tau_{\gamma,\mathrm{form}} \gg y^- - y_1^- $, the photon is not able to resolve the two different scatterings (coherent) and the contribution in Eq.\,\eqref{eq:M3_CentralCut} vanishes, while in the limit $ \tau_{\gamma,\mathrm{form}} \ll y^- - y_1^- $ the phases do not cancel each other out and there remains a net (incoherent) contribution\cite{Guo:2000nz,Wang:2001ifa}. 

We observe that the first term in Eq.\,\eqref{eq:M3_CentralCut} is identical to Eq.\,\eqref{eq:M1_CentralCut}, except for their perturbative factors. This is unsurprising as both contributions correspond to setting the first quark propagator on-shell. The same identification holds for the second term in Eq.\,\eqref{eq:M3_CentralCut} and Eq.\,\eqref{eq:M2_CentralCut}, and they correspond to the case in which the second quark propagator in on-shell. To make this identification more apparent we separate Eq.\,\eqref{eq:M3_CentralCut} as a sum of two contributions $\Mcal_{D3} =  \Mcal_{D3a} + \Mcal_{D3b}$ \footnote{In the definition of $\Mcal_{D3b}$, we swapped the variables $y_1 \leftrightarrow y$, and $\lt \leftrightarrow  \pt + \pgammat -\lt$.}:
\begin{align}
    &\Mcal_{D3a}^{\lambda \sigma \sigma'}
    = - e e_q g^2 \int_{y,y_1} A^{+}(y)  A^{+}(y_1) \theta(y^- - y_1^-) \\
    & \times \int \frac{\der^2 \lt}{(2\pi)^2} e^{-i \lt \cdot \yt}    e^{-i (\pt + \pgammat -\lt) \cdot \ytone}  \Ncal^{\lambda\sigma\sigma'}_{D3a}  \nonumber \\
    & \times   e^{i\left[ \frac{ \xi \pt^2 + (1-\xi)\pgammatU{2} - \xi (1-\xi)(\pt + \pgammat -\lt)^2}{\pgammatU{2}}\right] y^- x P_A^+}  \nonumber \\
    & \times e^{i \left[\frac{\xi (1-\xi) (\pt + \pgammat -\lt)^2}{\pgammatU{2}}\right] y_1^- x P_A^+}   \,, \nonumber  \\
    & \Mcal_{D3b}^{\lambda \sigma \sigma'} 
    = - e e_q g^2 \int_{y_1,y} A^{+}(y_1)  A^{+}(y) \theta(y_1^- - y^-)  \\
    & \times \int \frac{\der^2 \lt}{(2\pi)^2} e^{-i (\pt + \pgammat -\lt) \cdot \ytone}  e^{-i \lt \cdot \yt}     \Ncal^{\lambda\sigma\sigma'}_{D3b}  \nonumber \\
    & \times  e^{i \left[\frac{\xi \pt^2 - \xi( \pgammat-\lt)^2}{\pgammatU{2}}\right] y_1^- x P_A^+  } e^{i\left[ \frac{(1-\xi)\pgammatU{2} + \xi (\pgammat-\lt)^2}{\pgammatU{2}} \right] y^- x P_A^+  }   \,, \nonumber
\end{align}
where the hard factors read
\begin{align}
    \Ncal^{\lambda\sigma\sigma'}_{D3a} &= \frac{\left[ \xi \lt^{\alpha}- (\xi \ptU{\alpha} -(1-\xi)\pgammatU{\alpha} )\right] }{  \left[ \xi\lt - (\xi\pt - (1-\xi)\pgammat )\right]^2} \GammatLU{\alpha}{\lambda\sigma\sigma'}\,, \nonumber \\
    \Ncal^{\lambda\sigma\sigma'}_{D3b} &= \frac{\left[ \xi \ltU{\alpha} - \pgammatU{\alpha}\right] }{  \left[ \xi \lt - \pgammat \right]^2} \GammatLU{\alpha}{\lambda\sigma\sigma'}  \,.
    \label{eq:double-pert-factor-N2}
\end{align}
This observation suggests that we should define the amplitudes:
\begin{align}
    \Mcal_{D,I} =& \Mcal_{D1}+ \Mcal_{D3a} \,, \label{eq:double-initial-state}\\
    \Mcal_{D,F} =& \Mcal_{D2} + \Mcal_{D3b} \label{eq:double-final-state}\,,
\end{align}
then we have
\begin{align}
    &\Mcal_{D,I}^{\lambda \sigma \sigma'} =  -e e_q g^2 \nonumber \\
    & \times \int_{y,y_1} \!\!\!\!\!\! \Acal^{\lambda\sigma\sigma'}_{D,I}(y,y_1) A^{+}(y)  A^{+}(y_1)  \theta(y^- - y_1^-)    \,, \\
    &\Mcal_{D,F}^{\lambda \sigma \sigma'} = -e e_q g^2 \nonumber \\
    & \times \int_{y_1,y} \!\!\!\!\!\!\Acal^{\lambda\sigma\sigma'}_{D,F}(y_1,y) A^{+}(y_1)  A^{+}(y)  \theta(y_1^- - y^-)    \,,
\end{align}
where the perturbative factors are:
\begin{align}
    & \Acal^{\lambda\sigma\sigma'}_{D,I}(y,y_1) = \int \frac{\der^2 \lt}{(2\pi)^2} \left\{ \Ncal^{\lambda\sigma\sigma'}_{D1} +\Ncal^{\lambda\sigma\sigma'}_{D3a} \right\}  \nonumber \\
    &\times e^{-i\lt \cdot \yt} e^{-i (\pt + \pgammat -\lt) \cdot \ytone}  \nonumber \\
    & \times  e^{i \left[\frac{ \xi \pt^2 + (1-\xi) \pgammatU{2} - \xi(1-\xi)(\pt + \pgammat -\lt)^2} {\pgammatU{2}} \right]x P_A^+ y^-}  \nonumber \\
    & \times e^{i \left[\frac{\xi (1-\xi)(\pt + \pgammat -\lt)^2}{\pgammatU{2}} \right] x P_A^+ y_1^-}\,,
    \label{eq:perturbative-ADI} \\
    & \Acal^{\lambda\sigma\sigma'}_{D,F}(y_1,y) = \int \frac{\der^2 \lt}{(2\pi)^2}  \left\{ \Ncal^{\lambda\sigma\sigma'}_{D2}+\Ncal^{\lambda\sigma\sigma'}_{D3b} \right\} \nonumber \\
    & \times e^{-i (\pt + \pgammat -\lt) \cdot \ytone}  e^{-i \lt  \cdot \yt}  \nonumber \\
    & \times  e^{i \left[\frac{\xi\pt^2-\xi (\pgammat-\lt)^2}{\pgammatU{2}}\right] x P_A^+  y_1^- } \nonumber \\
    & \times e^{i \left[\frac{ (1-\xi)\pgammatU{2} + \xi(\pgammat-\lt)^2 }{\pgammatU{2}}\right] x P_A^+ y^-} \,.
    \label{eq:perturbative-ADF}
\end{align}
We note that $\Ncal^{\lambda\sigma\sigma'}_{D1} +\Ncal^{\lambda\sigma\sigma'}_{D3a} $ and $\Ncal^{\lambda\sigma\sigma'}_{D2} +\Ncal^{\lambda\sigma\sigma'}_{D3b} $ vanish in the limit $\lt \to 0$, a property which will be exploited when we make the correspondence to the high-twist formalism.

\subsubsection{Differential cross-section}
\label{subsec:double-cross-section}
Employing Eq.\,\eqref{eq:A2_to_XSec}, the differential cross-section for direct photon production has four contributions from double scattering:
\begin{align}
    & E_{\gamma} \frac{\der^3\sigma^{\rm CGC_{sub}}_{D}}{\der^3 \boldsymbol{p_{\gamma}}}    
    =\int  \frac{\der x_p f(x_p)}{(4\pi)^3 (1-\xi)} \int \frac{\der^2 \pt}{(2\pi)^2} \frac{1}{2N_c}  \label{eq:double-scattering-cross-section} \\
    & \times \left[\sum_{\lambda \sigma \sigma' i j} \left\langle  \left(\overline{\Mcal}_{D,I} + \overline{\Mcal}_{D,F} \right) \left( \Mcal_{D,I} + \Mcal_{D,F} \right)  \right \rangle_x \right]\,. \nonumber 
\end{align}
After we carry out the $1/p_{\gamma\perp}^2$ expansion in the next subsection, we will observe that the amplitudes $\Mcal_{D,I}$ and $\Mcal_{D,F}$ correspond to the cases in which the initial and final gluon are soft, respectively. This explains the subscripts $I$ and $F$ when defining these amplitudes. Let us illustrate the contribution proportional to $\left\langle  \overline{\Mcal}_{D,I}\Mcal_{D,I} \right \rangle $ explicitly:
\begin{align}
    & E_{\gamma} \frac{\der^3\sigma^{\mathrm{CGC_{sub}}}_{C,I}}{\der^3 \boldsymbol{p_{\gamma}}} 
    = \frac{\alpha_{em} \alpha_{s}^2}{N_c}\sum_q e_q^2 \int \der x_q f_{q/p}(x_q)  \label{eq:double_scattering_XSec_Initial}  \\
    & \times \int_{y,y',y_1,y_2} \!\!\!\!\!\!\!\!\!\!\!\!\!\!\!\!\! \Theta_{C,I}  \left \langle  \Tr\left[  A^{+}(y_2)  A^{+}(y') A^{+}(y)  A^{+}(y_1)     \right] \right \rangle_x \Hcal_{C,I} \,, \nonumber 
\end{align}
where $\Theta_{C,I} = \theta(y'^- - y_2^-) \theta(y^- - y_1^-) $ enforces the ordering of the gauge field insertions.
\begin{widetext}
The perturbative factor for the initial-initial scattering is defined as:
\begin{align}
    & \Hcal_{C,I}
    = \frac{1}{4\pi} \int \frac{\der^2 \pt}{(2\pi)^2}  \frac{1}{1-\xi} \frac{1}{2} \sum_{\lambda \sigma \sigma'} \overline{\Acal}_{D,I}^{\lambda\sigma\sigma'}(y',y_2)\Acal_{D,I}^{\lambda\sigma\sigma'}(y,y_1) \,.
    \label{eq:double_scattering_XSec_Initial_Hard}
\end{align} 
This expression can be computed explicitly from Eqs.\,\eqref{eq:double-pert-factor-N}\,,\eqref{eq:double-pert-factor-N2}\,,\eqref{eq:perturbative-ADI}, and the identity in Eq.\,\eqref{eq:identity1}:
    \begin{align}
    & \mathcal{H}_{C,I} = 8 \left[1 + (1-\xi)^2 \right] \!\! \int \!\! \frac{\der^2 \Lt}{(2\pi)^2} \!\! \int \!\! \frac{\der^2 \lt}{(2\pi)^2} \!\! \int \!\! \frac{\der^2 \lt'}{(2\pi)^2} e^{i \left[ (\Lt -\lt') \cdot \yttwo + \lt'  \cdot \yt'  - \lt  \cdot \yt - (\Lt -\lt) \cdot \ytone \right]}  \nonumber \\
    & \times \left\{\frac{\left[ \xi \ltCL{\alpha}  - (\xi \LtL{\alpha}  -\pgammatL{\alpha}) \right]}{\left[\xi\ltC -( \xi \Lt - \pgammat ) \right]^2} + \frac{\left[ \xi \LtL{\alpha} - \pgammatL{\alpha} \right]}{\left[ \xi \Lt -\pgammat \right]^2}  \right\} \left\{\frac{\left[ \xi \lt^{\alpha}  - (\xi \LtU{\alpha}  -\pgammatU{\alpha}) \right]}{\left[\xi\lt -( \xi \Lt - \pgammat ) \right]^2} + \frac{\left[ \xi \LtU{\alpha} - \pgammatU{\alpha} \right]}{\left[ \xi \Lt -\pgammat \right]^2}  \right\} \nonumber \\
    &\times e^{-i \left[ \frac{\xi(1-\xi)(\Lt -\ltC)^2}{\pgammatU{2}} \right]x P_A^+ y_2^-} e^{-i \left[\frac{ \xi (\Lt-\pgammat)^2 + (1-\xi)\pgammatU{2} - \xi(1-\xi) (\Lt -\ltC)^2}{\pgammatU{2}}\right] x P_A^+ y'^-}  \nonumber \\
    & \times e^{i \left[\frac{ \xi (\Lt-\pgammat)^2 + (1-\xi)\pgammatU{2} - \xi(1-\xi)(\Lt -\lt)^2}{\pgammatU{2}}\right]x P_A^+ y^-}  e^{i \left[\frac{ \xi(1-\xi)(\Lt -\lt)^2}{\pgammatU{2}} \right] x P_A^+ y_1^-} \,,
    \label{eq:hard_factor_CGC_sub_CI}
\end{align}
\end{widetext}
where we performed the change of variables $\pt \to \Lt -\pgammat$ in the integration. Eqs.\,\eqref{eq:double_scattering_XSec_Initial} and \eqref{eq:hard_factor_CGC_sub_CI} are the main results of this section. The momenta $\lt$, $\lt'$, and $\Lt$ are unconstrained by the kinematics and must be integrated over. In the next section, we will carry out the integration in the limit of large $\pgammatU{2}$.

The other three contributions are computed in a similar fashion: 
\begin{align}
    & E_{\gamma} \frac{\der^3\sigma^{\mathrm{CGC_{sub}}}_{C,F}}{\der^3 \boldsymbol{p_{\gamma}}}
    = \frac{\alpha_{em} \alpha_{s}^2}{N_c}\sum_q e_q^2  \int \der x_q  f_{q/p}(x_q) \label{eq:double_scattering_XSec_Final} \\
    & \times \int_{y,y',y_1,y_2} \!\!\!\!\!\!\!\!\!\!\!\!\!\!\! \Theta_F \left \langle  \Tr\left[  A^{+}(y')  A^{+}(y_2) A^{+}(y_1)  A^{+}(y)     \right] \right \rangle_x \Hcal_{C,F} \,, \nonumber 
\end{align}
\begin{align}
    & E_{\gamma} \frac{\der^3\sigma^{\mathrm{CGC_{sub}}}_{C,IF}}{\der^3 \boldsymbol{p_{\gamma}}} 
    =  \frac{\alpha_{em} \alpha_{s}^2}{N_c}\sum_q e_q^2  \int \der x_q  f_{q/p}(x_q)  \label{eq:double_scattering_XSec_InitialFinal} \\
    & \times \int_{y,y',y_1,y_2} \!\!\!\!\!\!\!\!\!\!\!\!\!\!\! \Theta_{IF} \left \langle  \Tr\left[  A^{+}(y')  A^{+}(y_2) A^{+}(y)  A^{+}(y_1)     \right] \right \rangle_x \Hcal_{C,IF} \,, \nonumber 
\end{align}
\begin{align}
    & E_{\gamma} \frac{\der^3\sigma^{\mathrm{CGC_{sub}}}_{C,FI}}{\der^3 \boldsymbol{p_{\gamma}}}
    = \frac{\alpha_{em} \alpha_{s}^2}{N_c}\sum_q e_q^2  \int \der x_q  f_{q/p}(x_q)  \label{eq:double_scattering_XSec_FinalInitial} \\
    & \times \int_{y,y',y_1,y_2} \!\!\!\!\!\!\!\!\!\!\!\!\!\!\! \Theta_{FI} \left \langle  \Tr\left[  A^{+}(y_2)  A^{+}(y') A^{+}(y_1)  A^{+}(y)     \right] \right \rangle_x \Hcal_{C,FI} \,, \nonumber 
\end{align}
where $\Theta_F = \theta(y_2^- - y'^-) \theta(y_1^- - y^-) $, $\Theta_{IF}=\theta(y_2^- - y'^-) \theta(y^- - y_1^-)$ and $\Theta_{FI}=\theta(y'^- - y_2^-) \theta(y_1^- - y^-)$.
The explicit results for the perturbative factors $\Hcal_{C,F}$, $\Hcal_{C,IF}$ and $\Hcal_{C,FI}$ can be found in appendix\,\ref{app:Hard_factors}. 

Lastly, the same procedure is followed to compute the interference contribution from triple-single scattering. For the interested reader, the details are shown in the Appendix \ref{app:triple-single-interference}.

\subsection{Correspondence between CGC and HT formalism}
\label{sec:CGC-HT-correspondence}

In this subsection, we perform a series expansion in $1/p_{\gamma\perp}^2$ of the single and double scattering contribution from the CGC with sub-eikonal phase and show that the results are consistent with those obtained in the high-twist formalism. 
\begin{align}
      \left. E_{\gamma} \frac{\der \sigma^{{\rm CGC_{sub}}}}{\der ^3 \boldsymbol{p_{\gamma}}}\right|_{p_{\gamma\perp}> Q_s}
        = E_{\gamma} \frac{\der \sigma^{{\rm LT}}}{\der ^3 \boldsymbol{p_{\gamma}}} +E_{\gamma} \frac{\der \sigma^{{\rm T4}}}{\der ^3 \boldsymbol{p_{\gamma}}}+\cdots.
\end{align}

\subsubsection{Twist-2}
We start from the expression for the single scattering contribution in Eqs.\,\eqref{eq:Xsec_single_scattering} and \eqref{eq:hard_factor_singlescattering_2}. Expanding the perturbative factor in inverse powers of  $\pgammatU{2}$ we find: 
\begin{align}
    & \Hcal_{S} = \frac{2}{\pi} \frac{\xi^2 \left[1 + (1-\xi)^2 \right]}{\pgammatU{4}}   e^{i x P_A^+ (y^- - y'^-)} \nonumber \\
    & \times \int \frac{\der^2 \lt}{(2\pi)^2} \lt^2 e^{-i \lt \cdot (\yt-\yt')} + \dots \,.
\end{align}
Trading $\lt^2$ by derivatives with respect to $\yt$ and $\yt'$:
\begin{align}
    & \Hcal_{S} = \frac{2}{\pi} \frac{\xi^2 \left[1 + (1-\xi)^2 \right]}{\pgammatU{4}}  e^{i x P_A^+ (y^- - y'^-)} \nonumber \\
    & \times \frac{\partial^2}{\partial \yt \cdot \partial \yt'} \int \frac{\der^2 \lt}{(2\pi)^2} e^{-i \lt \cdot (\yt-\yt')} + \dots \,.
\end{align}
The integral over the phase turns into a delta function in the transverse coordinates, thus the leading contribution to $\Hcal_{S}$ is
\begin{align}
    &\mathcal{H}_{S} = \frac{2 }{\pi} \frac{\xi^2 \left[1 + (1-\xi)^2 \right]}{\pgammatU{4}}   e^{i x P_A^+ (y^- - y'^-)}    \nonumber \\
    &\times \delta^{(2)}(\yt-\vect{y'}) (\partial_{\yt} \cdot \partial_{\ytC}) + \mathcal{O}(1/\pgammatU{6})    \,.
     \label{eq:hard_factor_singlescattering_leading}
\end{align}
The leading power contribution to direct photon production is obtained by inserting Eq.\,\eqref{eq:hard_factor_singlescattering_leading} into Eq.\,\eqref{eq:Xsec_single_scattering}, the derivatives act on the gauge field, and using\footnote{Strictly speaking, we should also consider the transverse components of the gauge field in Eq.\,\eqref{eq:fullQCD_vertex}. Thanks to gauge invariance the net effect is to replace $\partial_{\ytU{\alpha}} A^{+} \to \partial_{\ytU{\alpha}} A^{+}- \partial_{y^-} \vect{A}^i + ig [A^{+},\vect{A}^i]$, see e.g. \cite{Eguchi:2006mc}.}
\begin{align}
    F^{\alpha+}(y^-,\yt) = \partial_{\ytU{\alpha}} A^{+}(y^-,\yt)\,.
    \label{eq:A_to_F}
\end{align}
We obtain:
\begin{align}
    & E_{\gamma} \frac{\der^3\sigma^{\mathrm{CGC_{sub}}}_{S}}{\der^3 \boldsymbol{p_{\gamma}}} \Bigg |_{\mathrm{LT}}
    \!\!\!\!\!\! =  \frac{\alpha_{em} \alpha_{s}}{N_c} \sum_q e_q^2 \int \der x_q f_{q/p}(x_q)   \nonumber \\
    & \times \frac{\xi^2 \left[1 + (1-\xi)^2 \right]}{\pgammatU{4}} x f_{g/A}(x)  \,,
    \label{eq:CGC_sub_twist2}
\end{align}
where we identified the twist-2 collinear gluon distribution:
\begin{align}
    x f_{g/A}(x) &=  4 V  \int\der{y^-} \frac{e^{i x P_A^+ y^- }}{2\pi}  \left \langle \Tr\left[ 
 F^{\ +}_{\alpha}(0^-) F^{\alpha+}(y^-) \right] \right \rangle_x \,,
 \label{eq:twist-2-distirbution-singlescattering}
\end{align}
where $V = \int \der y^- \int \der^2  \ytC$ is a trivial volume factor due to the translational invariance of the correlator $\left \langle \Tr\left[ 
 F^{\ +}_{\alpha}(y'^-) F^{\alpha+}(y^-) \right] \right \rangle$. Using the relation between the CGC average and the nuclear matrix element: 
\begin{align}
    \langle  \Ocal \rangle_x = \frac{\langle P_A |\Ocal| P_A \rangle}{\langle P_A | P_A \rangle}  = \frac{\langle P_A |\Ocal| P_A \rangle}{2 P_A^+ V} \,,
    \label{eq:CGC-to-ME}
\end{align}
 we can write Eq.\,\eqref{eq:twist-2-distirbution-singlescattering} as
\begin{align}
    &x f_{g/A}(x) = \int \der{y^-} \frac{e^{i x P_A^+ y^- }}{ \pi P_A^+} \nonumber \\
    &\times \left \langle P_A|   
\Tr\left[ 
 F^{\ +}_{\alpha}(0^-) F^{\alpha+}(y^-) \right]  | P_A \right \rangle\,,
\end{align} 
or in the more conventional form:
\begin{align}
    f_{g/A}(x) =  \int \der{y^-} \frac{e^{i x P_A^+ y^- }}{2\pi x P_A^+}  \left \langle P_A|   
 F^{\ +}_{\alpha,a}(0^-) F^{\alpha+,a}(y^-)  | P_A \right \rangle \,,
\end{align}
where we use $\Tr\left[t^a t^b \right] = \frac{1}{2} \delta^{ab}$. The results in Eq.\,\eqref{eq:CGC_sub_twist2} exactly match the standard collinear factorization result in Eq.\,\eqref{eq:twist-2-coll-fact-stu}. Unlike the leading power expansion of Eq.\,\eqref{eq:CGC_photon_prod_LT_2}, the result in Eq.\,\eqref{eq:twist-2-distirbution-singlescattering} keeps track of the phase $e^{i x P_A^+ y^-}$ in the gluon distribution.

\subsubsection{Twist-4}
\label{sec:twist-2-expansion}
To extract the twist-4 contribution, we carry out the expansion of the double scattering contribution in Eq.\,\eqref{eq:double-scattering-cross-section}, as well as the single-triple scattering contribution in Eqs.\,\eqref{eq:single-triple-contribution} and \eqref{eq:triple-single-contribution} \footnote{One should also take into account the twist-4 expansion of the single scattering contribution Eq.\,\eqref{eq:Xsec_single_scattering}, as well as the interference between single and double scattering contribution. However, these lead to higher derivative terms of the strength field tensor, and such terms are not enhanced by the nuclear size; hence, they are neglected in the high-twist approach. Thus we will not take them into account in this work.}. It is sufficient to expand the perturbative factors. We illustrate this explicitly for the contribution in Eq.\,\eqref{eq:hard_factor_CGC_sub_CI}, the expansions of the other factors are carried out similarly, and the results are collected in the Appendix \ref{app:Hard_factors_collinear}.

It is convenient to first expand in powers of $\lt$ and $\lt'$, then we find, the leading contribution yields:
\begin{align}
    &\Hcal_{C,I}   =  8 \xi^2 \left[1 + (1-\xi)^2 \right] \int \frac{\der^2 \Lt}{(2\pi)^2} \frac{1}{( \xi \Lt - \pgammat )^4}\nonumber \\
    & \times  \int \frac{\der^2 \lt}{(2\pi)^2} \int \frac{\der^2 \lt'}{(2\pi)^2} (\lt \cdot \lt') \nonumber \\
    & \times  e^{-i \lt  \cdot (\yt-\ytone)} e^{-i \Lt \cdot (\ytone-\yttwo)}  e^{i \lt'  \cdot (\yt'-\yttwo)} \nonumber \\
    & \times e^{i \left[\frac{ \xi(1-\xi) \Lt^2}{\pgammatU{2}} \right] x P_A^+ (y_1^- - y_2^-)} \nonumber \\
    &\times e^{i \left[\frac{ \xi (\Lt-\pgammat)^2 + (1-\xi)\pgammatU{2} - \xi(1-\xi) \Lt^2}{\pgammatU{2}}\right]x P_A^+ (y^- - y'^-)} \,.
\end{align}
The integrals over $\lt$ and $\lt'$ can be carried out which lead to derivatives of the delta function:
\begin{align}
    &\Hcal_{C,I}   = 8 \xi^2 \left[1 + (1-\xi)^2 \right] \nonumber \\
    & \times \frac{\partial \delta^{(2)}(\yt-\ytone)}{\partial \yt} \cdot \frac{\partial \delta^{(2)}(\ytC-\yttwo)}{\partial \ytC} \nonumber \\
    & \times \int \frac{\der^2 \Lt}{(2\pi)^2} \frac{e^{-i \Lt \cdot (\ytone-\yttwo)}}{( \xi \Lt - \pgammat )^4} e^{i \left[\frac{ \xi(1-\xi) \Lt^2}{\pgammatU{2}} \right] x P_A^+ (y_1^- - y_2^-)}    \nonumber \\
    & \times e^{i \left[\frac{ \xi (\Lt-\pgammat)^2 + (1-\xi)\pgammatU{2} - \xi(1-\xi) \Lt^2}{\pgammatU{2}}\right]x P_A^+ (y^- - y'^-)} \,.
\end{align}
Lastly, we carry out the expansion in $\Lt$ up to quadratic order:
\begin{align}
    &\mathcal{H}_{C,I} = \frac{8 \xi^2 \left[1 + (1-\xi)^2 \right]}{\pgammatU{4}}  e^{i x P_A^+ (y^- - y'^-)}  \nonumber \\
    & \times  \delta^{(2)}(\yt-\ytone ) \delta^{(2)}(\ytC-\yttwo ) \delta^{(2)}(\ytone-\yttwo )  \nonumber \\
    & \times  \left[1 + \frac{\mathcal{D}_{C,I}}{\pgammatU{2}} \ (\partial_{ \ytone} \cdot \partial_{\yttwo}) \right] (\partial_{\yt} \cdot \partial_{\ytC}) + \mathcal{O}\left(1/\pgammatU8\right)
    \label{eq-hard-factor-CGC-twist-4-collinear}
    \,,
\end{align}
where
\begin{align}
    & \mathcal{D}_{C,I} =  4\xi^2 +\xi (1-\xi)(ix P_A^+ \Delta y_{12}^-) \nonumber \\
    &-3 \xi^2 (ix P_A^+ \Delta y^-) +\xi^2 (ix P_A^+ \Delta y^-)^2 \,.
    \label{eq:DCI-1}
\end{align}
We insert Eq.\,\eqref{eq-hard-factor-CGC-twist-4-collinear} into Eq.\,\eqref{eq:double_scattering_XSec_Initial}, the derivatives act on the gauge fields, and using Eq.\,\eqref{eq:A_to_F}, we find
\begin{align}
    & E_{\gamma} \frac{\der^3\sigma^{\mathrm{CGC_{sub}}}_{C,I}}{\der^3 \boldsymbol{p_{\gamma}}}
    \!\! = E_{\gamma} \frac{\der^3\sigma^{\mathrm{CGC_{sub}}}_{C,I}}{\der^3 \boldsymbol{p_{\gamma}}} \Bigg|_{\mathrm{gauge}} \!\!\!\! +  E_{\gamma} \frac{\der^3\sigma^{\mathrm{CGC_{sub}}}_{C,I}}{\der^3 \boldsymbol{p_{\gamma}}}\Bigg|_{\mathrm{T}4}   \,, \label{eq:eq:double_scattering_XSec_Initial_collinear}
\end{align}
where
\begin{align}
    & E_{\gamma} \frac{\der^3\sigma^{\mathrm{CGC_{sub}}}_{C,I}}{\der^3 \boldsymbol{p_{\gamma}}} \Bigg|_{\mathrm{T}4} =  \frac{(2\pi)^2 \alpha_{em} \alpha_{s}^2}{N_c^2} \int \der x_q f_{q/p}(x_q)    \\
    &  \times  \frac{\xi^2 \left[1 + (1-\xi)^2 \right]}{\pgammatU{6}} \mathcal{D}_{C,I} \nonumber  T_{C,I} (x_1,x_2,x_3) \Big |_{\substack{x_1=x \\x_2=x_3=0}} \,.
\end{align}
The first term in Eq.\,\eqref{eq:eq:double_scattering_XSec_Initial_collinear} corresponds to a contribution to the gauge link of the collinear twist-2 gluon distribution \cite{Qiu:1990xxa,Luo:1994np}, while the second term is power suppressed by inverse power of $\pgammatU{2}$ and it corresponds to the twist-4 contribution. This expansion is analogous to the one carried out in Eq.\,\eqref{eq:collinear-expansion-soft} in the high twist formalism. The distribution $T_{C,I}$ is defined as:
\begin{align}
    &T_{C,I}(x_1,x_2,x_3) = 8N_c V \int \frac{\der y^-}{2\pi} \frac{\der y^-_1 \der y^-_2}{2\pi} \nonumber\\
    & \times e^{i x_1 P_A^+ y^- } e^{i x_2 P_A^+ (y_1^- - y_2^-)} e^{i x_3 P_A^+ y_2^-}  \theta(y^- - y_1^-) \theta(-y_2^-) \nonumber \\
    & \times  \left \langle  \Tr\left[   F^{\beta+}(y_2^-) F^{\alpha+}(0^-) F^{\ +}_{\alpha}(y^-) F^{\ +}_{\beta}(y_1^-) \right] \right \rangle_x \,.
    \label{eq:T4-initial-state-distr-CGC}
\end{align}
The function $\mathcal{D}_{C,I}$ in Eq.\,\eqref{eq:DCI-1} can be interpreted as derivative operator:
\begin{align}
    \mathcal{D}_{C,I}=& \left[4 \xi^2 + \xi(1-\xi) x \frac{\partial}{\partial x_2} - 3\xi^2 x \frac{\partial}{\partial x_1} + \xi^2 x \frac{\partial^2}{\partial x_1^2} \right] \,,
\end{align}
where we exploited the fact that derivatives of the twist-4 distribution act as: 
\begin{align}
    \frac{\partial}{\partial x_1}  \to ix P_A^+ \Delta y^- \,, \quad \frac{\partial}{\partial x_2} \to ix P_A^+ \Delta y_{12}^- \,.
\end{align}
Lastly, we can express the twist-4 contribution as
\begin{widetext}
    \begin{align}
    & E_{\gamma} \frac{\der^3\sigma^{\mathrm{CGC_{sub}}}_{C,I}}{\der^3 \boldsymbol{p_{\gamma}}} \Bigg|_{\mathrm{T}4} \!\!\!\!\! =  \frac{4\pi^2 \alpha_{\mathrm{s}}^2 \alpha_{em} }{N_c^2}  \sum_q e_q^2 \int \der x_q f_{q/p}(x_q)  \frac{\left[1 + (1-\xi)^2 \right]}{ \pgammatU{6}} \left[4 \xi^4 + \xi^3(1-\xi) x \frac{\partial}{\partial x_2} - 3\xi^4 x \frac{\partial}{\partial x_1} + \xi^2 x \frac{\partial^2}{\partial x_1^2} \right]  T_{C,I} \Big |_{\substack{x_1=x \\x_2=x_3=0}} \,.
\end{align}

\end{widetext}
This result is consistent with the high-twist formalism obtained in Eq.\,\eqref{eq:initial-central} provided the distributions defined in Eqs.\,\eqref{eq:T4-initial-state-distr} and \eqref{eq:T4-initial-state-distr-CGC} are equivalent. To explicitly show this equivalence we again use the relation in Eq.\,\eqref{eq:CGC-to-ME}:
\begin{align}
    &T_{C,I}(x_1,x_2,x_3) = \frac{4N_c}{P_A^+}\int \frac{\der y^-}{2\pi} \frac{\der y^-_1 \der y^-_2}{2\pi} \nonumber\\
    & \times e^{i x_1 P_A^+ y^- } e^{i x_2 P_A^+ (y_1^- - y_2^-)} e^{i x_3 P_A^+ y_2^-}  \theta(y^- - y_1^-) \theta(-y_2^-) \nonumber \\
    & \times  \left \langle  P_A | \Tr\left[   F^{\beta+}(y_2^-) F^{\alpha+}(0^-) F^{\ +}_{\alpha}(y^-) F^{\ +}_{\beta}(y_1^-) \right] | P_A \right \rangle \,.
    \label{eq:T4-initial-state-distr-CGC-2}
\end{align}
Next, we express the field strength tensor in the adjoint representation:
\begin{align}
    & \Tr \left[ F^{\beta+}(y_2^-) F^{\alpha+}(0^-) F^{\ +}_{\alpha}(y^-) F^{\ +}_{\beta}(y_1^-) \right]  \nonumber \\
    &  = \Tr\left[  t^a t^b t^c t^d  \right] F^{\beta+}_a(y_2^-) F^{\alpha+}_b(0^-) F^{\ +}_{\alpha,c}(y^-) F^{\ +}_{\beta,d}(y_1^-)  \,.
    \label{eq:FCI_fundamental}
\end{align}
If we assume that the color correlation follows the same structure as the Lorentz indices, then we have
\begin{align}
    & \left \langle P_A|   F^{\beta+}_{a}(y_2^-) F^{\alpha+}_{b}(0^-) F^{\ +}_{\alpha,c}(y^-) F^{\ +}_{\beta,d}(y_1^-) | P_A \right \rangle \nonumber 
    \\
    & = \left \langle P_A|   F^{\beta+}_{b'}(y_2^-) F^{\alpha+}_{a'}(0^-) F^{\ +,a'}_{\alpha}(y^-) F^{\ +,b'}_{\beta}(y_1^-) | P_A \right \rangle \nonumber \\
    & \times \frac{\delta_{ad}\delta_{bc}}{(N_c^2-1)^2}  \,,
\end{align}
where the factor $(N_c^2-1)^2$ follows from normalization. Then we have
\begin{align}
    &\left \langle  P_A | \Tr\left[   F^{\beta+}(y_2^-) F^{\alpha+}(0^-) F^{\ +}_{\alpha}(y^-) F^{\ +}_{\beta}(y_1^-) \right] | P_A \right \rangle \nonumber \\
    & = \frac{1}{4 N_c} \left \langle P_A|   F^{\beta+}_{b'}(y_2^-) F^{\alpha+}_{a'}(0^-) F^{\ +,a'}_{\alpha}(y^-) F^{\ +,b'}_{\beta}(y_1^-) | P_A \right \rangle \,,
    \label{eq:fundamental-adjoint-correlator}
\end{align}
where we used $\frac{\delta_{ad}\delta_{bc}}{(N_c^2-1)^2} \Tr\left[  t^a t^b t^c t^d  \right] =  \frac{N_c C_F^2}{{(N_c^2-1)^2}} = \frac{1}{4 N_c}$\,. Inserting Eq.\,\eqref{eq:fundamental-adjoint-correlator} into Eq.\,\eqref{eq:T4-initial-state-distr-CGC-2} concludes with the proof of the equivalence between 
Eqs.\,\eqref{eq:T4-initial-state-distr} and \eqref{eq:T4-initial-state-distr-CGC}, and thus shows the consistency between CGC with sub-eikonal phases and high-twist formalism. The same procedure can be carried out to show the equivalence for all the other contributions shown in Table\,\ref{tb::xsection-HT}.

\section{Summary and Outlook}
\label{sec:conclusion-perspectives}

In this manuscript, by studying direct photon production in proton-nucleus collisions, we elucidated the correspondence between two well-known frameworks for describing the scattering processes in nuclear media: the High-Twist expansion formalism and the Color Glass Condensate (CGC) effective theory. We investigated the kinematic regime where they share a common domain of validity where the transverse momentum of the photon $p_{\gamma\perp} \gtrsim Q_s$, and $Q_s$ is the saturation scale which characterizes the typical transverse momentum of partons in the nuclear medium. In Sec.\,\ref{sec:high-twist formalism} we gave a detailed account of the calculation of direct photon production in the high-twist expansion formalism. Specifically, we have outlined the calculation of the differential cross section, starting from the leading twist contribution involving single scattering. Furthermore, we have presented a complete calculation at the next-to-leading twist, where we presented a detailed account of the initial-state scattering contribution with a central cut. Our findings are encapsulated in Eqs.\, (\ref{eq::full-T4},\ref{eq::TX}) and Table.\,\ref{tb::xsection-HT}. Notably, these results include various types of interactions, including initial- and final-state interactions, as well as their interferences, unlike previous studies that only considered initial-state interactions. These results serve as a crucial benchmark for the subsequent sections of our analysis.  In Sec.\,\ref{CGC-formalism} we reviewed the computation of direct photon production in the CGC framework and performed an expansion of the differential cross-section in inverse powers of $p_{\gamma\perp}^2$. At leading twist, we found that this expansion is consistent with the twist-2 collinear result in the strict limit $x \to 0$ as is well-known in the literature. However, the next-to-leading twist expansion of the CGC result does not match the twist-4 result in the high twist expansion formalism, unless all twist-4 distributions have the same behavior at small $x$ and their derivatives obey Eq.\,\eqref{eq:derivative-twist-4-condition}. 

We identify that the culprit of the mismatch between both frameworks is the absence of the longitudinal sub-eikonal phases in the CGC formalism which mediate the transition between incoherent and coherent scattering, the so-called LPM effect. In Sec.\,\ref{sec:CGC-sub-eik} we revisit the direct photon production in the CGC by bringing back sub-eikonal phases that keep explicit the dependence on the longitudinal momentum $x$ carried by each gluon from the target. This required the expansion of the CGC effective vertices (light-like Wilson lines) in terms of the classical gauge field and re-inserting the phases neglected in the eikonal approximation. We then consider single, double and the interference of single-triple scattering contributions to direct photon production, and explicitly show that the twist expansion of these results perfectly matches the results of the high twist expansion at twist-4. A more comprehensive approach would entail considering the next-to-next-to-eikonal (NNEik) corrections (e.g. finite size of the shockwave, sub-leading components of the gauge field, dependence of the light-cone time) to the CGC and perform the collinear expansion. By including sub-eikonal phases we have taken into account the finite extent of the shockwave, namely, that emissions are allowed between nuclear scatterings (LPM effect). Furthermore, the full NNEik calculation will take into account the transverse component of the gauge field (in addition to $A^+$) allowing us to explicitly recover the full field strength tensor. We defer the full NNEik to future studies.

The approach presented in this paper can be applied to other processes in proton-nucleus collisions at RHIC and the LHC, as well as deep inelastic scattering off nuclei at the Electron-Ion Collider. In the future, it would be interesting to explore the matching at all twists by constructing an effective propagator that resum multiple interactions to all orders while keeping the sub-eikonal phase, potentially allowing us to bridge the gap between dilute and dense regions of hadronic matter. Furthermore, we leave for future work the exploration of the interplay between different QCD evolution equations of the CGC, the HT formalism, as well as their transition regime (for a recent development at leading twist see \cite{Mukherjee:2023snp}).  We hope the present work can serve as a basis to incorporate saturation effects in the initial state into event generators such as HIJING \cite{Gyulassy:1994ew} and eHIJING \cite{Ke:2023xeo} based on the high-twist expansion formalism and capture final state interaction effects.

\section*{Acknowledgments}

This work is supported by the NSFC under Grants No. 12035007, No. 12475139 (H. X.), and No. 1935007 (Y. F., X.-N. W.); by the U.S. DOE under Grants No. DE-FG02-05ER41367 (Y. F.), No. DE-AC02-05CH11231 (F. S., X.-N. W.), and within the framework of the SURGE Collaboration (Z.-B. K., X.-N. W.); by the U.S. NSF under Grants No. PHY-1945471 (Z.-B. K., F. S.) and No. OAC-2004571 within the X-SCAPE Collaboration (F. S., X.-N. W.). F. S. is also supported by the Institute for Nuclear Theory of the U.S. DOE under Grant No. DE-FG02-00ER41132.

\begin{widetext}

\appendix

\section{Useful identities}
\label{app:useful-identities}
In this section, we compile some useful identities for the calculations of the cross-section in the CGC, and the computation of the Dirac traces.

\subsection{Differential cross-section from amplitude in the CGC}
\label{app:xsec-amplitude-CGC}

The generic expression for the cross-section for the scattering of a fast-moving particle (moving along the {\it{minus}} light-cone direction) with the background field of a rapidly moving nucleus (moving along the {\it{plus}} light-cone direction), and producing two particles in the final state is given by the cross-section in the CGC
\begin{align}
    \der \sigma^{a+A \to b+c+X} = (2\pi) \delta(p_a^- - p_b^- - p_c^-) \frac{1}{2p_a^-}\left[ \frac{1}{N_{\rm initial}}\sum_{\substack{\rm quantum \\
    \rm numbers}} \left\langle \overline{\Mcal}  \Mcal \right \rangle_x \right] \frac{\der^3 \boldsymbol{p_b}}{(2 E_b )(2\pi)^3} \frac{\der^3 \boldsymbol{p_c}}{(2 E_c)(2\pi)^3} \,,
\end{align}
where $p_a$ is the momentum of the incoming particle, and $p_b$ and $p_c$ are the momenta of the two outgoing particles. $N_{\rm initial}$ represents the degeneracy of the quantum number of the incoming particle. The reduced amplitude $\Mcal$ is defined as
\begin{align}
    (2\pi) \delta(p_a^- - p_b^- - p_c^-) \Mcal = \Scal - \Scal[A_{\rm cl}=0] \,,
\end{align}
where $\Scal$ is the usual scattering matrix computed using standard Feynman rules and the CGC effective vertices.

If we specialize to $q(k) + A \to q(p) + \gamma(p_{\gamma}) + X$:
\begin{align}
    E_{p} E_{\gamma} \frac{\der \sigma^{qA \to q\gamma X}}{\der^3 \boldsymbol{p} \ \der^3 \boldsymbol{p_{\gamma}} } =\frac{1}{(2\pi)^5} \frac{1}{8k^-} \left[\frac{1}{2N_c} \sum_{\lambda \sigma \sigma' ij} \left\langle  \overline{\Mcal}^{\lambda\sigma\sigma'}_{ij} \Mcal^{\lambda\sigma\sigma'}_{ij}    \right \rangle_x \right] \delta(k^- - p^- - p_{\gamma}^-) \,. \label{eq:qgamma-diff-Xsec-CGC}
\end{align}
The factor $N_{\rm initial}= 2 N_c$ comes from averaging over incoming quark spin and color.

Integrating over the phase space of the quark, we find the differential cross-section for $q(k) + A \to \gamma(p_\gamma) + X$
\begin{align}
    E_{\gamma} \frac{\der \sigma^{qA \to \gamma X}}{\der^3 \boldsymbol{p_{\gamma}}}
    &=\frac{1}{(4\pi)^3}   \frac{1}{(k^-)^2 (1-\xi)} \int \frac{\der^2 \pt}{(2\pi)^2} \left[ \frac{1}{2N_c} \sum_{\lambda \sigma \sigma' i j} \left\langle  \overline{\Mcal}^{\lambda\sigma\sigma'}_{ij} \Mcal^{\lambda\sigma\sigma'}_{ij}   \right \rangle_x \right] \,,
\end{align}
where $\xi =p_{\gamma}^-/k^-$. 

Lastly, to evaluate the differential cross-section for photon production in proton-nucleus collisions $p(P_p) + A \to \gamma(p_\gamma) + X$, we convolute with collinear parton distribution function:
\begin{align}
    E_{\gamma} \frac{\der \sigma^{p+A \to \gamma+X}}{\der^3 \boldsymbol{p_{\gamma}}}    
    &=\sum_{q} \int  \frac{\der x_q f_{q/p}(x_q)}{(4\pi)^3 (k^-)^2 (1-\xi)} \int \frac{\der^2 \pt}{(2\pi)^2} \left[\frac{1}{2N_c}   \sum_{\lambda \sigma \sigma' i j} \left\langle  \overline{\Mcal}^{\lambda\sigma\sigma'}_{ij} \Mcal^{\lambda\sigma\sigma'}_{ij}   \right \rangle_x \right]\,,
    \label{eq:A2_to_XSec}
\end{align}
where $x_q  = p_{\gamma}^-/ (\xi P_p^-)$, and $P_p^-$ is the (large) {\it{minus}} light-cone momentum of the proton, and we summed over the light quark flavors.

\subsection{Trace of gamma matrices}
We define the following spinor structure which appears repeatedly in the calculation of the amplitude photon+quark production in $pA$ collision:
\begin{align}
    \GammatLU{\alpha}{\lambda\sigma\sigma'} & =  \bar{u}(p,\sigma)\left[  \gammatL{\alpha} \gammatL{\beta} + (1-\xi) \gammatL{\beta} \gammatL{\alpha} \right] \frac{\gamma^-}{k^-} u(k,\sigma') \etU{\lambda*,\beta} \,,
\end{align}
where we used the polarization vector for the photon
\begin{align}
    \epsilon^{\mu}(p_\gamma) = \left( \frac{\etU{\lambda} \cdot \pgammat}{p_\gamma^-} , 0, \etU{\lambda}\right) \,,
\end{align}
with the two-dimensional transverse vector $\etU{\lambda} = \frac{1}{\sqrt{2}}(1, i \lambda)$.

To compute the cross-section we need the square summed over the quark helicities and the photon polarization:
\begin{align}
    \sum_{\lambda\sigma\sigma'} \GammatLU{\alpha}{\lambda\sigma\sigma'} \GammatCLU{\rho}{\lambda\sigma\sigma'} & = \Tr\left[\slashed{p} \left[  \gammatL{\alpha} \gammatL{\beta} + (1-\xi) \gammatL{\beta} \gammatL{\alpha}  \right] \frac{\gamma^-}{k^-} \slashed{k} \frac{\gamma^-}{k^-} \left[  \gammatL{\delta} \gammatL{\rho} + (1-\xi) \gammatL{\rho} \gammatL{\delta}  \right] \right] \sum_{\lambda} \etU{\lambda*,\beta} \etU{\lambda,\delta} \nonumber \\ 
    & = \Tr\left[\slashed{p} \left[  \gammatL{\alpha} \gammatL{\beta} + (1-\xi) \gammatL{\beta} \gammatL{\alpha}  \right] \frac{\gamma^-}{k^-} \slashed{k} \frac{\gamma^-}{k^-} \left[  \gammatU{\beta} \gammatL{\rho} + (1-\xi) \gammatL{\rho} \gammatU{\beta}  \right] \right] \nonumber \\
    & = 2 (1-\xi)\Tr\left[ \left[  \gammatL{\alpha} \gammatL{\beta} + (1-\xi) \gammatL{\beta} \gammatL{\alpha}  \right]\left[  \gammatU{\beta} \gammatL{\rho} + (1-\xi) \gammatL{\rho} \gammatU{\beta}  \right]   \right] \nonumber \\
    & = -4  (1-\xi)\Tr\left[  \gammatL{\alpha} \gammatL{\rho} + (1-\xi)^2 \gammatL{\alpha} \gammatL{\rho}  \right] \nonumber \\
    & = 16 (1-\xi) \left[  1+ (1-\xi)^2 \right] \delta_{\perp \alpha \rho} \,.
\end{align}
Thus we have:
\begin{align}
    \frac{1}{2} \sum_{\lambda\sigma\sigma'} \GammatLU{\alpha}{\lambda\sigma\sigma'} \GammatCLU{\rho}{\lambda\sigma\sigma'} 
    & = 8 (1-\xi) \left[  1+ (1-\xi)^2 \right] \delta_{\perp \alpha \rho} \,.
    \label{eq:identity1}
\end{align}

\section{On the relation between moments of the CGC dipole, twist-2 and twist-4 collinear distributions}
\label{app:CGC-collinear-dist-correspondance}

We begin this section by reviewing some useful identities of the derivatives of light-like Wilson lines. Then we provide a detailed derivation of the relation between the second and fourth moments of the CGC dipole distribution and the twist-2 and twist-4 distributions at small-$x$:
\begin{align}
     \frac{N_c}{2 \pi^2 \alpha_s }  \int  \frac{\der^2 \lt}{(2\pi)^2} \lt^2 F(x,\lt) \simeq \lim_{x \to 0} x f_{g/A}(x) \,, 
\end{align}
\begin{align}
    \frac{N_c^2}{2(2\pi)^4 \alpha_{s}^2 }  \int \frac{\der^2 \lt}{(2\pi)^2} \lt^4  F(x,\lt) \Bigg{|}_{\mathrm{T4}} \simeq  \lim_{x \to 0} T_{gg}(x,0,0)\,.
\end{align}

\subsection{Derivatives of light-like Wilson lines}
Let us define the gauge link along the light cone:
\begin{align}
    \left[y^-, x^-; \xt \right] =\Pcal \left \{ \exp \left[ ig \int_{x^-}^{y^-} \der z^- A^+(z^-,\xt) \right] \right \} \,,
\end{align}
where $\Pcal$ is the path ordering operator such
\begin{align}
    \Pcal\left[ \Ocal(y_1^-) \Ocal(y_2^-) \right] =  \Ocal(y_1^-) \Ocal(y_2^-) \theta(y_1^- - y_2^-) + \Ocal(y_2^-) \Ocal(y_1^-) \theta(y_2^- - y_1^-) \,.
\end{align}
The infinite light-like Wilson lines that appear in the CGC are simply
\begin{align}
    V(\yt) = \left[\infty, -\infty; \yt \right] \,. \label{eq:Vinfinite}
\end{align}
The derivative of Wilson lines is obtained by commuting the derivative and the path ordering operator, one obtains:
\begin{align}
    \frac{\partial V(\yt)}{\partial \ytL{\alpha}} = ig \int_{-\infty}^{\infty} \der y_1^-  \left[\infty, y_1^-; \yt \right] F^{\alpha+}(y_1^-,\yt) \left[y_1^-, -\infty; \yt \right] \,,
    \label{eq:dVinfinite}
\end{align}
for the first derivative, and 
\begin{align}
    & \frac{\partial^2 V(\yt)}{\partial \ytL{\alpha} \partial \ytL{\beta}} = +ig \int_{-\infty}^{\infty} \der y_1^- \left[\infty, y_1^-; \yt \right] \frac{\partial F^{\beta+}(y_1^-,\yt)}{\partial \ytL{\alpha}} \left[y_1^-, -\infty; \yt \right] \nonumber \\
    & -g^2 \left\{  \int_{-\infty}^{\infty} \der y_1^- \int_{-\infty}^{\infty} \der y_2^- \left[ \infty, y_1^-; \yt \right]F^{\alpha+}(y_1^-,\yt) \left[ y_1^-, y_2^-; \yt \right] F^{\beta+}(y_2^-,\yt) \left[ y_2^-, -\infty; \yt \right] \theta(y_1^- - y_2^-) + (\alpha \leftrightarrow \beta) \right\} \,,
    \label{eq:ddVinfinite}
\end{align}
for the second derivative, where $F^{\alpha+}(y^-,\yt) = \frac{\partial A^+(y^-, \yt)}{\partial \ytL{\alpha}}$.

\subsection{Second moment of the CGC dipole and the twist-2 collinear distribution}
The second moment of the CGC dipole can be written in terms of derivatives of light-like Wilson lines at the same transverse location:
\begin{align}
    \int  \frac{\der^2 \lt}{(2\pi)^2} \lt^2 F(x,\lt) &=   \int  \frac{\der^2 \lt}{(2\pi)^2} \lt^2  \int \der^2 \yt \int \der^2 \yt'    e^{-i \lt \cdot (\yt -\yt')} \frac{1}{N_c} \left \langle \Tr\left[V(\yt) V^\dagger(\yt')\right] \right \rangle_{x} \nonumber \\
    & = \int  \frac{\der^2 \lt}{(2\pi)^2} \int \der^2 \yt \int \der^2 \yt'    \left[ \frac{\partial^2 e^{-i \lt \cdot (\yt -\yt')}}{\partial \ytU{\alpha} \partial \ytCL{\alpha}}  \right] \frac{1}{N_c}  \left \langle \Tr\left[V(\yt) V^\dagger(\yt')\right] \right \rangle_{x} \nonumber \\
    &  =  \int  \frac{\der^2 \lt}{(2\pi)^2}  \int \der^2 \yt \int \der^2 \yt'     e^{-i \lt \cdot (\yt -\yt')}  \frac{1}{N_c}  \left \langle \Tr\left[ \frac{\partial V(\yt)}{\partial \ytU{\alpha} }  \frac{\partial V^\dagger(\ytC)}{\partial \ytCL{\alpha} }  \right] \right \rangle_{x}  \nonumber \\
    &  = \int \der^2 \yt  \frac{1}{N_c}  \left \langle \Tr\left[ \frac{\partial V(\yt)}{\partial \ytU{\alpha} }  \frac{\partial V^\dagger(\yt)}{\partial \ytL{\alpha} }  \right] \right \rangle_{x} \,.
\end{align} 
Next inserting 
\begin{align}
    \int  \frac{\der^2 \lt}{(2\pi)^2} \lt^2 F(x,\lt) &= \frac{g^2}{N_c} \int_{-\infty}^{\infty} \der y_2^- \int_{-\infty}^{\infty} \der y_1^- \int \der^2 \yt   \left \langle \Tr\left[   F^{\alpha+}(y_1^-,\yt) \left[y_1^-, y_2^-; \yt \right] F^{\ +}_{\alpha}(y_2^-,\yt) \left[y_2^-, y_1^-; \yt \right]    \right] \right \rangle_{x}
\end{align}
where we used the cyclic property of the trace and the property. Furthermore, using the translational invariance of the CGC correlator: 
\begin{align}
     \int  \frac{\der^2 \lt}{(2\pi)^2} \lt^2 F(x,\lt) &= \frac{g^2 V}{N_c} \int_{-\infty}^{\infty} \der y_1^-   \left \langle \Tr\left[   F^{\alpha+}(y_1^-) \left[y_1^-,  0^- \right] F^{\ +}_{\alpha}(0^-)  \left[0^-, y_1^-\right]  \right] \right \rangle_{x}
\end{align}
where we used the relation between the CGC ensemble average and the hadronic matrix element
\begin{align}
    \left\langle \Ocal \right \rangle_x \leftrightarrow \frac{\left \langle P_A| \Ocal|P_A\right \rangle}{\left \langle P_A|P_A\right \rangle}  = \frac{\left \langle P_A| \Ocal|P_A\right \rangle}{(2P_A^+) V} 
    \label{eq:normalizationCGCtoPP} \,,
\end{align}
with $V$ stands for the three-volume.
\begin{align}
     \int  \frac{\der^2 \lt}{(2\pi)^2} \lt^2 F(x,\lt) &= \frac{\pi \alpha_s }{N_c P_A^+} \int_{-\infty}^{\infty} \der y_1^-   \left \langle P_A |     F_a^{\alpha+}(y_1^-) \left[y_1^-,  0^- \right] F^{\ +,a}_{\alpha}(0^-)  \left[0^-, y_1^-\right]   |P_A \right \rangle \nonumber \\
     & = \frac{2 \pi^2 \alpha_s }{N_c} x f_{g/A}(x) \,.
\end{align}
\subsection{Fourth moment of the dipole distribution and the twist-4 collinear distribution}
Following the same calculation as in the second moment, we find that the fourth moment of the dipole can be written as
\begin{align}
    \int  \frac{\der^2 \lt}{(2\pi)^2} \lt^4 F(x,\lt) \Bigg{|}_{\mathrm{T4}}
    &  =\int \der^2 \yt  \frac{1}{N_c}  \left \langle \Tr\left[ \frac{\partial^2 V(\yt)}{\partial \ytU{\alpha} \partial \ytU{\beta} }  \frac{\partial^2 V^\dagger(\yt)}{\partial \ytL{\alpha} \partial \ytL{\beta}}   \right] \right \rangle_{x} \,.
\end{align} 
Inserting, 
\begin{align}
    &\int  \frac{\der^2 \lt}{(2\pi)^2} \lt^4 F(x,\lt) \Bigg{|}_{\mathrm{T4}} \nonumber \\
    & = \frac{g^4}{2 P^+_A N_c}  \int_{-\infty}^{\infty} \der y_1^- \int_{-\infty}^{\infty} \der y_2^- \int_{-\infty}^{\infty} \der y_3^- \ \left \langle P_A \big| \Tr\left[ F^{\alpha+}(y_1^-)  F^{\beta+} (y_2^-) F_{\beta}^{\ +}(y_3^-)  F_{\alpha}^{\ +} (0^-)  \right] \big| P_A \right \rangle \theta(y_1^- - y_2^-) \theta(- y_3^-) \nonumber \\
    & + \frac{g^4}{2 P^+_A N_c}  \int_{-\infty}^{\infty} \der y_1^- \int_{-\infty}^{\infty} \der y_2^- \int_{-\infty}^{\infty} \der y_3^- \ \left \langle P_A \big|  \Tr\left[   F^{\beta+} (y_2^-) F^{\alpha+}(y_1^-) F_{\beta}^{\ +}(y_3^-)  F_{\alpha}^{\ +} (0^-)  \right] \big| P_A \right \rangle \theta(y_2^- - y_1^-) \theta(- y_3^-) \nonumber \\
    & + \frac{g^4}{2 P^+_A N_c}  \int_{-\infty}^{\infty} \der y_1^- \int_{-\infty}^{\infty} \der y_2^- \int_{-\infty}^{\infty} \der y_3^- \ \left \langle P_A \big| \Tr\left[ F^{\alpha+}(y_1^-)  F^{\beta+} (y_2^-)   F_{\alpha}^{\ +} (0^-)  F_{\beta}^{\ +}(y_3^-) \right] \big| P_A \right \rangle  \theta(y_1^- - y_2^-) \theta(y_3^-) \nonumber \\
    & + \frac{g^4}{2 P^+_A N_c}  \int_{-\infty}^{\infty} \der y_1^- \int_{-\infty}^{\infty} \der y_2^- \int_{-\infty}^{\infty} \der y_3^- \ \left \langle P_A \big| \Tr\left[   F^{\beta+} (y_2^-) F^{\alpha+}(y_1^-)  F_{\alpha}^{\ +} (0^-)  F_{\beta}^{\ +}(y_3^-) \right] \big| P_A \right \rangle  \theta(y_2^- - y_1^-) \theta(y_3^-) \nonumber \\
    & -\frac{ig^3}{2 P_A^+ N_c}  \int_{-\infty}^{\infty} \der y_1^- \int_{-\infty}^{\infty} \der y_2^-  \ \left \langle P_A \big| \Tr\left[ \partial_\perp^{\alpha} F^{\beta +}(y_1^-)   F_{\beta}^{\ +} (y_2^-)  F_{\alpha}^{\ +}(0^-) \right] \big| P_A \right \rangle  \theta(-y_2^-) \nonumber \\
    & -\frac{ig^3}{2 P_A^+ N_c}  \int_{-\infty}^{\infty} \der y_1^- \int_{-\infty}^{\infty} \der y_2^-  \ \left \langle P_A \big| \Tr\left[ \partial_\perp^{\alpha} F^{\beta +}(y_1^-)   F_{\alpha}^{\ +} (0^-)  F_{\beta}^{\ +}(y_2^-) \right] \big| P_A \right \rangle  \theta(y_2^-) \nonumber \\
    & +\frac{ig^3}{2 P_A^+ N_c}  \int_{-\infty}^{\infty} \der y_1^- \int_{-\infty}^{\infty} \der y_2^-  \ \left \langle P_A \big| \Tr\left[ F_{\alpha}^{\ +} (y_2^-)  F_{\beta}^{\ +}(0^-) \partial_\perp^{\alpha} F^{\beta +}(y_1^-) \right] \big| P_A \right \rangle 
 \theta(y_2^-) \nonumber\\ 
    & +\frac{ig^3}{2 P_A^+ N_c}  \int_{-\infty}^{\infty} \der y_1^- \int_{-\infty}^{\infty} \der y_2^-  \ \left \langle P_A \big| \Tr\left[  F_{\beta}^{\ +} (0^-)  F_{\alpha}^{\ +}(y_2^-) \partial_\perp^{\alpha} F^{\beta +}(y_1^-)   \right] \big| P_A \right \rangle  \theta(-y_2^-) \nonumber \\
    & + \frac{g^2}{2 P^+_A N_c} \int_{-\infty}^{\infty} \der y_1^-   \ \left \langle P_A \big| \Tr\left[   \partial_{\perp \alpha} F^{\beta+}(y_1^-) \partial_\perp^{\alpha} F^{\ +}_{\beta}(0^-) \right] \big| P_A \right \rangle  \,,
\end{align} 
where we have kept the gauge links implicit to keep the expression in a compact form.

We observe that the four first terms correspond to the four twist-4 distributions for the central cut. Defining
\begin{align}
    T_{gg}(x_1,x_2,x_3) = \frac{1}{4}\left[ T_{C,I}(x_1,x_2,x_3) + T_{C,IF}(x_1,x_2,x_3) + T_{C,FI}(x_1,x_2,x_3) + T_{C,F}(x_1,x_2,x_3)\right]
\end{align}
Then we have:
\begin{align}
    \int  \frac{\der^2 \lt}{(2\pi)^2} \lt^4 F(x,\lt) \Bigg{|}_{\mathrm{T4}}  = \frac{2(2\pi)^4\alpha_s^2 }{N_c^2} T_{gg}(x,0,0) + \dots
\end{align}
where $\dots$ indicate that we have neglected terms involving derivatives of the strength field tensor. These terms are not enhanced by the nuclear size; hence, they are neglected in the high-twist approach. They would correspond to contributions in which the single scattering cross-section (and interference with double scattering) is expanded to twist-4.

\section{Interference contribution: Triple-single scattering}
\label{app:triple-single-interference}
In this section, we compute the single-triple scattering interference contribution to direct photon production in photon nucleus collisions in the CGC with the sub-eikonal phase. We present the calculation of the amplitudes for triple scattering in Sec.\,\ref{sec:amplitude-triple}, and the differential cross-section is computed in Sec.\,\ref{sec:differential-Xsec-triple-single}.

\subsection{Amplitudes for triple scattering}
\label{sec:amplitude-triple}
At the amplitude level, we must evaluate the four diagrams shown in Fig.\,\ref{fig:asymmetric_cut_diagrams}. Two gluon momenta $l$ and $L$ are unconstrained and must be integrated over. As in the double scattering case, we integrate over the $-$ and $+$ and light-cone components of $l$ and $L$. The integration over the minus components is carried out with the help of the delta functions, and the integration over $+$ components is done via Contour integration, the results for the amplitudes are:
\begin{figure}[H]
    \centering
    \includegraphics[width=0.48\textwidth]{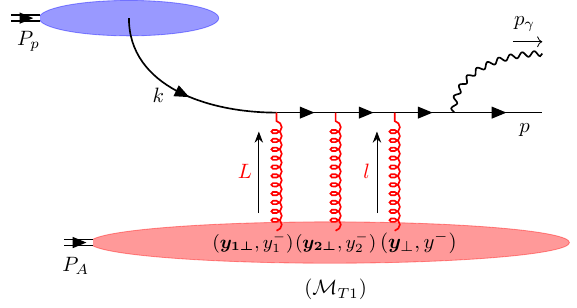}\includegraphics[width=0.48\textwidth]{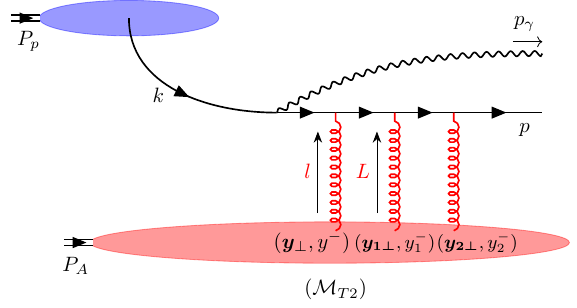}
    \includegraphics[width=0.48\textwidth]{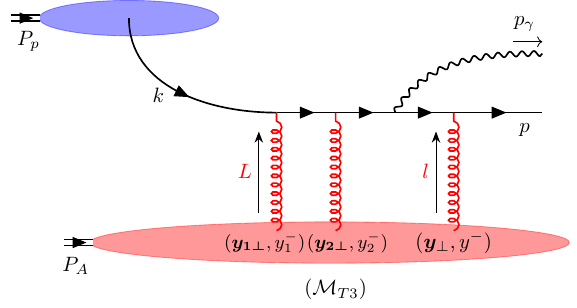}\includegraphics[width=0.48\textwidth]{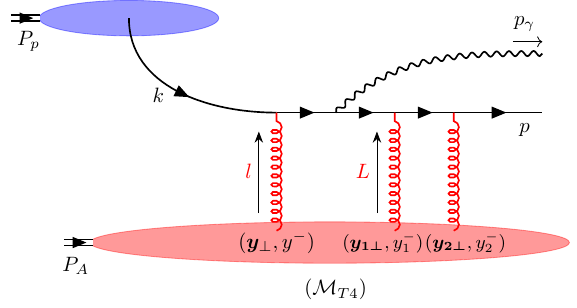}
    \caption{Triple scattering diagrams to quark + photon production in proton-nucleus collisions. The momenta $l$ and $L$ of two gluons are unconstrained and must be integrated over. By momentum conservation, the other gluon carries momentum $p+p_\gamma -k -l-L$.}
    \label{fig:asymmetric_cut_diagrams}
\end{figure}
\begin{align}
    &\Mcal_{T1}^{\lambda \sigma \sigma'}  = -i e g^3 \int_{y,y_1,y_2} \!\!\!\!\!\!\!\!\!\!\!\!\! A^{+}(y)  A^{+}(y_2) A^{+}(y_1) \theta(y^- - y_2^-) \theta(y_2^- - y_1^-) \int \frac{\der^2 \lt}{(2\pi)^2} \frac{\der^2 \Lt}{(2\pi)^2} e^{-i \lt \cdot \yt} e^{-i (\pt + \pgammat -\lt-\Lt) \cdot \yttwo} \Ncal^{\lambda\sigma\sigma'}_{T1}\nonumber \\
    & \times e^{-i \Lt \cdot \ytone } e^{i \left[ \frac{\xi \pt^2 +(1-\xi)\pgammatU{2} -\xi(1-\xi)(\pt + \pgammat -\lt)^2}{\pgammatU{2}}\right] x P^+_A y^-}  e^{i \left[ \frac{\xi(1-\xi)((\pt + \pgammat -\lt)^2 - \Lt^2)}{\pgammatU{2}} \right] x P^+_A y_2^- } e^{i \frac{\xi(1-\xi)\Lt^2}{\pgammatU{2}} x P^+_A y_1^-}   \,,
    \label{eq:M1_AsymmetricCut}
\end{align}
\begin{align}
    &\Mcal_{T2}^{\lambda \sigma \sigma'} = -ie g^3 \int_{y_2,y_1,y} \!\!\!\!\!\!\!\!\!\!\!\!\! A^{+}(y_2)  A^{+}(y_1) A^{+}(y) \theta(y_2^- - y_1^-) \theta(y_1^- - y^-) \int \frac{\der^2 \lt}{(2\pi)^2} \frac{\der^2 \Lt}{(2\pi)^2}
    e^{-i (\pt + \pgammat -\lt-\Lt) \cdot \yttwo} e^{-i \Lt \cdot \ytone}  \Ncal^{\lambda\sigma\sigma'}_{T2} \nonumber \\
    & \times e^{-i \lt \cdot \yt} e^{i \left[ \frac{\xi(\pt^2 - (\pgammat-\lt -\Lt)^2 )}{\pgammatU{2}} \right] x P^+_A y_2^-} e^{i\left[ \frac{\xi((\pgammat - \lt -\Lt)^2 - (\pgammat-\lt)^2 )}{\pgammatU{2}}  \right] x P^+_A  y_1^- } e^{i \left[ \frac{(1-\xi)\pgammatU{2} + \xi (\pgammat-\lt)^2 }{\pgammatU{2}} \right] x P^+_A  y^-} \,,
    \label{eq:M2_AsymmetricCut}
\end{align}
\begin{align}
    & \Mcal_{T3}^{\lambda \sigma \sigma'} = -ie g^3 \int_{y,y_2,y_1} \!\!\!\!\!\!\!\!\!\!\!\!\! A^{+}(y)  A^{+}(y_2) A^{+}(y_1) \theta(y^- - y_2^-) \theta(y_2^- - y_1^-) \int \frac{\der^2 \lt}{(2\pi)^2} \frac{\der^2 \Lt}{(2\pi)^2} e^{-i \lt \cdot \yt} e^{-i (\pt + \pgammat -\lt-\Lt)  \cdot \yttwo} \Ncal^{\lambda\sigma\sigma'}_{T3} \nonumber \\
    & \times e^{-i \Lt \cdot \ytone} \Bigg \{  e^{i \left\{\left[\frac{\xi \pt^2+ (1-\xi) \pgammatU{2} - \xi (1-\xi)(\pt + \pgammat -\lt)^2}{\pgammatU{2}}\right] y^- + \left[\frac{\xi(1-\xi)(\pt + \pgammat -\lt)^2 - \xi(1-\xi) \Lt^2}{\pgammatU{2}} \right] y_2^- + \frac{\xi(1-\xi) \Lt^2}{\pgammatU{2}} y_1^- \right\} x P_A^+ } \nonumber \\
    & - e^{i  \left\{ \left[\frac{\xi \pt^2  - \xi (\pt-\lt)^2}{\pgammatU{2}}\right] y^- + \left[ \frac{\xi (\pt-\lt)^2 + (1-\xi) \pgammatU{2} - \xi (1-\xi) \Lt^2}{\pgammatU{2}} \right] y_2^-  + \frac{\xi (1-\xi) \Lt^2}{\pgammatU{2}} y_1^- \right\} x P_A^+ }  \Bigg \}\,,
    \label{eq:M3_AsymmetricCut}
\end{align}
\begin{align}
    &\Mcal_{T4}^{\lambda \sigma \sigma'} = -ie g^3 \int_{y_2,y_1,y} \!\!\!\!\!\!\!\!\!\!\!\!\! A^{+}(y_2)  A^{+}(y_1) A^{+}(y) \theta(y_2^- - y_1^-) \theta(y_1^- - y^-) \int \frac{\der^2 \lt}{(2\pi)^2} \frac{\der^2 \Lt}{(2\pi)^2}  e^{-i (\pt + \pgammat -\lt-\Lt) \cdot \yttwo} e^{-i \Lt \cdot \ytone}   \Ncal^{\lambda\sigma\sigma'}_{T4} \nonumber \\
    & \times e^{-i\lt \cdot \yt} \Bigg \{  e^{i \left\{  \left[ \frac{\xi \pt^2 - \xi (\pgammat-\lt-\Lt)^2}{\pgammatU{2}}\right] y_2^- + \left[ \frac{\xi (\pgammat-\lt-\Lt)^2-\xi (\pgammat-\lt)^2}{\pgammatU{2}}\right] y_1^- + \left[\frac{(1-\xi) \pgammatU{2}+\xi(\pgammat-\lt)^2}{\pgammatU{2}}\right] y^- \right\} x P_A^+ }  \nonumber \\
    & - e^{i  \left\{ \left[\frac{\xi \pt^2 - \xi (\pgammat-\lt-\Lt)^2}{\pgammatU{2}} \right] y_2^- + \left[\frac{ (1-\xi) \pgammatU{2} +\xi (\pgammat-\lt-\Lt)^2 -\xi(1-\xi) \lt^2}{\pgammatU{2}} \right] y_1^- + \frac{\xi (1-\xi) \lt^2}{\pgammatU{2}} y^- \right\} x P_A^+ }  \Bigg \}\,.
    \label{eq:M4_AsymmetricCut}
\end{align}
The corresponding perturbative factors are:
\begin{align}
    \Ncal^{\lambda\sigma\sigma'}_{T1}&=\frac{\left[ \xi \ptU{\alpha} -(1-\xi)\pgammatU{\alpha}\right]}{\left[ \xi \pt -(1-\xi)\pgammat\right]^2} \GammatLU{\alpha}{\lambda\sigma\sigma'} \,, \nonumber \\
    \Ncal^{\lambda\sigma\sigma'}_{T2}&=\frac{ \pgammatU{\alpha}}{ \pgammatU{2}} \GammatLU{\alpha}{\lambda\sigma\sigma'} \,, \nonumber \\
    \Ncal^{\lambda\sigma\sigma'}_{T3} & = \frac{\left[ \xi \lt^{\alpha}- (\xi \ptU{\alpha} -(1-\xi)\pgammatU{\alpha} )\right] }{  \left[ \xi\lt - (\xi\pt - (1-\xi)\pgammat )\right]^2} \GammatLU{\alpha}{\lambda\sigma\sigma'} \,, \nonumber \\
    \Ncal^{\lambda\sigma\sigma'}_{T4}&= \frac{\left[ \xi \ltU{\alpha} - \pgammatU{\alpha}\right] }{  \left[ \xi \lt - \pgammat \right]^2} \GammatLU{\alpha}{\lambda\sigma\sigma'} \,.
    \label{eq:double-pert-factor-TN}
\end{align}
Each of the amplitudes $\Mcal_{T3}$ and $\Mcal_{T4}$ are the sum of two terms corresponding to enclosing two different poles, or equivalently to setting on-shell the different quark propagators adjacent the photon emission. In the sub-eikonal limit, these contributions vanish exactly. The amplitudes $\Mcal_{T3}$ and $\Mcal_{T4}$ corresponding to emissions between scatterings display the LPM effect:
\begin{align}
     \Mcal_{T3}^{\lambda \sigma \sigma'}  \propto &\ e^{i \frac{\xi(1-\xi) \Lt^2}{\pgammatU{2}} y_1^-  x P_A^+} e^{i \left[\frac{\xi \pt^2+ (1-\xi) \pgammatU{2} - \xi (1-\xi)(\pt + \pgammat -\lt)^2}{\pgammatU{2}}\right] y^- x P_A^+} e^{i  \left[\frac{\xi(1-\xi)(\pt + \pgammat -\lt)^2 - \xi(1-\xi) \Lt^2}{\pgammatU{2}} \right] y_2^- x P_A^+ }  \nonumber \\
    & \times \Bigg \{ 1 - e^{-i \frac{(\pgammat - \xi \ellt)^2}{\pgammatU{2}} (y^- - y_2^-)   x P_A^+ }  \Bigg \} \,, \\
    \Mcal_{T4}^{\lambda \sigma \sigma'}  \propto  &\  e^{i \left\{  \left[ \frac{\xi \pt^2 - \xi (\pgammat-\lt-\Lt)^2}{\pgammatU{2}}\right] y_2^- + \left[ \frac{\xi (\pgammat-\lt-\Lt)^2-\xi (\pgammat-\lt)^2}{\pgammatU{2}}\right] y_1^- + \left[\frac{(1-\xi) \pgammatU{2}+\xi(\pgammat-\lt)^2}{\pgammatU{2}}\right] y^- \right\} x P_A^+ } \nonumber \\
    & \times \Bigg \{ 1 - e^{i \frac{(\pgammat - \xi \lt)^2}{\pgammatU{2}} (y_1^- - y^-)   x P_A^+ }  \Bigg \} \,.
\end{align}
As we did before, we can separate Eqs.\,\eqref{eq:M3_AsymmetricCut} and \eqref{eq:M4_AsymmetricCut} into two contributions $\Mcal_{T2} =  \Mcal_{T3a} + \Mcal_{T3b}$ and $\Mcal_{T4} =  \Mcal_{T4a} + \Mcal_{T4b}$ respectively \footnote{In the definition of $\Mcal_{T3b}$, we swapped the variables $y_2 \leftrightarrow y$, and $\lt \leftrightarrow  \pt + \pgammat -\lt -\Lt$. While in the definition of $\Mcal_{T4b}$, we swapped the variables $y_1 \leftrightarrow y$, and $\lt \leftrightarrow \Lt$.}:
\begin{align}
    & \Mcal_{T3a}^{\lambda \sigma \sigma'} = -ie g^3 \int_{y,y_2,y_1}\!\!\!\!\!\!\!\!\!\!\!\! A^{+}(y)  A^{+}(y_2) A^{+}(y_1) \theta(y^- - y_2^-) \theta(y_2^- - y_1^-) \int \frac{\der^2 \lt}{(2\pi)^2} \frac{\der^2 \Lt}{(2\pi)^2} e^{-i\lt \cdot \yt} e^{-i (\pt + \pgammat -\lt-\Lt) \cdot \yttwo} \Ncal^{\lambda\sigma\sigma'}_{T3a} \nonumber \\
    & \times e^{-i \Lt \cdot \ytone} e^{i \left[ \frac{\xi \pt^2 +(1-\xi)\pgammatU{2} -\xi(1-\xi)(\pt + \pgammat -\lt)^2}{\pgammatU{2}}\right] x P^+_A y^-}  e^{i \left[ \frac{\xi(1-\xi)((\pt + \pgammat -\lt)^2 - \Lt^2)}{\pgammatU{2}} \right] x P^+_A y_2^- } e^{i \frac{\xi (1-\xi)\Lt^2}{\pgammat{2}} x P^+_A y_1^-}  \,,
\end{align}
\begin{align}
    & \Mcal_{T3b}^{\lambda \sigma \sigma'} = -ie g^3 \int_{y_2,y,y_1}\!\!\!\!\!\!\!\!\!\!\!\! A^{+}(y_2)  A^{+}(y) A^{+}(y_1) \theta(y_2^- - y^-) \theta(y^- - y_1^-) \int \frac{\der^2 \lt}{(2\pi)^2} \frac{\der^2 \Lt}{(2\pi)^2} e^{-i (\pt + \pgammat -\lt-\Lt) \cdot \yttwo} e^{-i \lt \cdot \yt} \Ncal^{\lambda\sigma\sigma'}_{T3b} \nonumber \\
    & \times e^{-i \Lt \cdot \ytone} e^{i \left[ \frac{\xi(\pt^2 - (\pgammat-\lt -\Lt)^2 )}{\pgammatU{2}} \right] x P^+_A y_2^-}  e^{i \left[ \frac{\xi(\pgammat -\lt -\Lt)^2 + (1-\xi)\pgammatU{2} - \xi(1-\xi)\Lt^2 }{\pgammatU{2}}  \right] x P_A^+ y^- } e^{i \frac{\xi (1-\xi)\Lt^2}{\pgammatU{2}} x P^+_A y_1^-} \,,
\end{align}
\begin{align}
    &\Mcal_{T4a}^{\lambda \sigma \sigma'}= -ie g^3 \int_{y_2,y_1,y} \!\!\!\!\!\!\!\!\!\!\!\! A^{+}(y_2)  A^{+}(y_1) A^{+}(y) \theta(y_2^- - y^-) \theta(y^- - y_1^-) \int \frac{\der^2 \lt}{(2\pi)^2} \frac{\der^2 \Lt}{(2\pi)^2} e^{i (\pt + \pgammat -\lt-\Lt) \cdot \yttwo} e^{i \Lt \cdot \ytone}   \Ncal^{\lambda\sigma\sigma'}_{T4a} \nonumber \\
    & \times e^{i\lt \cdot \yt} e^{i \left[ \frac{\xi(\pt^2 - (\pgammat-\lt -\Lt)^2 )}{\pgammatU{2}} \right] x P^+_A y_2^-} e^{i\left[ \frac{\xi((\pgammat - \lt -\Lt)^2 - (\pgammat-\lt)^2 )}{\pgammatU{2}}  \right] x P^+_A  y_1^- } e^{i \left[ \frac{(1-\xi)\pgammatU{2} + \xi (\pgammat-\lt)^2 }{\pgammatU{2}} \right] x P^+_A  y^-}  \,,
\end{align}
\begin{align}
    & \Mcal_{T4b}^{\lambda \sigma \sigma'} = -ie g^3 \int_{y_2,y,y_1} \!\!\!\!\!\!\!\!\!\!\!\! A^{+}(y_2)  A^{+}(y) A^{+}(y_1) \theta(y_2^- - y^-) \theta(y^- - y_1^-) \int \frac{\der^2 \lt}{(2\pi)^2} \frac{\der^2 \Lt}{(2\pi)^2} e^{i (\pt + \pgammat -\lt-\Lt) \cdot \yttwo} e^{i \Lt \cdot \yt}  \Ncal^{\lambda\sigma\sigma'}_{T4b} \nonumber \\
    & \times e^{i\lt \cdot \ytone}  e^{i \left[ \frac{\xi(\pt^2 - (\pgammat-\lt -\Lt)^2 )}{\pgammatU{2}} \right] x P^+_A y_2^-}  e^{i \left[ \frac{\xi(\pgammat -\lt -\Lt)^2 + (1-\xi)\pgammatU{2} - \xi(1-\xi)\Lt^2 }{\pgammatU{2}}  \right] x P_A^+ y^- } e^{i \frac{\xi (1-\xi)\Lt^2}{\pgammatU{2}} x P^+_A y_1^-}  \,,
\end{align}
where the perturbative factors read
\begin{align}
    \Ncal^{\lambda\sigma\sigma'}_{T3a} & = \frac{\left[ \xi \lt^{\alpha}- (\xi \ptU{\alpha} -(1-\xi)\pgammatU{\alpha} )\right] }{  \left[ \xi\lt - (\xi\pt - (1-\xi)\pgammat )\right]^2} \GammatLU{\alpha}{\lambda\sigma\sigma'} \,, \nonumber \\    \Ncal^{\lambda\sigma\sigma'}_{T3b} &= \frac{\left[ \xi \lt^{\alpha}  - (\pgammatU{\alpha} -\xi \LtU{\alpha}) \right] }{  \left[ \xi\lt -(\pgammat -\xi \Lt)\right]^2} \GammatLU{\alpha}{\lambda\sigma\sigma'} \,, \nonumber \\
    \Ncal^{\lambda\sigma\sigma'}_{T4a} &= \frac{\left[ \xi \ltU{\alpha} - \pgammatU{\alpha}\right] }{  \left[ \xi \lt - \pgammat \right]^2} \GammatLU{\alpha}{\lambda\sigma\sigma'} \,, \nonumber \\    \Ncal^{\lambda\sigma\sigma'}_{T4b} &= \frac{\left[  \pgammatU{\alpha} - \xi \LtU{\alpha}\right] }{  \left[ \pgammat -\xi \Lt  \right]^2} \GammatLU{\alpha}{\lambda\sigma\sigma'} \,.
    \label{eq:double-pert-factor-TN2}
\end{align}
This decomposition suggests that we should define
\begin{align}
    \Mcal_{T,I} =& \Mcal_{T1} + \Mcal_{T3a} \,, \\
    \Mcal_{3,F} =& \Mcal_{T2} + \Mcal_{T4a} \,, \\
    \Mcal_{T,FI} =& \Mcal_{T3b} + \Mcal_{T4b} \,.
\end{align}
Then we have
\begin{align}
    \Mcal_{T,I}^{\lambda \sigma \sigma'}
    &= -ie g^3 \int_{y,y_2,y_1} \!\!\!\!\!\!\!\! \Acal^{\lambda\sigma\sigma'}_{T,I}(y,y_2,y_1) \theta(y^- - y_2^-) \theta(y_2^- - y_1^-)   A^{+}(y)  A^{+}(y_2) A^{+}(y_1)  \,,
\end{align}
\begin{align}
    \Mcal_{T,F}^{\lambda \sigma \sigma'} 
    &= -ie g^3 \int_{y_2,y_1,y} \!\!\!\!\!\!\!\!  \Acal^{\lambda\sigma\sigma'}_{T,F}(y_2,y_1,y) \theta(y_2^- - y_1^-) \theta(y_1^- - y^-) A^{+}(y_2)  A^{+}(y_1) A^{+}(y)   \,,
\end{align}
\begin{align}
    \Mcal_{T,FI}^{\lambda \sigma \sigma'} 
    &= -ie g^3 \int_{y_2,y,y_1} \!\!\!\!\!\!\!\! \Acal^{\lambda\sigma\sigma'}_{T,FI}(y_2,y,y_1) \theta(y_2^- - y^-) \theta(y^- - y_1^-)   A^{+}(y_2)  A^{+}(y) A^{+}(y_1) \,,
\end{align}
where we define the perturbative factors:
\begin{align}
     \Acal^{\lambda\sigma\sigma'}_{T,I}(y,y_2,y_1)
    = &\int \frac{\der^2 \lt}{(2\pi)^2} \frac{\der^2 \Lt}{(2\pi)^2} \left\{\Ncal^{\lambda\sigma\sigma'}_{T1} + \Ncal^{\lambda\sigma\sigma'}_{T3a} \right\} e^{-i\lt \cdot \yt} e^{-i (\pt + \pgammat -\lt-\Lt) \cdot \yttwo} e^{-i \Lt \cdot \ytone}\nonumber \\
    & \times  e^{i \left[ \frac{\xi \pt^2 +(1-\xi)\pgammatU{2} -\xi(1-\xi)(\pt + \pgammat -\lt)^2}{\pgammatU{2}}\right] x P^+_A y^-}  e^{i \left[ \frac{\xi(1-\xi)((\pt + \pgammat -\lt)^2 - \Lt^2)}{\pgammatU{2}} \right] x P^+_A y_2^- } e^{i \frac{\xi (1-\xi)\Lt^2}{\pgammatU{2}} x P^+_A y_1^-} \,,
    \label{eq:perturbative-ATI}
\end{align}
\begin{align}
    \Acal^{\lambda\sigma\sigma'}_{T,F}(y_2,y_1,y) = & \int \frac{\der^2 \lt}{(2\pi)^2} \frac{\der^2 \Lt}{(2\pi)^2} \left\{\Ncal^{\lambda\sigma\sigma'}_{T2} + \Ncal^{\lambda\sigma\sigma'}_{T4a} \right\} e^{-i (\pt + \pgammat -\lt-\Lt) \cdot \yttwo} e^{-i \Lt \cdot \ytone} e^{-i\lt \cdot \yt}  \nonumber \\
    & \times e^{i \left[ \frac{\xi(\pt^2 - (\pgammat-\lt -\Lt)^2 )}{\pgammatU{2}} \right] x P^+_A y_2^-} e^{i\left[ \frac{\xi((\pgammat - \lt -\Lt)^2 - (\pgammat-\lt)^2 )}{\pgammatU{2}}  \right] x P^+_A  y_1^- } e^{i \left[ \frac{(1-\xi)\pgammatU{2} + \xi (\pgammat-\lt)^2 }{\pgammatU{2}} \right] x P^+_A  y^-} \,,
    \label{eq:perturbative-ATF}
\end{align}
\begin{align}
    \Acal^{\lambda\sigma\sigma'}_{T,FI}(y_2,y,y_1) =& \int \frac{\der^2 \lt}{(2\pi)^2} \frac{\der^2 \Lt}{(2\pi)^2} \left\{\Ncal^{\lambda\sigma\sigma'}_{T3b} + \Ncal^{\lambda\sigma\sigma'}_{T4b} \right\} e^{-i (\pt + \pgammat -\lt-\Lt) \cdot \yttwo} e^{-i\lt \cdot \yt} e^{-i \Lt \cdot \ytone} \nonumber \\
    & \times e^{i \left[ \frac{\xi(\pt^2 - (\pgammat-\lt -\Lt)^2 )}{\pgammatU{2}} \right] x P^+_A y_2^-}  e^{i \left[ \frac{\xi(\pgammat -\lt -\Lt)^2 + (1-\xi)\pgammatU{2} - \xi(1-\xi)\Lt^2 }{\pgammatU{2}}  \right] x P_A^+ y^- } e^{i \frac{\xi (1-\xi)\Lt^2}{\pgammatU{2}} x P^+_A y_1^-} \,.
    \label{eq:perturbative-ATFI}
\end{align}
\subsection{Differential cross-section}
\label{sec:differential-Xsec-triple-single}
The differential cross-section has three contributions for the triple-single interference ``left cut" \footnote{When constructing the contributions for the left cut we have interchanged $y_1$ and $y_2$ (and their corresponding momenta) to follow the same convention as in the high-twist expansion calculation.}:
\begin{align}
    E_{\gamma} \frac{\der^3\sigma^{\rm CGC_{sub}}_{TS,\rm{left\ cut}}}{\der^3 \boldsymbol{p_{\gamma}}}    
    =\int  \frac{\der x_q f_{q/p}(x_q)}{(4\pi)^3 (1-\xi)} \int \frac{\der^2 \pt}{(2\pi)^2} \frac{1}{2N_c} \left[\sum_{\lambda \sigma \sigma' i j} \left\langle  \left(\overline{\Mcal}_{T,I} + \overline{\Mcal}_{T,F} + \overline{\Mcal}_{T,FI} 
 \right) \Mcal_{S}   \right \rangle_x \right]\,.
 \label{eq:triple-single-contribution}
\end{align}
These three contributions are:
\begin{align}
    & E_{\gamma} \frac{\der^3\sigma^{\mathrm{CGC_{sub}}}_{L,I}}{\der^3 \boldsymbol{p_{\gamma}}}
    = \frac{\alpha_{em} \alpha_{s}^2}{N_c}\sum_q e_q^2  \int \der x_q  f_{q/p}(x_q) \int_{y,y',y_1,y_2} \!\!\!\!\!\!\!\!\!\!\!\!\!\!\! \theta(y'^- - y_1^-) \theta(y_1^- - y_2^-)  \left \langle  \Tr\left[  A^{+}(y_2)  A^{+}(y_1) A^{+}(y')  A^{+}(y)     \right] \right \rangle_x \Hcal_{L,I} \,, \\
    & E_{\gamma} \frac{\der^3\sigma^{\mathrm{CGC_{sub}}}_{L,F}}{\der^3 \boldsymbol{p_{\gamma}}}
    = \frac{\alpha_{em} \alpha_{s}^2}{N_c}\sum_q e_q^2  \int \der x_q  f_{q/p}(x_q) \int_{y,y',y_1,y_2} \!\!\!\!\!\!\!\!\!\!\!\!\!\!\! \theta(y_1^- - y_2^-) \theta(y_2^- - y'^-)  \left \langle  \Tr\left[  A^{+}(y')  A^{+}(y_2) A^{+}(y_1)  A^{+}(y)     \right] \right \rangle_x \Hcal_{L,F} \,, \\
    & E_{\gamma} \frac{\der^3\sigma^{\mathrm{CGC_{sub}}}_{L,FI}}{\der^3 \boldsymbol{p_{\gamma}}}
    = \frac{\alpha_{em} \alpha_{s}^2}{N_c}\sum_q e_q^2  \int \der x_q  f_{q/p}(x_q) \int_{y,y',y_1,y_2} \!\!\!\!\!\!\!\!\!\!\!\!\!\!\! \theta(y_1^- - y'^-) \theta(y'^- - y_2^-) \left \langle  \Tr\left[  A^{+}(y_2)  A^{+}(y') A^{+}(y_1)  A^{+}(y)     \right] \right \rangle_x \Hcal_{L,FI}  \,,
    \label{eq:single-triple_scattering_XSec_Initial_Hard-Left}
\end{align}
where
\begin{align}
    & \Hcal_{L,I}
    = -\frac{1}{(4\pi)}  \int \frac{\der^2 \pt}{(2\pi)^2}  \frac{1}{(1-\xi)} \frac{1}{2} \sum_{\lambda \sigma \sigma'} \overline{\Acal}_{T,I}^{\lambda\sigma\sigma'}(y',y_1,y_2)\Acal_{S}^{\lambda\sigma\sigma'}(y) \,,
\end{align}
and similar expressions for other perturbative factors. For the explicit expressions of the perturbative factors see Sec.\,\ref{app:pert-factor-LC}. 

The differential cross-section has three contributions to the single-triple interference ``right cut":
\begin{align}
    E_{\gamma} \frac{\der^3\sigma^{\rm CGC_{sub}}_{ST,\rm{right\ cut}}}{\der^3 \boldsymbol{p_{\gamma}}}    
    =\int  \frac{\der x_q f_{q/p}(x_q)}{(4\pi)^3 (1-\xi)} \int \frac{\der^2 \pt}{(2\pi)^2} \frac{1}{2 N_c} \left[\sum_{\lambda \sigma \sigma' i j} \left\langle  \overline{\Mcal}_{S} \left(\Mcal_{T,I} + \Mcal_{T,F} + \Mcal_{T,FI} 
 \right)    \right \rangle_x \right]\,.
 \label{eq:single-triple-contribution}
\end{align}
These three contributions are:
\begin{align}
    & E_{\gamma} \frac{\der^3\sigma^{\mathrm{CGC_{sub}}}_{R,I}}{\der^3 \boldsymbol{p_{\gamma}}}
    = \frac{\alpha_{em} \alpha_{s}^2}{N_c}\sum_q e_q^2  \int \der x_q  f_{q/p}(x_q) \int_{y,y',y_1,y_2} \!\!\!\!\!\!\!\!\!\!\!\!\!\!\! \theta(y^- - y_2^-) \theta(y_2^- - y_1^-)   \left \langle  \Tr\left[  A^{+}(y')  A^{+}(y) A^{+}(y_2)  A^{+}(y_1)     \right] \right \rangle_x \Hcal_{R,I} \,, \\
    & E_{\gamma} \frac{\der^3\sigma^{\mathrm{CGC_{sub}}}_{R,F}}{\der^3 \boldsymbol{p_{\gamma}}}
    = \frac{\alpha_{em} \alpha_{s}^2}{N_c}\sum_q e_q^2  \int \der x_q  f_{q/p}(x_q) \int_{y,y',y_1,y_2} \!\!\!\!\!\!\!\!\!\!\!\!\!\!\! \theta(y_2^- - y_1^-) \theta(y_1^- - y^-)  \left \langle  \Tr\left[  A^{+}(y')  A^{+}(y_2) A^{+}(y_1)  A^{+}(y)     \right] \right \rangle_x \Hcal_{R,F} \,, \\
    & E_{\gamma} \frac{\der^3\sigma^{\mathrm{CGC_{sub}}}_{R,IF}}{\der^3 \boldsymbol{p_{\gamma}}}
    = \frac{\alpha_{em} \alpha_{s}^2}{N_c}\sum_q e_q^2  \int \der x_q  f_{q/p}(x_q) \int_{y,y',y_1,y_2} \!\!\!\!\!\!\!\!\!\!\!\!\!\!\! \theta(y_2^- - y^-) \theta(y^- - y_1^-)  \left \langle  \Tr\left[  A^{+}(y')  A^{+}(y_2) A^{+}(y)  A^{+}(y_1)     \right] \right \rangle_x \Hcal_{R,IF} \,,
\end{align}
where
\begin{align}
    & \Hcal_{R,I}
    = -\frac{1}{(4\pi)} \int \frac{\der^2 \pt}{(2\pi)^2}  \frac{1}{(1-\xi)} \frac{1}{2} \sum_{\lambda \sigma \sigma'} \overline{\Acal}_{S}^{\lambda\sigma\sigma'}(y')\Acal_{T,I}^{\lambda\sigma\sigma'}(y,y_2,y_1)\,,
    \label{eq:single-triple_scattering_XSec_Initial_Hard-Right}
\end{align}
and similar expressions for other perturbative factors. For the explicit expressions of the perturbative factors see Sec.\,\ref{app:pert-factor-RC}. We note that compared to the double scattering contribution (c.f. Eq.\,\eqref{eq:double_scattering_XSec_Initial_Hard}), the perturbative factors for triple-single interference in Eq.\,\eqref{eq:single-triple_scattering_XSec_Initial_Hard-Right} possess an additional overall minus sign, which is characteristic of interference contributions.

\section{Perturbative factors and their collinear expansions}
\label{app:Hard_factors}
In this section, we collect all the perturbative factors in calculating the double-scattering (c.f. Sec.\,\ref{sec:double_scattering_central_phase}) and triple-single-scattering (c.f. Sec.\,\ref{app:triple-single-interference}) contributions (left and right cuts). 
\subsection{Central-cut}
\label{app:pert-factor-CC}
The perturbative factors for double scattering are defined:
\begin{align}
    & \left\{ \Hcal_{C,I}, \Hcal_{C,F}, \Hcal_{C,IF}, \Hcal_{C,FI} \right\} = \frac{1}{4\pi} \int \frac{\der^2 \pt}{(2\pi)^2}  \frac{1}{1-\xi} \frac{1}{2}  \nonumber \\
    & \times \sum_{\lambda \sigma \sigma'} \left\{ \overline{\Acal}_{D,I}^{\lambda\sigma\sigma'}(y',y_2)\Acal_{D,I}^{\lambda\sigma\sigma'}(y,y_1), \overline{\Acal}_{D,F}^{\lambda\sigma\sigma'}(y_2,y')\Acal_{D,F}^{\lambda\sigma\sigma'}(y_1,y), \overline{\Acal}_{D,I}^{\lambda\sigma\sigma'}(y',y_2)\Acal_{D,F}^{\lambda\sigma\sigma'}(y_1,y), \overline{\Acal}_{D,F}^{\lambda\sigma\sigma'}(y_2,y')\Acal_{D,I}^{\lambda\sigma\sigma'}(y,y_1)    \right\}\,,
\end{align}
where $\Acal_{D,I}$ and $\Acal_{D,F}$ are defined in Eqs.\,(\eqref{eq:perturbative-ADI},\eqref{eq:perturbative-ADF}), together with Eqs.\,(\eqref{eq:double-pert-factor-N},\eqref{eq:double-pert-factor-N2}). The sum over $\lambda, \sigma$ and $\sigma'$ is performed using the identity in Eq.\,\eqref{eq:identity1}, and performed the change of variables $\pt \to \Lt = \pt + \pgammat$. We find the following results for the double scattering contribution (central cut):
\begin{align}
    \mathcal{H}_{C,X} &= 8 \left[1 + (1-\xi)^2 \right] \!\! \int \!\! \frac{\der^2 \Lt}{(2\pi)^2} \!\! \int \!\! \frac{\der^2 \lt}{(2\pi)^2} \!\! \int \!\! \frac{\der^2 \lt'}{(2\pi)^2} e^{-i \left[\lt  \cdot \yt - \lt'  \cdot \yt' + (\Lt -\lt) \cdot \ytone -(\Lt -\lt') \cdot \yttwo    \right]}  \mathcal{N}_{C,X}\,,
\end{align}
where
\begin{align}
    & \mathcal{N}_{C,I} = \left\{\frac{\left[ \xi \ltCL{\alpha}  - (\xi \LtL{\alpha}  -\pgammatL{\alpha}) \right]}{\left[\xi\ltC -( \xi \Lt - \pgammat ) \right]^2} + \frac{\left[ \xi \LtL{\alpha} - \pgammatL{\alpha} \right]}{\left[ \xi \Lt -\pgammat \right]^2}  \right\} \left\{\frac{\left[ \xi \lt^{\alpha}  - (\xi \LtU{\alpha}  -\pgammatU{\alpha}) \right]}{\left[\xi\lt -( \xi \Lt - \pgammat ) \right]^2} + \frac{\left[ \xi \LtU{\alpha} - \pgammatU{\alpha} \right]}{\left[ \xi \Lt -\pgammat \right]^2}  \right\} \nonumber \\
    &\times e^{-i \left[ \frac{\xi(1-\xi)(\Lt -\ltC)^2}{\pgammatU{2}} \right]x P_A^+ y_2^-} e^{-i \left[\frac{ \xi (\Lt-\pgammat)^2 + (1-\xi)\pgammatU{2} - \xi(1-\xi) (\Lt -\ltC)^2}{\pgammatU{2}}\right] x P_A^+ y'^-} \nonumber \\
    & \times e^{i \left[\frac{ \xi (\Lt-\pgammat)^2 + (1-\xi)\pgammatU{2} - \xi(1-\xi)(\Lt -\lt)^2}{\pgammatU{2}}\right]x P_A^+ y^-}  e^{i \left[\frac{ \xi(1-\xi)(\Lt -\lt)^2}{\pgammatU{2}} \right] x P_A^+ y_1^-} \,, \\
    & \mathcal{N}_{C,F} = \left\{\frac{(\xi \ltCL{\alpha} -\pgammatL{\alpha})}{\left( \xi \ltC -\pgammat\right)^2}+\frac{\pgammatL{\alpha}}{\pgammatU{2}}  \right \} \left\{\frac{(\xi \ltU{\alpha} -\pgammatU{\alpha})}{\left( \xi \lt -\pgammat\right)^2}+\frac{\pgammatU{\alpha}}{\pgammatU{2}}  \right \}  \nonumber \\
    & \times  e^{-i \left[\frac{ (1-\xi)\pgammatU{2} +\xi(\ltC-\pgammat)^2}{\pgammatU{2}} \right] x P_A^+ y'^-} e^{-i\left[\frac{\xi((\Lt-\pgammat)^2 - (\ltC - \pgammat)^2)}{\pgammatU{2}} \right] x P_A^+ y_2^- }  \nonumber \\
    & \times e^{i\left[\frac{\xi((\Lt-\pgammat)^2 - (\lt - \pgammat)^2)}{\pgammatU{2}} \right] x P_A^+ y_1^- } e^{i \left[\frac{ (1-\xi)\pgammatU{2} +\xi(\lt-\pgammat)^2}{\pgammatU{2}} \right] x P_A^+ y^-} \,, \\
    & \mathcal{N}_{C,IF} = \left\{\frac{(\xi \ltCL{\alpha} -\pgammatL{\alpha})}{\left( \xi \ltC -\pgammat\right)^2}+\frac{\pgammatL{\alpha}}{\pgammatU{2}}  \right \} \left\{\frac{\left[ \xi \lt^{\alpha}  - (\xi \LtU{\alpha}  -\pgammatU{\alpha}) \right]}{\left[\xi\lt -( \xi \Lt - \pgammat ) \right]^2} + \frac{\left[ \xi \LtU{\alpha} - \pgammatU{\alpha} \right]}{\left[ \xi \Lt -\pgammat \right]^2}  \right\} \nonumber \\
    &\times e^{-i \left[\frac{ (1-\xi)\pgammatU{2} +\xi(\ltC-\pgammat)^2}{\pgammatU{2}} \right] x P_A^+ y'^-} e^{-i\left[\frac{\xi((\Lt-\pgammat)^2 - (\ltC - \pgammat)^2)}{\pgammatU{2}} \right] x P_A^+ y_2^- }   \nonumber \\
    & \times e^{i \left[\frac{ \xi (\Lt-\pgammat)^2 + (1-\xi)\pgammatU{2} - \xi(1-\xi)(\Lt -\lt)^2}{\pgammatU{2}}\right]x P_A^+ y^-} e^{i \left[\frac{ \xi(1-\xi)(\Lt -\lt)^2}{\pgammatU{2}} \right] x P_A^+ y_1^-} \,, \\
    &\mathcal{N}_{C,FI} = \left\{\frac{\left[ \xi \ltCL{\alpha}  - (\xi \LtL{\alpha}  -\pgammatL{\alpha}) \right]}{\left[\xi\ltC -( \xi \Lt - \pgammat ) \right]^2} + \frac{\left[ \xi \LtL{\alpha} - \pgammatL{\alpha} \right]}{\left[ \xi \Lt -\pgammat \right]^2}  \right\} \left\{\frac{(\xi \ltU{\alpha} -\pgammatU{\alpha})}{\left( \xi \lt -\pgammat\right)^2}+\frac{\pgammatU{\alpha}}{\pgammatU{2}}  \right \} \nonumber \\
    & \times e^{-i \left[ \frac{\xi(1-\xi)(\Lt -\ltC)^2}{\pgammatU{2}} \right]x P_A^+ y_2^-}  e^{-i \left[\frac{ \xi (\Lt-\pgammat)^2 + (1-\xi)\pgammatU{2} - \xi(1-\xi) (\Lt -\ltC)^2}{\pgammatU{2}}\right] x P_A^+ y'^-} \nonumber \\
    & \times e^{i\left[\frac{\xi((\Lt-\pgammat)^2 - (\lt - \pgammat)^2)}{\pgammatU{2}} \right] x P_A^+ y_1^- } e^{i \left[\frac{ (1-\xi)\pgammatU{2} +\xi(\lt-\pgammat)^2}{\pgammatU{2}} \right] x P_A^+ y^-}  \,.
\end{align}
\subsection{Left-cut}
\label{app:pert-factor-LC}
Similarly, the perturbative factors for the triple-single scattering contribution (left cut) are defined as
\begin{align}
    &\left\{ \Hcal_{L,I}, \Hcal_{L,F}, \Hcal_{L,FI} \right\}
    = -\frac{1}{(4\pi)}  \int \frac{\der^2 \pt}{(2\pi)^2}  \frac{1}{(1-\xi)} \frac{1}{2} \nonumber \\
    & \times \sum_{\lambda \sigma \sigma'} \left\{ \overline{\Acal}_{T,I}^{\lambda\sigma\sigma'}(y',y_1,y_2)\Acal_{S}^{\lambda\sigma\sigma'}(y), \overline{\Acal}_{T,F}^{\lambda\sigma\sigma'}(y_1,y_2,y')\Acal_{S}^{\lambda\sigma\sigma'}(y), \overline{\Acal}_{T,FI}^{\lambda\sigma\sigma'}(y_1,y',y_2)\Acal_{S}^{\lambda\sigma\sigma'}(y) \right\} \,,
    \label{eq:pert-factors-left-cut}
\end{align}
where $\Acal_{T,I}$, $\Acal_{T,F}$ and $\Acal_{T,FI}$ are defined in Eqs.\,(\eqref{eq:perturbative-ATI},\eqref{eq:perturbative-ATF},\eqref{eq:perturbative-ATFI}), together with Eqs.\,(\eqref{eq:double-pert-factor-TN},\eqref{eq:double-pert-factor-TN2}). We conveniently performed the change of variables: $\pt \to \ltC = \pt + \pgammat$. The results read:
\begin{align}
    \mathcal{H}_{L,X} &= 8 \left[1 + (1-\xi)^2 \right] \!\! \int \!\! \frac{\der^2 \Lt}{(2\pi)^2} \!\! \int \!\! \frac{\der^2 \lt}{(2\pi)^2} \!\! \int \!\! \frac{\der^2 \lt'}{(2\pi)^2} e^{-i \left[\lt'  \cdot \yt - \lt  \cdot \yt' - \Lt \cdot \ytone -(\lt' -\lt -\Lt) \cdot \yttwo    \right]}  \mathcal{N}_{L,X}\,,
\end{align}
where
\begin{align}
    & \mathcal{N}_{L,I} = -\left\{\frac{\left[ \xi \lt^{\alpha} - ( \xi \ltCU{\alpha} -\pgammatU{\alpha}) \right]}{\left[\xi\lt -( \xi \ltC - \pgammat ) \right]^2} + \frac{\left[ \xi \ltCU{\alpha} - \pgammatU{\alpha} \right]}{\left[ \xi \ltC - \pgammat \right]^2}  \right\} \left\{\frac{\left[ \xi  \ltCL{\alpha}-\pgammatL{\alpha} \right] }{\left[ \xi  \ltC- \pgammat \right]^2} + \frac{ \pgammatL{\alpha} }{\pgammatU{2}} \right\}
    \nonumber \\
    & \times e^{-i \frac{\xi (1-\xi)\Lt^2}{\pgammatU{2}} x P^+_A y_2^-} e^{-i \left[ \frac{\xi(1-\xi)((\ltC -\lt)^2 - \Lt^2)}{\pgammatU{2}} \right] x P^+_A y_1^- }  e^{-i \left[ \frac{\xi (\ltC-\pgammat)^2 +(1-\xi)\pgammatU{2} -\xi(1-\xi)(\ltC -\lt)^2}{\pgammatU{2}}\right] x P^+_A y'^-}  e^{i x P_A^+ y^-} \,, \\
    & \mathcal{N}_{L,F} = -\left\{\frac{(\xi \ltU{\alpha} -\pgammatU{\alpha})}{\left( \xi \lt -\pgammat\right)^2}+\frac{\pgammatU{\alpha}}{\pgammatU{2}}  \right \} \left\{\frac{\left[ \xi  \ltCL{\alpha}-\pgammatL{\alpha} \right] }{\left[ \xi  \ltC- \pgammat \right]^2} + \frac{ \pgammatL{\alpha} }{\pgammatU{2}} \right\} \nonumber \\
    & \times  e^{-i \left[ \frac{(1-\xi)\pgammatU{2} + \xi (\pgammat-\lt)^2 }{\pgammatU{2}} \right] x P^+_A  y'^-} e^{-i\left[ \frac{\xi((\pgammat - \lt -\Lt)^2 - (\pgammat-\lt)^2 )}{\pgammatU{2}}  \right] x P^+_A  y_2^- }  e^{-i \left[ \frac{\xi( (\ltC-\pgammat)^2 - (\pgammat-\lt -\Lt)^2 )}{\pgammatU{2}} \right] x P^+_A y_1^-} e^{i x P_A^+ y^-} \,, \\
    & \mathcal{N}_{L,FI} = -\left\{ \frac{\left[ \xi \lt^{\alpha}  - (\pgammatU{\alpha} -\xi \LtU{\alpha}) \right] }{  \left[ \xi\lt -(\pgammat -\xi \Lt)\right]^2} + \frac{\left[  \pgammatU{\alpha} - \xi \LtU{\alpha}\right] }{  \left[ \pgammat -\xi \Lt  \right]^2} \right\} \left\{\frac{\left[ \xi  \ltCL{\alpha}-\pgammatL{\alpha} \right] }{\left[ \xi  \ltC- \pgammat \right]^2} + \frac{ \pgammatL{\alpha} }{\pgammatU{2}} \right\}  \nonumber \\
    & \times e^{-i \frac{\xi (1-\xi)\Lt^2}{\pgammatU{2}} x P^+_A y_2^-}  e^{-i \left[  \frac{\xi(\pgammat -\lt -\Lt)^2 + (1-\xi)\pgammatU{2} - \xi(1-\xi)\Lt^2}{\pgammatU{2}} \right] x P_A^+ y'^- }  e^{-i\left[ \frac{\xi (\ltC-\pgammat)^2 - \xi (\pgammat -\lt -\Lt)^2 }{\pgammatU{2}}  \right] x P_A^+ y_1^-}  e^{i x P_A^+ y^-} \,.
\end{align}
\subsection{Right-cut}
\label{app:pert-factor-RC}
Lastly, for the triple-single scattering contribution (right cut) the perturbative factors are defined as
\begin{align}
    & \left\{ \Hcal_{R,I}, \Hcal_{R,F}, \Hcal_{R,IF}
    \right \} = -\frac{1}{(4\pi)} \int \frac{\der^2 \pt}{(2\pi)^2}  \frac{1}{(1-\xi)} \frac{1}{2} \nonumber \\
    & \times \sum_{\lambda \sigma \sigma'} \left\{ \overline{\Acal}_{S}^{\lambda\sigma\sigma'}(y')\Acal_{T,I}^{\lambda\sigma\sigma'}(y,y_2,y_1), \overline{\Acal}_{S}^{\lambda\sigma\sigma'}(y')\Acal_{T,F}^{\lambda\sigma\sigma'}(y_2,y_1,y), \overline{\Acal}_{S}^{\lambda\sigma\sigma'}(y')\Acal_{T,FI}^{\lambda\sigma\sigma'}(y_2,y,y_1) \right\} \,.
\end{align}
These are simply the complex conjugate of the perturbative factors in Eq.\,\eqref{eq:pert-factors-left-cut}. To follow the same convention as the high-twist formalism expansion, we interchange the coordinates ($y \leftrightarrow y'$ and $y_1 \leftrightarrow y_2$), namely:
\begin{align}
    \Hcal_{R,I}(y,y',y_1,y_2) & =  (\Hcal_{L,I}(y',y,y_2,y_1))^{*} \,, \\
    \Hcal_{R,F}(y,y',y_1,y_2) & =  (\Hcal_{L,F}(y',y,y_2,y_1))^{*} \,, \\
    \Hcal_{R,IF}(y,y',y_1,y_2) & =  (\Hcal_{L,FI}(y',y,y_2,y_1))^{*} \,.
\end{align}
For completeness, we state the results:
\begin{align}
    \mathcal{H}_{R,X} &= 8 \left[1 + (1-\xi)^2 \right] \!\! \int \!\! \frac{\der^2 \Lt}{(2\pi)^2} \!\! \int \!\! \frac{\der^2 \lt}{(2\pi)^2} \!\! \int \!\! \frac{\der^2 \lt'}{(2\pi)^2} e^{-i \left[\lt  \cdot \yt - \lt'  \cdot \yt' + \Lt \cdot \ytone +(\lt' -\lt -\Lt) \cdot \yttwo    \right]}  \mathcal{N}_{R,X}\,,
\end{align}
where
\begin{align}
    &\mathcal{N}_{R,I} = -\left\{\frac{\left[ \xi  \ltCL{\alpha} -\pgammatL{\alpha} \right] }{\left[ \xi  \ltC - \pgammat \right]^2} + \frac{ \pgammatL{\alpha} }{\pgammatU{2}} \right\} \left\{\frac{\left[ \xi \lt^{\alpha} - ( \xi \ltCU{\alpha} -\pgammatU{\alpha}) \right]}{\left[\xi\lt -( \xi \ltC - \pgammat ) \right]^2} + \frac{\left[ \xi \ltCU{\alpha} - \pgammatU{\alpha} \right]}{\left[ \xi \ltC - \pgammat \right]^2}  \right\}  
    \nonumber \\
    & \times e^{-i x P_A^+ y'^-} e^{i \left[ \frac{\xi (\ltC-\pgammat)^2 +(1-\xi)\pgammatU{2} -\xi(1-\xi)(\ltC -\lt)^2}{\pgammatU{2}}\right] x P^+_A y^-}  e^{i \left[ \frac{\xi(1-\xi)((\ltC -\lt)^2 - \Lt^2)}{\pgammatU{2}} \right] x P^+_A y_2^- } e^{i \frac{\xi (1-\xi)\Lt^2}{\pgammatU{2}} x P^+_A y_1^-} \,, \\
    &\mathcal{N}_{R,F} = -\left\{\frac{\left[ \xi  \ltCL{\alpha}-\pgammatL{\alpha} \right] }{\left[ \xi  \ltC- \pgammat \right]^2} + \frac{ \pgammatL{\alpha} }{\pgammatU{2}} \right\} \left\{\frac{(\xi \ltU{\alpha} -\pgammatU{\alpha})}{\left( \xi \lt -\pgammat\right)^2}+\frac{\pgammatU{\alpha}}{\pgammatU{2}}  \right \}  \nonumber \\
    & \times e^{-i x P_A^+ y'^-} e^{i \left[ \frac{\xi( (\ltC-\pgammat)^2 - (\pgammat-\lt -\Lt)^2 )}{\pgammatU{2}} \right] x P^+_A y_2^-} e^{i\left[ \frac{\xi((\pgammat - \lt -\Lt)^2 - (\pgammat-\lt)^2 )}{\pgammatU{2}}  \right] x P^+_A  y_1^- } e^{i \left[ \frac{(1-\xi)\pgammatU{2} + \xi (\pgammat-\lt)^2 }{\pgammatU{2}} \right] x P^+_A  y^-} \,, \\
    &\mathcal{N}_{R,IF} = -\left\{\frac{\left[ \xi  \ltCL{\alpha}-\pgammatL{\alpha} \right] }{\left[ \xi  \ltC- \pgammat \right]^2} + \frac{ \pgammatL{\alpha} }{\pgammatU{2}} \right\} \left\{ \frac{\left[ \xi \lt^{\alpha}  - (\pgammatU{\alpha} -\xi \LtU{\alpha}) \right] }{  \left[ \xi\lt -(\pgammat -\xi \Lt)\right]^2} + \frac{\left[  \pgammatU{\alpha} - \xi \LtU{\alpha}\right] }{  \left[ \pgammat -\xi \Lt  \right]^2} \right\}   \nonumber \\
    & \times e^{-i x P_A^+ y'^-} e^{i \left[ \frac{\xi (\ltC-\pgammat)^2-\xi(\pgammat -\lt -\Lt)^2}{\pgammatU{2}}\right] x P_A^+ y_2^-} e^{i \left[ \frac{\xi(\pgammat -\lt -\Lt)^2 + (1-\xi)\pgammatU{2} - \xi (1-\xi) \Lt^2}{\pgammatU{2}} \right] x P_A^+ y^- } e^{i \frac{\xi(1-\xi)\Lt^2}{\pgammatU{2}} x P_A^+ y_1^-} \,.
\end{align}

\subsection{Collinear expansion of perturbative factors}
\label{app:Hard_factors_collinear}
We end this section by providing the results for the collinear expansion of the perturbative factors. We follow the same procedure as described in Sec.\,\ref{sec:twist-2-expansion}. The results are:
\begin{align}
    & \mathcal{H}_{X} = \frac{8 \xi^2 \left[1 + (1-\xi)^2 \right]}{p_{\gamma\perp}^4} e^{i x P_A^+ (y^- - y'^-)}   \delta^{(2)}(\yt-\ytone ) \delta^{(2)}(\ytC-\yttwo ) \delta^{(2)}(\ytone-\yttwo )  \nonumber \\
    & \times  \left[1 + \frac{\mathcal{D}_{X}}{\pgammatU{2}} \ (\partial_{ \ytone} \cdot \partial_{\yttwo}) \right] (\partial_{\yt} \cdot \partial_{\ytC}) + \mathcal{O}(1/\pgammatU{8})
    \,,
\end{align}
where
\begin{align}
    \mathcal{D}_{C,I} &=     \xi^2 (ix P_A^+ \Delta y^-)^2 -3 \xi^2 (ix P_A^+ \Delta y^-) + \xi (1-\xi)(ix P_A^+ \Delta y_{12}^-)  + 4\xi^2 \,, \nonumber \\
    \mathcal{D}_{C,F} & = \xi^2 (ix P_A^+ \Delta y_{12}^-)^2  +  \xi (i x P_A^+ \Delta y_{12}^-) \,, \nonumber  \\
    \mathcal{D}_{C,IF} & =  \xi^2 \left( ix P_A^+ (\Delta y^- - y_2^-) \right)^2 -\xi^2 \left( ix P_A^+ (\Delta y^- - y_2^-) \right)  +\xi (1-\xi) (i x P_A^+ \Delta y_{12}^- ) \,, \nonumber \\
    \mathcal{D}_{C,FI} & =  \xi^2 \left( ix P_A^+ (\Delta y_{12}^- + y_2^-) \right)^2 -\xi^2 \left( ix P_A^+ y_2^- \right)  +\xi (1-2\xi) (i x P_A^+ \Delta y_{12}^- ) \,, \nonumber  \\
    \mathcal{D}_{L,I} & = \mathcal{D}_{R,I} = -\xi (1-\xi)(ix P_A^+ \Delta y_{12}^-) \,, \nonumber  \\
    \mathcal{D}_{L,F} & = \mathcal{D}_{L,I} =  -\xi^2 (ix P_A^+ \Delta y_{12}^-)^2  -  \xi (i x P_A^+ \Delta y_{12}^-) \,, \nonumber  \\ 
    \mathcal{D}_{L,IF} & = -\xi^2 \left( ix P_A^+ (\Delta y^- - y_2^-) \right)^2 + \xi^2 \left( ix P_A^+ (\Delta y^- - y_2^-) \right)  - \xi (1-\xi) (i x P_A^+ \Delta y_{12}^- ) \,, \nonumber  \\
    \mathcal{D}_{R,FI} & = -\xi^2 \left( ix P_A^+ (\Delta y_{12}^- + y_2^-) \right)^2   +\xi^2 \left( ix P_A^+ y_2^- \right) - \xi (1-2\xi) (i x P_A^+ \Delta y_{12}^- ) \,.
\end{align}
\end{widetext}

\bibliographystyle{JHEP-2modlong.bst}

\bibliography{refs}

\end{document}